\documentclass[aps, prd, onecolumn, tightenlines, notitlepage, superscriptaddress, nofootinbib, preprintnumbers, floatfix,showkeys,11pt,altaffilletter]{revtex4-2}

\usepackage{upgreek}

\usepackage[official]{eurosym}
\usepackage[normalem]{ulem}
\usepackage{amstext}
\usepackage{graphicx}
\graphicspath{{figures}}
\usepackage{url}
\usepackage{color}
\usepackage{ulem}
\usepackage[version=4]{mhchem}
\usepackage[utf8]{inputenc}
\usepackage{fontawesome}
\usepackage{yfonts}
\usepackage{slashed}

\pdfoutput=1
\usepackage{textcomp}
\usepackage{comment}
\usepackage{yfonts}
\usepackage{epsfig,amsfonts,mathrsfs,graphicx,color,slashed,multirow}
\usepackage{latexsym,graphicx,slashed,color,enumerate,url,cancel,gensymb}
\usepackage{textcomp}

\usepackage[x11names]{xcolor}
\usepackage[colorlinks,pdfstartview=FitV,breaklinks=true]{hyperref}
\usepackage{booktabs}
\usepackage{adjustbox}
\usepackage{textgreek} 
\usepackage{caption, subcaption} 
\usepackage{booktabs} 
\usepackage{float}

\usepackage{lmodern}
\usepackage{ae,aecompl}
\usepackage{appendix}
\usepackage{orcidlink}
\usepackage{nccmath} 
\usepackage{textcomp}
\allowdisplaybreaks 


\def\cevns{CE\textnu NS}
\def\eves{E\textnu ES}
\def\d{\mathrm{d}}
\newcommand{\qtransfer}{\left|\mathbf{q}\right|}

\definecolor{vdrgreen}{rgb}{0.0, 0.6, 0.0}

\definecolor{hanblue}{rgb}{0.27, 0.42, 0.81}
\definecolor{byzantium}{rgb}{0.44, 0.16, 0.39}

\makeatletter
    \newcommand{\colorboxed}[3][white]{\fcolorbox{#2}{#1}{\m@th$\displaystyle#3$}}
\makeatother

\AtBeginDocument{\hypersetup{citecolor=hanblue,linkcolor=hanblue,urlcolor=hanblue}}
\usepackage{appendix}

\begin{document}

\title{{\LARGE Implications of the first CONUS+ measurement\\ of coherent elastic neutrino-nucleus scattering }}

\author{V. De Romeri~\orcidlink{0000-0003-3585-7437}}
\email{deromeri@ific.uv.es}
\affiliation{Instituto de F\'{i}sica Corpuscular (CSIC-Universitat de Val\`{e}ncia), Parc Cient\'ific UV C/ Catedr\'atico Jos\'e Beltr\'an, 2 E-46980 Paterna (Valencia), Spain}

\author{D. K. Papoulias~\orcidlink{0000-0003-0453-8492}}\email{dipapou@ific.uv.es}
\affiliation{Instituto de F\'{i}sica Corpuscular (CSIC-Universitat de Val\`{e}ncia), Parc Cient\'ific UV C/ Catedr\'atico Jos\'e Beltr\'an, 2 E-46980 Paterna (Valencia), Spain}

\author{G. Sanchez Garcia~\orcidlink{0000-0003-1830-2325}}\email{gsanchez@ific.uv.es}
\affiliation{Instituto de F\'{i}sica Corpuscular (CSIC-Universitat de Val\`{e}ncia), Parc Cient\'ific UV C/ Catedr\'atico Jos\'e Beltr\'an, 2 E-46980 Paterna (Valencia), Spain}
\affiliation{Departament de F\'isica  Te\'orica,  Universitat  de  Val\`{e}ncia, Spain}


\begin{abstract}
The CONUS+ collaboration has reported their first observation of coherent elastic neutrino-nucleus scattering (\cevns).
The experiment uses reactor electron antineutrinos and germanium detectors with recoil thresholds as low as $160~\mathrm{eV_\text{ee}}$. With an exposure of $327$ kg $\times$ d, the measurement was made with a statistical significance of $3.7 \sigma$. We explore several physics implications of this observation, both within the standard model and in the context of new physics. We focus on a determination of the weak mixing angle, nonstandard and generalized neutrino interactions both with heavy and light mediators, neutrino magnetic moments, and the up-scattering of neutrinos into sterile fermions through the sterile dipole portal and new mediators. 
Our results highlight the role of reactor-based \cevns~experiments in probing a vast array of neutrino properties and new physics models.
\end{abstract}

\maketitle


\section{Introduction}
Coherent elastic neutrino-nucleus scattering (\cevns) is a neutral-current process in which a neutrino interacts with an entire nucleus. The scattering occurs by the exchange of a mediating Z boson within the standard model (SM), as first theoretically predicted~\cite{Freedman:1973yd,Kopeliovich:1974mv}. 
The coherence condition requires a $\sim$MeV neutrino source and implies a very low momentum transfer, which, however, translates into an enhanced cross section that scales with the square of the number of target neutrons ($N$).
This feature allows for more compact detectors compared to standard neutrino experiments. On the other hand, the expected signature of \cevns~is a nuclear recoil of very low momentum, making its detection challenging from an experimental point of view~\cite{Freedman:1973yd,Drukier:1984vhf}. The advent of detectors with keV thresholds has now allowed \cevns~to be realistically included among neutrino detection channels, in addition to the long-standing channels inverse beta decay and elastic neutrino-electron scattering (\eves). \cevns~measurements have so far inspired a wave of phenomenological studies, addressing both SM and new physics (see, for instance, the reviews~\cite{Abdullah:2022zue,Cadeddu:2023tkp}).

The first \cevns~detection was made in 2017 by the COHERENT experiment~\cite{COHERENT:2017ipa}, which exploited an intense spallation source that produces neutrinos from pions decaying at rest. Other observations by the same collaboration have followed the pioneering one, using different target materials, with the goal of exploring the expected $N^2$dependence of the cross section: CsI (in 2017~\cite{COHERENT:2017ipa} and in 2021~\cite{COHERENT:2021xmm}), LAr (in 2020~\cite{COHERENT:2020ybo}) and Ge (in 2024~\cite{Adamski:2024yqt}). 
Among natural sources, the Sun generates large neutrino fluxes; in particular those from the $^8$B chain meet the required features to induce sizeable \cevns~events at dark matter (DM) direct detection facilities~\cite{Goodman:1984dc,Schumann:2019eaa,Billard:2021uyg}. The sensitivity and low thresholds reached in the latest generation of low-threshold dual-phase liquid xenon experiments have recently allowed  XENONnT~\cite{XENON:2024ijk} and PandaX-4T~\cite{PandaX:2024muv} collaborations to report their first indications of nuclear recoils from $^8$B solar neutrinos via \cevns, with interesting phenomenological implications~\cite{AristizabalSierra:2024nwf,Li:2024iij,Xia:2024ytb,Maity:2024aji,DeRomeri:2024iaw,DeRomeri:2024hvc,Blanco-Mas:2024ale}. 

Nuclear reactors are another viable artificial source of low-energy neutrinos. They generate intense fluxes of electron antineutrinos with energies lower than those of spallation sources, thus ensuring that the coherence condition is fully preserved. In this regime, the uncertainties related to finite nuclear size effects are dramatically reduced, and the sensitivity to tiny nuclear recoils makes them specially suitable to probe spectral distortions that may be induced by new physics or nontrivial electromagnetic properties in the neutrino sector. For all these reasons, in recent years the community has devoted a great deal of experimental effort to try to observe \cevns~with reactor neutrinos, see,  e.g.,~\cite{Kerman:2016jqp,CONNIE:2021ggh,CONUS:2020skt,nGeN:2022uje,NUCLEUS:2019igx,MINER:2016igy,Ricochet:2021rjo,Akimov:2022xvr,NEON:2022hbk,NEWS-G:2021mhf,SBC:2021yal}. In 2022, a first indication of \cevns~from reactor neutrinos was reported, using an ultra-low-noise germanium detector and neutrinos from the Dresden-II reactor~\cite{Colaresi:2022obx}. 
Because this detection relies solely on the ionization signal, one major source of uncertainty in this measurement arises from to the quenching factor, which describes the reduction in ionization energy generated by nuclear recoils compared to electron recoils of the same energy. The Dresden-II result already inspired several phenomenological studies, e.g.,~\cite{AristizabalSierra:2022axl,Coloma:2022avw,Liao:2022hno,Majumdar:2022nby,AtzoriCorona:2022qrf,Denton:2022nol}.

Very recently, the CONUS+ collaboration has released their first result showing the observation of
a \cevns~signal, with a statistical significance of $3.7 \sigma$~\cite{Ackermann:2025obx}. CONUS+~\cite{CONUS:2024lnu} is the continuation of the CONUS experiment~\cite{CONUS:2020skt}, using electron antineutrinos generated by the Leibstadt nuclear power plant in Switzerland and four point contact high-purity
germanium (HPGe) detectors with optimized low-energy
thresholds of about $160$ eV$_\text{ee}$ [in electron-equivalent ($ee$) energy]. This result has extended physics implications, ranging from SM physics, nuclear physics, astrophysics, and the existence of physics beyond the standard model (BSM). Some of these implications had already been addressed with the results from the previous CONUS experiment~\cite{CONUS:2021dwh,CONUS:2022qbb,Lindner:2024eng}, and have recently been updated right after the announcement of the new CONUS+ result~\cite{Alpizar-Venegas:2025wor,Chattaraj:2025fvx,AtzoriCorona:2025ygn}.

In this paper, we explore the implications of the CONUS+ measurement~\cite{Ackermann:2025obx} in terms of a determination of the weak mixing angle at the $\sim 10$ MeV scale, neutrino nonstandard interactions (NSIs), new light scalar and vector mediators, neutrino generalized interactions (NGIs), neutrino magnetic moments, and the up-scattering production of a sterile fermion through the sterile dipole portal and through new interactions. We complement previous analyses presented in~\cite{Alpizar-Venegas:2025wor,Chattaraj:2025fvx} by exploring additional BSM scenarios. Specifically, we consider flavor-changing NSI, the simultaneous presence of multiple NGI, the sterile dipole scenario, and the neutrino up-scattering into a sterile fermion through a scalar or a vector mediator. We perform a statistical analysis based on spectral information. Since in some of the BSM scenarios the \eves~events are not negligible, we further include them in those analyses.

Our paper is organized as follows. In Sec.~\ref{sec:theory}, we provide the theoretical framework with all relevant cross sections. In Sec.~\ref{sec:analysis}, we describe the CONUS+ result and our statistical analysis. In Sec.~\ref{sec:results}, we present the results of our phenomenological studies, for all scenarios under consideration. Finally, in Sec.~\ref{sec:conclusions}, we draw our conclusions.

\section{Theoretical framework}
\label{sec:theory}
In this section, we present the relevant cross sections for both \cevns~and \eves~in the theoretical scenarios under consideration. We start with the SM cross sections, and then we consider the cases of NSIs, NGIs, light mediators, sterile neutrinos, and the up-scattering dipole portal.

\subsection{\cevns~cross section in the standard model}
\label{subsec:CEvNS-SM}

The \cevns~differential cross section within the SM, given in terms of nuclear recoil energy $T_\mathcal{N}$,  reads~\cite{Freedman:1973yd,Barranco:2005yy}
\begin{equation}
\label{eq:xsec_CEvNS_SM}
\left. \frac{\d \sigma_{\nu \mathcal{N}}}{\d T_\mathcal{N}}\right|^\mathrm{SM}_{\rm CE\nu NS}=\frac{G_F^2 m_\mathcal{N}}{\pi}\left({Q_V^\mathrm{SM}}\right)^2 F_{W}^2(\qtransfer^2)\left(1-\frac{m_\mathcal{N} T_\mathcal{N}}{2E_\nu^2} - \frac{T_\mathcal{N}}{E_\nu} \right) \, ,
\end{equation}
where $G_F$ is the Fermi constant, $E_\nu$ is the reactor antineutrino energy, and $m_\mathcal{N}$ denotes the nuclear mass. Note that we are neglecting subleading terms of order $T_\mathcal{N}/m_\mathcal{N}$ and $\mathcal{O}(T_\mathcal{N}^2)$. Moreover, $Q_V^\text{SM}$ is the SM weak charge, given by

\begin{equation}
\label{eq:CEvNS_SM_Qw}
    Q_V^\text{SM} = g_V^p Z + g_V^n N \, .
\end{equation}
Here, $Z~(N)$ is the proton (neutron) number, and we assume the following proton and neutron couplings (at tree level): $g_V^p = (1- 4 \sin^2 \theta_W)/2$ and $ g_V^n = -1/2$. Flavor-dependent corrections appearing at higher order mostly affect $g_V^p$~\cite{Cadeddu:2020lky} and can be safely neglected. The weak charge $Q_V^\text{SM}$ encodes both the $N^2$ dependence and the dependence on the weak mixing angle $\theta_W$, the latter through the---subdominant---proton contribution. The theoretically predicted value, from Renormalization Group Equation (RGE) running, in the low-energy regime is $\sin^2 \theta_W=0.23857(5)$~\cite{ParticleDataGroup:2024cfk}. 
Finally, we assume the Klein-Nystrand parametrization~\cite{Klein:1999qj} for the nuclear form factor $F_{W}^2(\qtransfer^2)$,

\begin{equation}
  \label{eq:KNFF}
   F_{W}(\qtransfer^2)=3\frac{j_1(\qtransfer R_A)}{\qtransfer R_A} \left(\frac{1}{1+\qtransfer^2a_k^2} \right)\ ,
\end{equation}
where $j_1(x)=\sin(x)/x^2-\cos(x)/x$ is the spherical Bessel function of order one, $a_k = 0.7$~fm and $R_A = 1.23 \, A^{1/3}$ is the root mean square radius (in fm), with $A$ being the atomic mass number.

\subsection{\eves~cross section in the standard model}
\label{subsec:EvES-SM}

Within the SM, the \eves~cross section on an atomic nucleus $\mathcal{A}$ with $Z$ protons is obtained as
$Z_\text{eff}(E_\mathrm{er})$ times the cross section of a neutrino scattering off a single electron.  $Z_\text{eff}(E_\mathrm{er})$ accounts for the effective number of protons seen by the neutrino for an energy deposition $E_\mathrm{er}$.
The flavor-dependent differential \eves~cross section then reads
\begin{align}
\label{eq:xsec_EvES_SM}
\begin{split}
\frac{\d \sigma_{\nu \mathcal{A}}}{\d E_\mathrm{er}}\Big|_\mathrm{E \nu ES}^\mathrm{SM}=& Z^\mathcal{A}_\mathrm{eff}(E_\mathrm{er}) \frac{G_F^2m_e}{2\pi}\left[(g_V + g_A)^2 +(g_V - g_A)^2\left(1-\frac{E_\mathrm{er}}{E_\nu}\right)^2 -\left( g_V^2-g_A^2 \right)\frac{m_e E_\mathrm{er}}{E_\nu^2}\right] \, ,
\end{split}
\end{align}
where $m_e$ is the mass of the electron, and $g_V^\mathrm{SM}=2 \sin^2 \theta_W + 1/2$ and $g_A^\mathrm{SM}=-1/2$, since both charged and neutral currents contribute to the process for electron antineutrinos.  We take the effective charges $Z_\text{eff}(E_\mathrm{er})$ for germanium from Ref.~\cite{AtzoriCorona:2022qrf}.

\subsection{Effective neutrino magnetic moment}
\label{subsec:xsec-magmom}

Neutrinos are considered as massless and neutral particles within the SM. However, extensions to this model that account for massive neutrinos can give rise to loop interactions that couple neutrinos to photons. As a result, several electromagnetic properties can be associated with neutrinos~\cite{Giunti:2014ixa}.  In the context of the recent CONUS+ result, implications for neutrino millicharges and neutrino charge radius have been studied in Ref.~\cite{Chattaraj:2025fvx}. Here, we focus on neutrino magnetic moments,  also explored in Refs.~\cite{Alpizar-Venegas:2025wor,Chattaraj:2025fvx}. Phenomenologically, neutrino magnetic moments are studied through a single parameter, which we refer to as effective neutrino magnetic moment, $\mu_{\nu_\ell}^{\textrm{eff}}$, written in terms of diagonal and transition magnetic moments, with different expressions depending on the neutrino source \cite{AristizabalSierra:2021fuc}. For short-baseline experiments, as is the case of reactor neutrinos, we have 
\begin{equation}
 \mu^\text{eff}_{\nu_\ell} =\sum_k \left | \sum_j U^*_{\ell k} \lambda_{jk} \right |^2,
\end{equation}
with $U$ denoting the lepton mixing matrix, and $\lambda_{jk}$ are the magnetic moment matrix elements in the fundamental parameter space.
The cross section associated with this contribution is given by~\cite{Vogel:1989iv}

\begin{equation}
   \frac{\d \sigma_{\nu \mathcal{N}}}{\d T_\mathcal{N}}\Big|_\mathrm{CE\nu NS}^\mathrm{MM}
=
\dfrac{ \pi \alpha^2_\mathrm{EM} }{ m_{e}^2 }
\left( \dfrac{1}{T_\mathcal{N}} - \dfrac{1}{E_\nu} \right)
Z^2 F_{W}^2(\qtransfer^2)
\left| \dfrac{\mu_{\nu_{\ell}}^\mathrm{eff}}{\mu_{\text{B}}} \right|^2 \, ,
\label{eq:cross:NMM}
\end{equation}
where $\alpha_{\textrm{EM}}$ is the fine structure constant,  $\mu_{\textrm{B}}$ is the Bohr magneton,   and $\ell$ denotes the neutrino flavor. Contrarily to the SM interaction, neutrino magnetic moment effects induce a chirality flip, and the associated cross section does not interfere with the SM one. Moreover, notice that the cross section shows a dependence on $1/T_\mathcal{N}$, specially enhancing the effects at low-energy thresholds, as is the case of the CONUS+ experiment.

\subsection{Neutrino nonstandard interactions}
\label{subsec:xsec-NSI}

NSI is a widely used parametrization to account for new physics effects at low energies in a model-independent way~\cite{Ohlsson:2012kf,Miranda:2015dra,Farzan:2017xzy}. For example, a variety of models for the generation of neutrino mass give rise to vector interactions that are weighted by some new Yukawa couplings in the presence of new mediators.
In general, both charged-current and neutral-current NSIs can exist. Since \cevns~is a neutral-current process, here we only focus on neutral-current NSIs~\cite{Schechter:1980gr,Valle:1987gv,Giunti:2019xpr}, which are described by the effective Lagrangian

\begin{equation}
{\cal{L}}^\mathrm{NSI}_\mathrm{NC}=-2\sqrt{2}G_F \sum\limits_{\ell,\ell'} \varepsilon_{\ell\ell'}^{f C}(\bar{\nu}_\ell\gamma^\mu P_L\nu_{\ell'})(\bar{f}\gamma_\mu P_C f) \, ,
\label{eq:NSI}
\end{equation}
where the indices $\ell$ and $\ell'$ are flavor indices, $C = L, R$ denotes the chirality, and $f$ stands for the SM charged fermions. The Lagrangian in Eq. \eqref{eq:NSI} is weighted by the Fermi constant, and $\varepsilon_{\ell \ell'}^{fC}$ are the so-called NSI parameters, which account for the strength of the interaction and are expected to be less than order unity since they are subdominant with respect to the weak scale. It is well known that for these interactions axial contributions are suppressed with respect to vector contributions \cite{Barranco:2005yy}, so here we focus on vector-type NSIs, which are given by

\begin{equation}
\varepsilon_{\ell\ell'}^{qV} = \varepsilon_{\ell\ell'}^{qR} + \varepsilon_{\ell\ell'}^{qL}\, .
\end{equation}

Notice that the main difference with respect to the SM is that, within this formalism, we allow for flavor transitions through the interaction when $\ell \neq \ell'$, in which case we refer to nonuniversal NSIs. On the other hand, there is also the freedom of choosing   $\varepsilon_{\ell \ell }^{fC} \neq \varepsilon_{\ell' \ell'}^{fC}$, which in contrast to the SM, allows for a nonuniversality of the interaction. The result of considering these contributions in the computation of the \cevns~cross section is a redefinition of the weak charge given in Eq.~\eqref{eq:CEvNS_SM_Qw}, which is now modified to

\begin{eqnarray}
\left(Q_{V}^{\textrm{NSI}}\right)^2 & = \left[Z\left(g_V^p + 2\varepsilon_{\ell\ell}^{uV}+\varepsilon_{\ell\ell}^{dV}\right) + N\left(g_V^n + \varepsilon_{\ell\ell}^{uV}+2\varepsilon_{\ell\ell}^{dV}\right)\right]^2 \nonumber \\
 & + \sum\limits_{\ell\neq \ell'} \left| Z\left(2 \varepsilon_{\ell\ell'}^{uV} + \varepsilon_{\ell\ell'}^{dV} \right)  + N\left(\varepsilon_{\ell\ell'}^{uV} + 2\varepsilon_{\ell\ell'}^{dV} \right) \right|^2
 \label{eq:weak:charge:NSI}\, .
\end{eqnarray}

From this equation, we explicitly see that diagonal NSIs can interfere with the SM cross section, while nondiagonal NSIs add up incoherently. 

\subsection{Neutrino generalized interactions}
\label{subsec:xsec-NGI}

Still working with an effective interaction approach, we consider an extension of neutrino NSIs that includes all Lorentz-invariant interactions. Concretely, we assume NGIs with heavy mediators~\cite{Lee:1956qn,Lindner:2016wff,AristizabalSierra:2018eqm}, described by the effective Lagrangian 

\begin{equation}
\label{eq:NGIlagr}
\mathscr{L}^\mathrm{NGI}_\mathrm{NC}  \supset \frac{G_F}{\sqrt{2}}
   \sum_{a=(S,V)} \mathcal{Q}_a  \left(\bar{\nu}_e \Gamma^a P_L \nu_e \right) \left(\bar{\mathcal{N}} \Gamma_a \mathcal{N}\right) \, ,
\end{equation}
where $\Gamma^a=\{\mathbb{I}, \gamma^\mu \}$, $P_L \equiv (1 - \gamma^5) / 2$ is the left-handed projector, and $\mathcal{N}$ indicates the nucleus.
The quantity $\mathcal{Q}_a$ denotes the corresponding neutrino-nucleus coupling for the scalar ($a = S$) and vector ($a = V$) interactions.  Note that the scalar interaction flips the neutrino chirality, thus requiring the presence of (final-state) right-handed neutrinos.
Moreover, in full generality, NGIs could also include pseudoscalar, axial, and tensor interactions. However, 
we will ignore them for the following reasons. First, because they are nuclear spin-suppressed and, therefore the expected bounds are less competitive. Second, only the $^{73}$Ge isotope of the CONUS+ detector has nonzero spin and would contribute to these interactions, which, however, has a very small abundance of 7.75\%. Finally, pseudoscalar interactions are of the order of $\mathcal{O}(T_\mathcal{N}/m_\mathcal{N})$ and hence very suppressed.

Under these considerations, the NGI \cevns~cross section reads

\begin{equation}
\left.\frac{\d \sigma_{\nu \mathcal{N}}}{\d T_\mathcal{N}}\right|_\mathrm{CE\nu NS}^\mathrm{NGI}=\frac{G_F^2 m_\mathcal{N}}{\pi}F_{W}^2(\qtransfer^2)\left[\mathcal{Q}_S^2 \frac{m_N T_\mathcal{N}}{8E_\nu^2}+\left(\frac{\mathcal{Q}_V}{2} + Q_V^\text{SM}\right)^2\left(1-\frac{m_N T_\mathcal{N}}{2E_\nu^2}-\frac{T_\mathcal{N}}{E_\nu}\right)\right]\, .
\label{eq:xsecNGI}
\end{equation}
The scalar and vector charges take the forms  ~\cite{Cirelli:2013ufw,DelNobile:2021wmp} 
\begin{align}
\label{eq:couplingsCa}
    \mathcal{Q}_S  & = C_{\nu S}\left( Z \sum_{q = u, d} C_{q S} \dfrac{m_p}{m_q} f_{T_q}^{(p)}  + N \sum_{q = u,d} C_{q S}\dfrac{m_n}{m_q} f_{T_q}^{(n)} \right), \\[4pt]
    \mathcal{Q}_V &= \kappa \, C_{\nu V} \left[Z (2 C_{u V} + C_{d V}) +  N (C_{u V} + 2 C_{d V}) \right] \label{eq:couplingsCV} \, .
\end{align}
Here, $m_p$ and $m_n$ are the proton and neutron masses, respectively,  $m_q$ are the quark $q$ masses, while $\kappa$ is a model-dependent parameter (see below), which for NGIs is set to $\kappa=1$. Moreover, $f_{T_q}^{(p)}$ and $f_{T_q}^{(n)}$ denote the contributions of the quark mass to the nucleon (proton and neutron) mass, with values~\cite{DelNobile:2021wmp}
\begin{equation*}
f_{T_u}^p = 0.026\, ,  ~~~~~ f_{T_d}^p =  0.038\, ,  ~~~~~ f_{T_u}^n = 0.018\, ,  ~~~~~ f_{T_d}^n =  0.056\, .
\end{equation*}  
For simplicity, in our calculations we will not discriminate between quark and neutrino couplings, and instead we will present our results in terms of the quark-dependent coupling $C_a^q = \sqrt{C_{\nu a} \cdot C_{qa}}$.

In our analyses, we will consider the scalar and vector contributions separately, as well as  allow for the simultaneous presence of both of them. Regarding \eves, note that the correspondent NGI cross section is not enhanced, and, therefore we do not expect any impact if we were to include the \eves~events in the analysis. For this reason, we will not include them in the NGI study.

\subsection{Light mediators}
\label{subsec:xsec-lightmed}

Next, we promote the generalized interactions previously considered to the case in which they occur through the exchange of a light mediator. By ``light" mediator, we mean a mediator with a mass comparable to the typical momentum transfer of the \cevns~process, $\qtransfer \sim \mathcal{O} (10)$ MeV. The heavy NGI couplings $C_a^q = \sqrt{C_{\nu a} \cdot C_{qa}}$, as  defined in Sec.~\ref{subsec:xsec-NGI}, are related to those in the light mediator case according to the expression  
\begin{equation}
   (C_{a}^q)^2 = \frac{\sqrt{2}}{G_F} \frac{(g_a^q)^2}{\qtransfer^2 + M_a^2} \, ,
\end{equation}
where the coupling $g^q_a=\sqrt{g_{\nu a} \cdot g_{q a}}$ is written in terms of the fundamental BSM couplings between the neutrinos and the light mediator $(g_{\nu a})$ and between the quarks and the light mediator $(g_{q a})$, respectively. When charged leptons are involved instead of quarks, the corresponding coupling is $g^\ell_a=\sqrt{g_{\nu a} \cdot g_{\ell a}}$, with $\ell$ being the charged lepton ($\ell=e$ for \eves).
In the following, we will assume that the new mediator $a$ couples with equal strength to neutrinos, charged leptons, and quarks, i.e., we take $g_a=\sqrt{g_{\nu a} \cdot g_{qa}}=\sqrt{g_{\nu a} \cdot g_{\ell a}}$.

The relevant \cevns~cross sections for light scalar and vector mediators are given by 

\begin{align}
    \left.\dfrac{\d \sigma_{\nu \mathcal{N}}}{\d T_\mathcal{N}}\right|_\mathrm{CE\nu NS}^{S} =& \, \dfrac{m_\mathcal{N} \mathcal{Q}_S^2}{4\pi (m_S^2 + 2m_\mathcal{N} T_\mathcal{N})^2} F_W^2(\qtransfer^2) \dfrac{m_\mathcal{N} T_\mathcal{N}}{E_\nu^2} , \label{eq:xsec_CEvNS_lightmedS} \\[4pt]
    \left.\dfrac{\d \sigma_{\nu \mathcal{N}}}{\d T_\mathcal{N}}\right|_\mathrm{CE\nu NS}^{V}  =& \, \left[1+ \kappa \frac{\mathcal{Q}_V}{\sqrt{2}G_F Q_V^\mathrm{SM}\left(m_{V}^2+2 m_\mathcal{N} T_\mathcal{N}\right)}\right]^2 \left.\frac{\d \sigma_{\nu \mathcal{N}}}{\d T_\mathcal{N}}\right|_\mathrm{CE\nu NS}^\mathrm{SM} \, ,
     \label{eq:xsec_CEvNS_lightmedV}
\end{align}
where it appears a clear dependence on the light mediator mass $m_a$. The factor $\kappa$ accounts for the typical particle charges under the specific $U(1)'$ extension. We will consider $\kappa=-1/3$ corresponding to the B-L model~\cite{Langacker:2008yv,Okada:2018ktp}, and $\kappa=1$ when considering the universal vector scenario. The latter choice corresponds to the universal light vector scenario, which is not anomaly-free, but it is a good phenomenological example whose results can be easily recast into more sophisticated models.

Concerning \eves, the cross section in the presence of a new light scalar reads~\cite{Link:2019pbm} 
\begin{equation}
\label{eq:xsec_EvES_S}
\left. \frac{\d \sigma_{\nu \mathcal{A}}}{\d E_\mathrm{er}}\right|_\mathrm{E \nu ES}^\mathrm{S}= Z^\mathcal{A}_\mathrm{eff}(E_\mathrm{er}) \left[\frac{g^4_S}{4\pi(2m_e E_{\mathrm{er}} + m_{S}^2)^2}\right]\frac{m_e^2 E_{\mathrm{er}}}{E_\nu^2}\, , 
\end{equation}
where, as before, $g_S = \sqrt{g_{\nu S}\cdot g_{eS}}$.
In the case of a light vector, the \eves~cross section can be obtained by modifying the couplings in the SM expression [Eq.~\eqref{eq:xsec_EvES_SM}] as follows:
\begin{equation}
    g_V \to g_V^{\mathrm{SM}} + \frac{g^2_{V} Q^e_{V} Q^{\nu_e}_{V}}{\sqrt{2}G_F (2 m_e E_{\mathrm{er}} + m_{V}^2)} \, .
    \label{eq:xsec_EvES_V}
\end{equation}
In the B-L model $Q^e_{V} = Q^{\nu_e}_{V}= -1$, instead, in the universal scenario, we fix $Q^e_{V} = Q^{\nu_e}_{V}= +1$.

\subsection{Light sterile neutrino oscillations}
\label{subsec:xsec-vsosc}

Given its neutral-current nature, \cevns~is sensitive to the total neutrino flux and, therefore, can be used to investigate oscillations into light sterile neutrinos.
Working in the minimal (3+1) model, with three active plus one light sterile neutrino, we can define the electron neutrinos' survival probability as  
 \begin{equation}
   P_{ee}(E_\nu)  \simeq 1 - \sin^2 2\theta_{1 4} \sin^2 \left( \frac{\Delta m^2_{41} L}{4E_\nu}\right)\, ,
   \label{eq:prob_nue_osc}
 \end{equation}
where  $\theta_{1 4}$ is the active-sterile mixing angle, $\Delta m^2_{41}$ is the active-sterile mass splitting, and $L = 20.7$ m is the baseline of the CONUS+ experiment. 
In this scenario, the \cevns~cross section is obtained by changing $Q_V^\mathrm{SM} \to Q_V^\text{SM}\times P_{ee}(E_\nu) $ in Eq.~\eqref{eq:xsec_CEvNS_SM}.

\subsection{Sterile dipole portal}
\label{subsec:xsec-dipole}

Sterile neutrinos are motivated BSM candidates that appear in several theoretical models that accommodate neutrino masses and mixings~\cite{Abdullahi:2022jlv,Minkowski:1977sc,Yanagida:1979as,GellMann:1980vs,Mohapatra:1979ia,Schechter:1980gr}. We consider the BSM scenario in which the SM particle content is extended by adding one sterile neutrino, $\nu_4$, with mass $m_4$. The presence of an active-sterile transition magnetic moment can give rise to an up-scattering process of the type $\nu_\ell + \mathcal{N} \to \nu_4 + \mathcal{N}$, a scenario usually referred to as the sterile dipole portal~\cite{McKeen:2010rx,Magill:2018jla}. 
The Lagrangian for this model reads~\cite{Gninenko:1998nn,Grimus:2000tq}
\begin{equation}
    \mathcal{L}_\mathrm{DP} = \bar{\nu}_4(i \slashed{\partial}- m_4) \nu_4 + \frac{\sqrt{\pi \alpha_\mathrm{EM}}}{2 m_e}\left| \dfrac{\mu_{\nu_{\ell}}^\mathrm{eff}}{\mu_{\text{B}}} \right|^2 \bar{\nu}_4 \sigma_{\mu \nu} \nu_\ell F^{\mu \nu} \,, 
\end{equation}
with $F^{\mu \nu}$ being the electromagnetic field tensor, and $\sigma_{\mu \nu} = i (\gamma^\mu \gamma^\nu - \gamma^\nu\gamma^\mu)/2$.
Note that the neutrino transition magnetic moment can also be expressed as $d_\ell = \frac{\sqrt{\pi \alpha_\mathrm{EM}}}{m_e}\left| \dfrac{\mu_{\nu_{\ell}}^\mathrm{eff}}{\mu_{\text{B}}} \right|^2 $ ([GeV$^{-1}$]) and can be connected to a new physics scale.

This process is chirality-violating, so its cross section, that reads~\cite{McKeen:2010rx}
\begin{equation}
  \label{eq:xsec_dipolesterile}
\begin{aligned}
 \left. \frac{\d \sigma_{\nu \mathcal{N}}}{\d  T_\mathcal{N}}\right|_\mathrm{CE\nu NS}^\mathrm{DP} = &
  \dfrac{ \pi \alpha^2_\mathrm{EM} }{ m_{e}^2 }\, Z^2 F_{W}^2(\qtransfer^2)
\left| \dfrac{\mu_{\nu_{\ell}}^\mathrm{eff}}{\mu_{\text{B}}} \right|^2 \\
 & \times \left[\frac{1}{T_\mathcal{N}} - \frac{1}{E_\nu} 
    - \frac{m_4^2}{2E_\nu T_\mathcal{N} m_\mathcal{N}}
    \left(1- \frac{T_\mathcal{N}}{2E_\nu} + \frac{m_\mathcal{N}}{2E_\nu}\right)
    + \frac{m_4^4(T_\mathcal{N}-m_\mathcal{N})}{8E_\nu^2 T_\mathcal{N}^2 m_\mathcal{N}^2}
  \right]  \,  ,
  \end{aligned}
\end{equation}
adds incoherently to the SM one. We ignore subdominant contributions arising from interference between magnetic and weak interactions~\cite{Grimus:1997aa,Miranda:2021kre}. Moreover, Eq.~\eqref{eq:xsec_dipolesterile} is obtained for a spin-1/2 nucleus, however, it can be used for a spin-zero target too, neglecting minor corrections~\cite{Miranda:2021kre}.

The corresponding cross section for \eves~in the sterile dipole scenario is easily obtained from Eq.~\eqref{eq:xsec_dipolesterile} through the following substitutions:
$T_\mathcal{N}\to E_\text{er}$, $m_\mathcal{N} \to m_e$, and $Z^2  F_{W}^2(\qtransfer^2) \to Z_\text{eff}^\mathcal{A}$.

\subsection{Up-scattering production of a sterile fermion}
\label{subsec:xsec-upscat}

The final BSM scenario that we consider is the possible production of an MeV-scale sterile fermion $\chi$ through the up-scattering of neutrinos on nuclei and atomic electrons, $
    \nu_\ell e \rightarrow \chi e$ and $ \nu_\ell \mathcal{N} \rightarrow \chi \mathcal{N}$. In this case, the up-scattering occurs through a new light scalar or vector mediator~\cite{Brdar:2018qqj,Chang:2020jwl,Chao:2021bvq,Candela:2023rvt,Alonso-Gonzalez:2023tgm,Candela:2024ljb}. The relevant \cevns~cross sections in this scenario are~\cite{Candela:2024ljb} 

\begin{align}
    \left.\dfrac{\d \sigma_{\nu \mathcal{N}}}{\d T_\mathcal{N}}\right|_\mathrm{CE \nu NS}^\mathrm{S \chi} =& \, \dfrac{m_\mathcal{N} \mathcal{Q}_S^2}{4\pi (m_S^2 + 2m_\mathcal{N} T_\mathcal{N})^2} F_W^2(\qtransfer^2) \left(1 + \dfrac{T_\mathcal{N}}{2 m_\mathcal{N}}\right) \left(\dfrac{m_\mathcal{N} T_\mathcal{N}}{E_\nu^2} + \dfrac{m_\chi^2}{2 E_\nu^2}\right), \label{eq:xsec-CEvNS_chi_S} \\[4pt]
    \left.\dfrac{\d \sigma_{\nu \mathcal{N}}}{\d T_\mathcal{N}}\right|_\mathrm{CE \nu NS}^\mathrm{V\chi} =& \, \dfrac{m_\mathcal{N} \mathcal{Q}_V^2}{2\pi (m_V^2 + 2m_\mathcal{N} T_\mathcal{N})^2} F_W^2(\qtransfer^2) \nonumber \\[4pt]
    &\times \left[
        \left(1 - \dfrac{m_\mathcal{N} T_\mathcal{N}}{2 E_\nu^2} - \dfrac{T_\mathcal{N}}{E_\nu} + \dfrac{T_\mathcal{N}^2}{2 E_\nu^2}\right) - \dfrac{m_\chi^2}{4 E_\nu^2}\left(1 + \dfrac{2 E_\nu}{m_\mathcal{N}} - \dfrac{T_\mathcal{N}}{m_\mathcal{N}}\right)
    \right], \label{eq:xsec-CEvNS_chi_V} 
\end{align}
where $m_\chi$ is the sterile fermion mass. The relevant scalar and vector couplings at the nuclear level are related to the fundamental ones (at the quark level), as in Eq.~\eqref{eq:couplingsCV}.

The corresponding \eves~cross sections  read~\cite{Candela:2024ljb} 

\begin{align}
    \left.\dfrac{\d \sigma_{\nu \mathcal{A}}}{\d E_\text{er}}\right|_\mathrm{\mathrm{E\nu ES}}^{\mathrm{S\chi}}  =& \, \dfrac{m_e g_S^4}{4\pi (m_S^2 + 2m_e E_\text{er})^2} Z_{\mathrm{eff}}^{\mathcal{A}}\left(E_\text{er}\right) \left(1 + \dfrac{E_\text{er}}{2 m_e}\right) \left(\dfrac{m_e E_\text{er}}{E_\nu^2} + \dfrac{m_\chi^2}{2 E_\nu^2}\right), \label{eq:xsec-EvES_chi_S} \\[4pt]
    \left.\dfrac{\d \sigma_{\nu \mathcal{A}}}{\d E_\text{er}}\right|_\mathrm{\mathrm{E\nu ES}}^\mathrm{V\chi}  =& \, \dfrac{m_e g_V^4}{2\pi (m_V^2 + 2m_e E_\text{er})^2} Z_{\mathrm{eff}}^{\mathcal{A}}\left(E_\text{er}\right) \nonumber \\[2pt]
    &\times \left[\left(1 - \dfrac{m_e E_\text{er}}{2 E_\nu^2} - \dfrac{E_\text{er}}{E_\nu} + \dfrac{E_\text{er}^2}{2 E_\nu^2}\right) - \dfrac{m_\chi^2}{4 E_\nu^2}\left(1 + \dfrac{2 E_\nu}{m_e} - \dfrac{E_\text{er}}{m_e}\right)\right]. \label{eq:xsec-EvES_chi_V} 
\end{align}

\section{CONUS+ data and statistical analysis}
\label{sec:analysis}

The CONUS+ experiment has been operating since November 2023. It is located at a distance of $20.7$ m from the nuclear power plant in Leibstadt, Switzerland, which provides an intense flux of electron antineutrinos, $\phi_{\bar{\nu}_e} = 1.5 \times 10^{13}~\bar{\nu}_e/({\rm cm^2 s})$. The experimental setup consists of four HPGe detectors, called C2, C3, C4, and C5, each with a mass of about 1 kg  and characterized by extremely low-energy thresholds. However, since the C4 detector showed instabilities in the rate, it was removed from the final dataset~\cite{Ackermann:2025obx}. The \cevns~measurement was conducted with
119 days of reactor-on operation (117 days for C2, 110 for C3, and 119 for C5), and a total fiducial mass of $\sim 2.83$ kg  (with individual masses $\mathrm{C2=0.95~kg}$, $\mathrm{C3=0.94~kg}$, and $\mathrm{C5=0.94~kg}$), translating into a total exposure of $327$ kg $\times$ d. Each of the three detectors has a different recoil threshold, $E_{\rm thr} = 160$ eV$_\text{ee}$ for C3,  $E_{\rm thr} = 170$ eV$_\text{ee}$ for C5, and  $E_{\rm thr} = 180$ eV$_\text{ee}$ for C2.\\

We now discuss the implementation of our statistical analysis. Note that even though we are interested in the \cevns~signal, in our analyses we will also include the \eves~rates. Indeed, while the \eves~events are negligible within the SM, compared to \cevns~ones, some BSM scenarios predict enhanced \eves~rates. The HPGe detectors are not capable of distinguishing between nuclear and electron recoils so the \eves~events must be taken into account.
The differential event rates are obtained through a convolution of the reactor neutrino flux with the differential cross section

\begin{eqnarray}
\label{eq:dRdEr_CEvNS}
    \left. \dfrac{\d R}{\d E_\text{ee}^\text{reco}}\right|_{\rm CE\nu NS} &= \mathcal{E} \int_{E_\mathrm{er}^\mathrm{min}}^{E_\mathrm{er}^\mathrm{max}} \d E_\text{er} \, \mathcal{G}(E_\text{ee}^{\rm reco},E_\text{er}) \mathcal{F}(E_\text{er})\int_{E_\nu^\mathrm{min}}^{E_\nu^\mathrm{max}}  \, \d E_\nu \, \,  \dfrac{\d \phi}{\d E_\nu} \left. \dfrac{\d \sigma_{\nu \mathcal{N}}}{\d E_\text{er}}\right|_{\rm CE\nu NS} \,,\\
    \label{eq:dRdEr_ES}
    \left. \dfrac{\d R}{\d E_\text{ee}^\text{reco}} \right|_{\rm E\nu ES}&= \mathcal{E}\int_{E_\mathrm{er}^\mathrm{min}}^{E_\mathrm{er}^\mathrm{max}} \d E_\text{er} \,  \mathcal{G}(E_\text{ee}^{\rm reco},E_\text{er}) \mathcal{F}(E_\text{er})\int_{E_\nu^\mathrm{min}}^{E_\nu^\mathrm{max}}  \, \d E_\nu \,  \dfrac{\d \phi}{\d E_\nu}\left. \dfrac{\d \sigma_{\nu \mathcal{A}}}{\d E_\mathrm{er}}\right|_{\rm E\nu ES}\, ,
\end{eqnarray}
where $\mathcal{E}$ is the exposure of the experiment,  $E_\text{er}$ denotes the true electron recoil energy, and $E_\text{ee}^\text{reco}$ the reconstructed one. In order to express the  \cevns~cross section in terms of the true electron recoil energy, we change variables using the quenching factor $\mathrm{Q_F}$ (see below), as
\begin{equation}
 \left.\dfrac{\d \sigma_{\nu \mathcal{N}}}{\d E_\text{er}}\right|_{\rm CE\nu NS} = \frac{\d T_\mathcal{N}}{\d E_\text{er}}
   \left.\dfrac{\d \sigma_{\nu \mathcal{N}}}{\d T_\mathcal{N}}\right|_{\rm CE\nu NS} = \frac{1}{\mathrm{Q_F}} \left( 1- \frac{E_\text{er}}{\mathrm{Q_F}} 
   \frac{\d \mathrm{Q_F}}{\d E_\text{er}} \right) \left.\dfrac{\d \sigma_{\nu \mathcal{N}}}{\d T_\mathcal{N}}\right|_{\rm CE\nu NS} \, .
   \label{eq:CEvNS_change_variable}
\end{equation} 
In the above expressions, $\left. \dfrac{\d \sigma_{\nu \mathcal{N}}}{\d T_\mathcal{N}}\right|_{\rm CE\nu NS}$ refers to any of the \cevns~expressions in Sec.~\ref{sec:theory}, depending on the scenario under consideration, while $\left. \dfrac{\d \sigma_{\nu \mathcal{A}}}{\d E_\mathrm{er}}\right|_{\rm E\nu ES}$ is the correspondent \eves~cross section. The integration limits are $E_\nu^\mathrm{min} = \sqrt{m_\mathcal{N}T_\mathcal{N}}/2$ for \cevns~and $E_\nu^\mathrm{min} = (E_\text{er} + \sqrt{2 m_e E_\text{er} + E_\text{er}^2})/2$ for \eves, while $E_\nu^\mathrm{max}\sim 10$ MeV. Moreover we take $E_\text{er}^\text{min}=2.96~\mathrm{eV_\text{ee}}$ corresponding to the minimum average ionization energy of germanium, while $E_\text{er}^\text{max}$ is obtained by inverting the $E_\nu^\mathrm{min}$ relations for \cevns~and \eves,~as dictated by the kinematics.
$\mathcal{F}$ is the CONUS efficiency, which is quoted to be approximately $100 \%$ above the threshold~\cite{Ackermann:2025obx}.  The reactor flux is evaluated using the spectral function from~\cite{Kopeikin:2012zz} for neutrino energies below 2 MeV, and from~\cite{Mueller:2011nm} for larger energies, and normalized over the entire energy range. We have checked that using another parametrization for the reactor neutrino flux would not sizably affect the results, as was also shown in Ref.~\cite{AtzoriCorona:2022qrf} in the context of Dresden-II data. 
As explained below, we include a $4.6\%$~\cite{Ackermann:2025obx} uncertainty on the neutrino flux in our analysis to take into account these fluctuations. In addition, in our calculations, we take into account the isotopic abundance of all stable isotopes comprising natural germanium, i.e., we assume 
$^{70}$Ge (20.57\%),
$^{72}$Ge (27.45\%),
$^{73}$Ge (7.75\%),
$^{74}$Ge (36.50\%), and
$^{76}$Ge (7.73\%), but we also note that the predicted signal is not altered if the average mass number of $A=72.6$ is instead considered.

The energy resolution function in Eqs.~\eqref{eq:dRdEr_CEvNS} and~\eqref{eq:dRdEr_ES} is defined as a Gaussian function
\begin{equation}
\mathcal{G}(E_\text{ee}^{\rm reco},E_{\rm er}) = \frac{1}{\sqrt{2 \pi}\sigma_{\rm res}} e^{-\frac{\left(E_\text{ee}^{\rm reco}-E_{\rm er}\right)^2}{2 \sigma_{\rm res}^2}} \,, 
\end{equation}
with its width given as~\cite{Lindner:2024eng,Chattaraj:2025fvx}
\begin{equation}
\sigma_{\rm res} = \sqrt{\left(\sigma_0^2 + \mathcal{F}_\mathrm{fano} \times  \eta \times  E_{\rm er}\right)}\, . 
\end{equation}
In this expression, $\eta = 2.96~\mathrm{eV_\text{ee}}$ is the energy needed to create an energy-hole pair in Ge, while $\mathcal{F}_\mathrm{fano} = 0.1096$ is the Fano factor for Ge~\cite{Ackermann:2025obx}. Finally, $\sigma_0$ is the pulser resolution, which has been reported in terms of the full-width at half-maximum (FWHM) in~\cite{CONUS:2024lnu}. For a Gaussian distribution, the two quantities are related by $\sigma_0 = \mathrm{FWHM}/\sqrt{8 \ln(2)}$, while taking $\mathrm{FWHM=48~eV_\text{ee}}$, we find $\sigma_0=20.38~\mathrm{eV_\text{ee}}$.

Going back to Eq.~\eqref{eq:CEvNS_change_variable}, the true electron recoil (ionization) energy $E_{\rm er}$ is related to the nuclear recoil energy $T_\mathcal{N}$ through the quenching factor  ($\mathrm{Q_F}$)
\begin{equation}
    E_{\rm er} = \mathrm{Q_F}(T_\mathcal{N}) \times T_\mathcal{N} = \frac{k~g(\epsilon)}{1+ k~g(\epsilon)}T_\mathcal{N} \, , 
\label{eq:quenching}    
\end{equation}
where $\epsilon = 11.5~Z^{-7/3} T_\mathcal{N}$ and $g(\epsilon) = 3 \epsilon^{0.15} + 0.7 \epsilon^{0.6} + \epsilon$.
As anticipated, the quenching factor is a relevant parameter and often a source of uncertainties, especially at low-recoil energies. The CONUS+ detection has been found to be in agreement with the predicted Ge quenching using the Lindhard theory~\cite{osti_4701226}, with $k = 0.162 \pm 0.004$~\cite{Bonhomme:2022lcz}. Note that for the \eves~signal, no translation is needed.

The CONUS+ result is presented in terms of excess counts, indicating the difference between the data in the reactor-on phase and the background model, per ionization energy bin. We, therefore, compute our predicted event rates as 
 
\begin{equation}
    R_i = \int_i  \dfrac{\d R}{\d E_\text{ee}^{\rm reco}} \d E_\text{ee}^{\rm reco}\, .
\label{eq:ev_rate}    
\end{equation}
The integral is performed over the reconstructed  ionization  energy bin $i$, and eventually our total prediction is given by the sum $R_i^{\rm th} = R_i^{\rm CE\nu NS} + R_i^{\rm E\nu ES}$.

Our total theoretical predictions are then compared to the CONUS+ excess counts $R^{\rm exp}_i$ through

\begin{equation}
\label{eq:chi2}
 \chi^2 (\vec{\beta}) =  \sum_i \frac{\left[R^{\rm exp}_i- (1 + \alpha)R_i^{\rm th}(\vec{\beta})\right]^2}{\sigma_i^2} + \left(\frac{\alpha}{\sigma_\alpha}\right)^2\, ,
\end{equation}
where $i$ is the number of reconstructed ionization energy bins (19 in total), $\alpha$ is a cumulative nuisance parameter with  $\sigma_\alpha = 16.9\%$~\cite{Ackermann:2025obx}, accounting for uncertainties on the reactor antineutrino flux ($ 4.6 \%$), the quenching factor ($ 7.3 \%$), the threshold ($14.1\%$), the active mass of Ge ($1.1\%$), the trigger efficiency ($0.7\%$), and the form factor ($3.2 \%$), while $\vec{\beta}$ are the scenario-dependent parameters that are varied in the analyses. The quenching factor and form factor uncertainties are included only in the analysis of \cevns~events, and not in \eves~events.
The $\sigma_i$ are given by the error bars provided in the experimental paper~\cite{Ackermann:2025obx}. For each analysis, the $\chi^2$ profile in Eq.~\eqref{eq:chi2} is minimized over the nuisance $\alpha$.

\begin{figure}[t]
    \centering
    \includegraphics[width=0.7\textwidth]{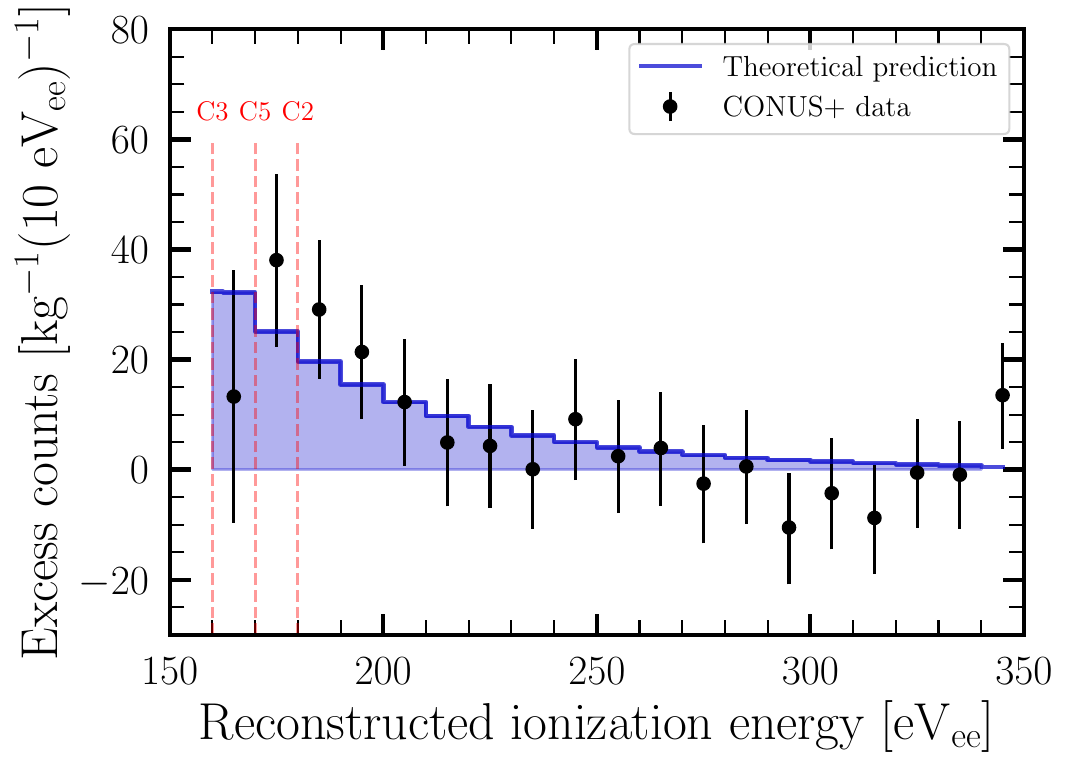}
    \caption{CONUS+ excess counts (black points with error bars) from~\cite{Ackermann:2025obx} as a function of the reconstructed ionization energy, together with our SM theoretical prediction for the sum of the three detectors (blue solid histogram). The red dashed vertical lines indicate the different energy thresholds of the three detectors.}
    \label{fig:datahisto}
\end{figure}

In Fig.~\ref{fig:datahisto}, we show our theoretical prediction for the number of SM \cevns~events at CONUS+. 
The result is given for the total number of events (blue solid histogram), including the three detectors. To compare with the excess counts reported by  CONUS+ collaboration, the rates are calculated assuming a single effective detector with a mass of 1~kg, a threshold of $160~\mathrm{eV_{ee}}$, and an exposure of 119 kg $\times$ d. Let us add that the theoretical number of events, after accounting for the individual detector masses and summing over all reconstructed ionization energy bins, are found to be 90, 143, and 113 for C2, C3, and C5, respectively, giving a total number of 346 \cevns~events, i.e., in agreement with the prediction of CONUS+ collaboration~\cite{Ackermann:2025obx}.

Finally, we have also verified that the expected number of \eves~events in the SM is negligible and of the order of $\mathcal{O}(0.1)$ and therefore can be safely neglected. However, as it will become evident in the remainder of this work, in certain BSM scenarios the \eves~rates are enhanced and, therefore, have to be included in the analysis.

\section{Results}
\label{sec:results}

In this section, we discuss the results of our analyses of CONUS+ data. We start with the SM implications, namely, a determination of the weak mixing angle. Then, we present results for a set of BSM scenarios, including neutrino magnetic moments, new neutrino interactions with heavy and light mediators, sterile neutrinos, and the neutrino up-scattering into a sterile neutral fermion.

\subsection{Weak mixing angle}
\label{subsec:res-thw}

\begin{figure}[!htb]
\centering
\includegraphics[width=0.65\textwidth]{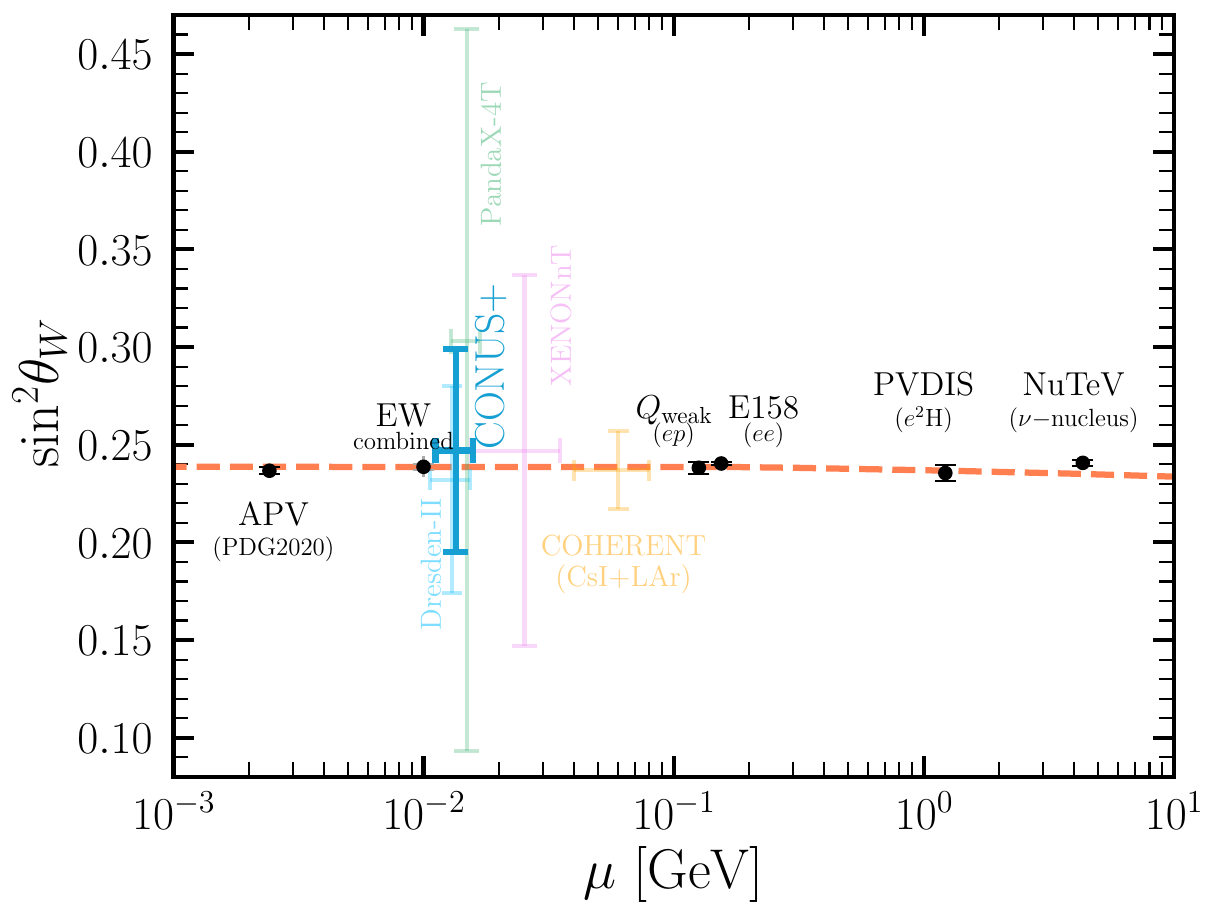}
\caption{$1 \sigma$ determination of the weak mixing angle from our CONUS+ analysis (light blue). The coral dashed line indicates the sin$^2\theta_W$ running in the SM for the $\overline{\text{MS}}$ renormalization scheme, as a function of the renormalization scale. Measurements from other experiments~\cite{Majumdar:2022nby,DeRomeri:2022twg,ParticleDataGroup:2024cfk,Qweak:2018tjf,SLACE158:2005uay,PVDIS:2014cmd,NuTeV:2001whx,AtzoriCorona:2024vhj,DeRomeri:2024iaw} are also shown for completeness.}
\label{fig:res_sw2}
\end{figure}

We begin our discussion by exploring the CONUS+ sensitivity on the weak mixing angle, noting that this type of analysis has been previously reported in Refs.~\cite{Alpizar-Venegas:2025wor,Chattaraj:2025fvx}. However, we find it useful to repeat it here since these two works followed a different analysis strategy, i.e., Ref.~\cite{Alpizar-Venegas:2025wor} conducted a single-bin analysis, while Ref.~\cite{Chattaraj:2025fvx} performed a spectral fit like that in the present work.  Our $1 \sigma$ determination of the weak mixing angle is 
\begin{equation}
    \sin^2 \theta_W= 0.247 \pm 0.052\, .
\end{equation}
 Our best fit is in overall agreement with the results obtained in Refs.~\cite{Alpizar-Venegas:2025wor,Chattaraj:2025fvx,AtzoriCorona:2025ygn}. However, we should note that a direct comparison with Ref.~\cite{Alpizar-Venegas:2025wor} is not possible since the authors followed a single-bin analysis strategy. 

We show in Fig.~\ref{fig:res_sw2} the running of the weak mixing angle as a function of the momentum transfer. Clearly, in the low-energy regime, the determination of $\sin^2 \theta_W$ is still poor, but \cevns~measurements, such as the CONUS+ one, are pushing toward it. The present result from the reactor data of the CONUS+ experiment is not as constraining as that coming from the analysis of COHERENT CsI+LAr data in Ref.~\cite{DeRomeri:2022twg}, however it is obtained at a slightly lower energy. Moreover, the CONUS+ limit reported here improves on recent constraints obtained in Refs.~\cite{DeRomeri:2024iaw} and \cite{Maity:2024aji} from the analysis of $^8$B solar neutrino-induced \cevns~data reported by PandaX-4T and XENONnT. Finally, the present result is significantly improved compared to the previous limit resulting from the analysis of the Dresden-II reactor experiment in Refs.~\cite{Majumdar:2022nby} and~\cite{AristizabalSierra:2022axl}, highlighting the importance of the quenching factor modeling. We close this discussion by noting that a combination of \cevns~data with other electroweak measurements can dramatically improve the precision of the $\sin^2 \theta_W$ measurement at low energy, as shown in~\cite{AtzoriCorona:2024vhj,Cadeddu:2024baq}.

\subsection{Effective neutrino magnetic moment}
\label{subsec:res-magmom}
\begin{figure}[t]
\centering
\includegraphics[width=0.48\textwidth]{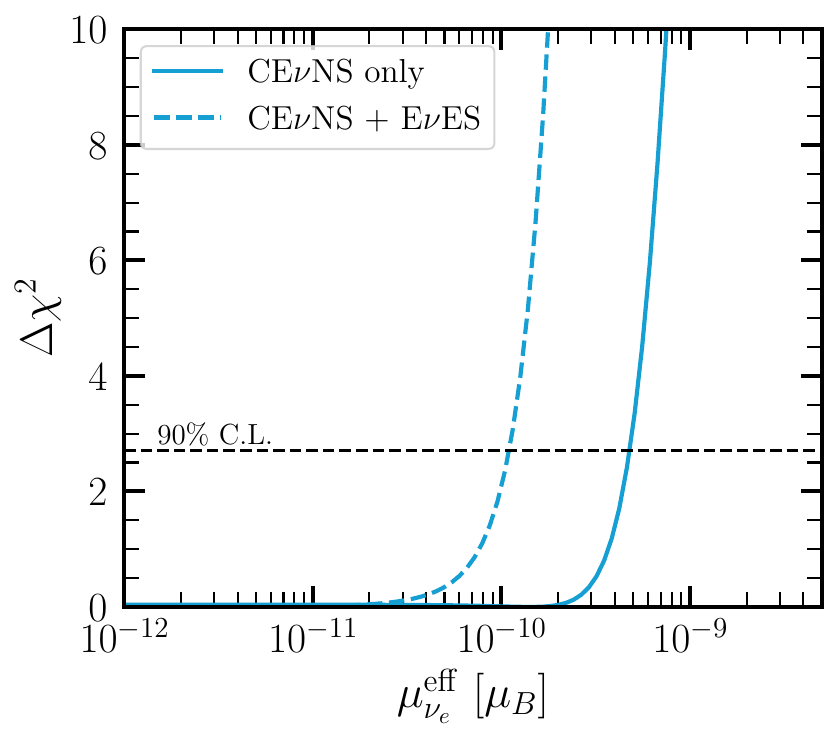}
\caption{$\Delta \chi^2$ profiles for the effective magnetic moment $\mu_{\nu_e}^\mathrm{eff}$ from the analysis of CONUS+ data. The illustrated results correspond to the \cevns-only analysis (plain curve) and the \cevns+\eves~analysis (dashed curve).}
\label{fig:res_magnetic_moment}
\end{figure}

Next, we focus our attention on BSM scenarios. We begin our discussion by exploring the implications of the CONUS+ result on the effective neutrino magnetic moment. For this scenario, the inclusion of \eves~events in the analysis is crucial and significantly improves the constraints. The corresponding $\Delta \chi^2$ profiles are shown in Fig.~\ref{fig:res_magnetic_moment}, where we can appreciate the impact of incorporating the \eves~events in the statistical analysis versus the \cevns-only case. The 90\% C.L.  resulting from our present analysis read

\begin{equation}
    \begin{aligned}
        \mu_{\nu_e}^\mathrm{eff} & \leq   4.78 \times 10^{-10}~\mu_B \qquad \mathrm{(CE\nu NS~ only)} \, ,\\
         \mu_{\nu_e}^\mathrm{eff} & \leq 1.10 \times 10^{-10}~\mu_B \qquad \mathrm{(CE\nu NS + E\nu ES)} \,.  
    \end{aligned}
\end{equation}

Taking into account that different analysis strategies have been followed, our results are found to be in reasonable agreement with those reported in~\cite{Alpizar-Venegas:2025wor, Chattaraj:2025fvx,AtzoriCorona:2025ygn}.

A comparison with other limits existing in the literature is given in Table~\ref{tab:magnetic_moment_comparison}.  It is interesting to note that the CONUS+ data provide the most stringent limits among all the \cevns~experiments, surpassing the existing ones from COHERENT (Dresden-II) by one (half) order of magnitude. The CONUS+ limits are also more stringent compared to those extracted from the analysis of Ref.~\cite{DeRomeri:2024hvc}, which focused on the $^8$B-induced \cevns~data reported by PandaX-4T and XENONnT experiments. However, as can be seen in the table, CONUS+ cannot yet compete with \eves-induced limits from TEXONO, Borexino, XENONnT, and LZ. Before closing this discussion, we should warn the reader that these comparisons should be made with special care, since the effective magnetic moment is not directly comparable when different neutrino sources are involved (see, e.g., the discussion in Ref.~\cite{AristizabalSierra:2021fuc}).

\begin{table}[ht!]
\centering
\begin{tabular}{|l|c|c|c|}
\hline
\textbf{Experiment} & $\boldsymbol{\mu^{\mathrm{eff}}_{\nu_e}}$ $\boldsymbol{(10^{-11}~\mu_B)}$ & \textbf{Process} & \textbf{Reference} \\ \hline
CONUS+ & $\leq 11$ & \cevns  +\eves & this work \\
COHERENT (CsI+LAr) & $\leq 360$ & \cevns+\eves & \cite{DeRomeri:2022twg} \\
DRESDEN-II & $\leq 19$ & \cevns+\eves & \cite{Coloma:2022avw} \\
XENONnT + PandaX-4T (combined) & $\leq 190$ & \cevns & \cite{DeRomeri:2024hvc} \\
CONUS & $\leq 7.5$ & \eves & \cite{CONUS:2022qbb} \\
Borexino & $\leq 3.7$ & \eves & \cite{Coloma:2022umy} \\
TEXONO & $\leq 7.4$ & \eves & \cite{TEXONO:2006xds} \\
GEMMA & $\leq 2.9$ & \eves & \cite{Beda:2012zz} \\
LZ & $\leq 1.4$ & \eves & \cite{A:2022acy} \\
XENONnT & $\leq 0.9$ & \eves & \cite{A:2022acy} \\ 
XENONnT+PandaX-4T+LZ (combined)  & $\leq 1.03$ & \eves & \cite{Giunti:2023yha} \\ \hline
\end{tabular}
    \caption{Comparison of constraints on the effective neutrino magnetic moment $\mu_{\nu_e}^\mathrm{eff}$ from different experiments.}
    \label{tab:magnetic_moment_comparison}
\end{table}

\subsection{Neutrino nonstandard interactions}
\label{subsec:res-NSI}

\begin{figure}[t]
\centering
\includegraphics[width=0.48\textwidth]{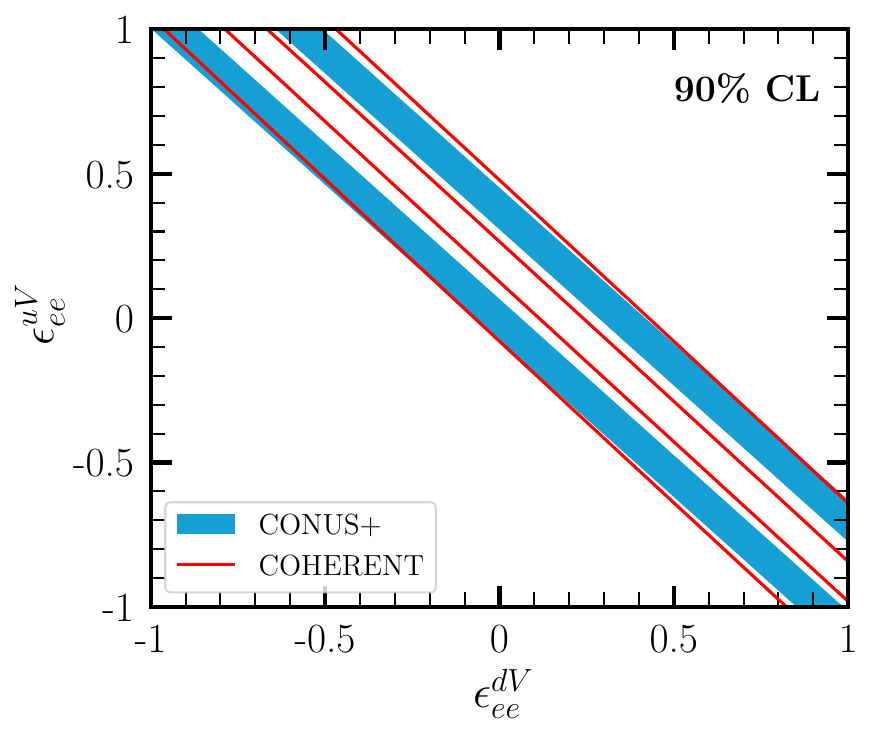}
\includegraphics[width=0.48\textwidth]{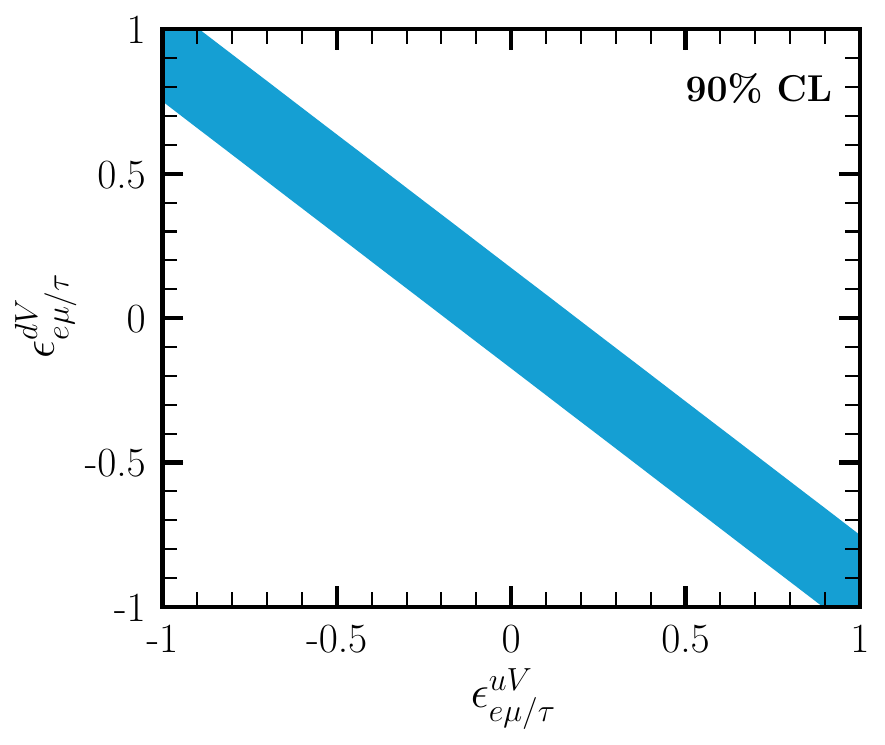}
 \includegraphics[width=0.48\textwidth]{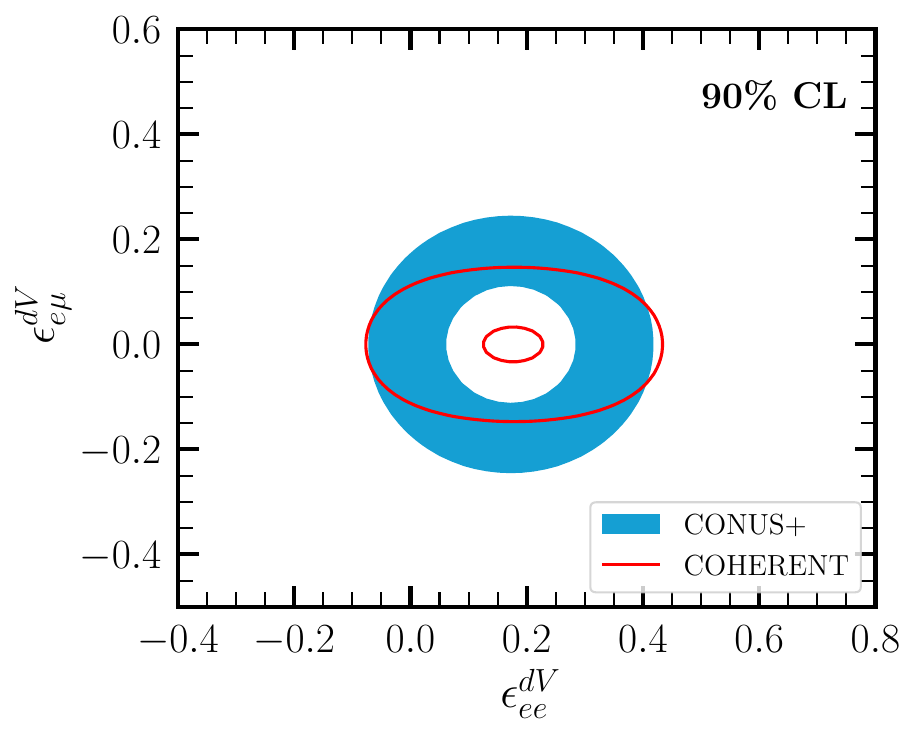}
\includegraphics[width=0.48\textwidth]{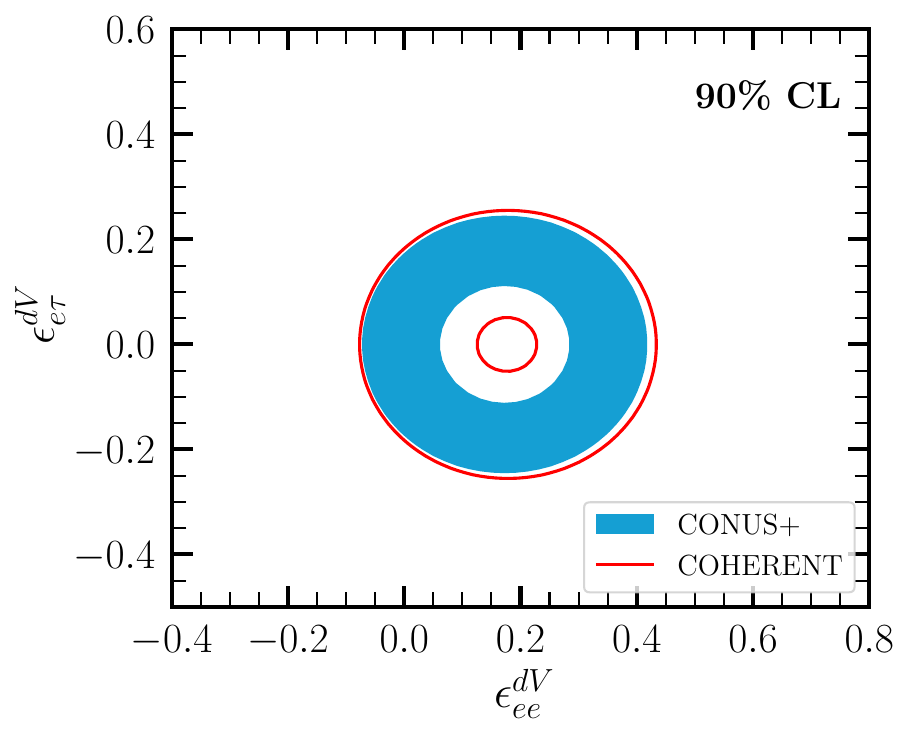}
\caption{$90 \%$ C.L. allowed regions assuming both nonuniversal and flavor-changing neutrino NSIs, from the CONUS+ analysis (blue), see the text for more details. The analysis  includes only \cevns~interactions. For comparison, we also show existing limits from COHERENT CsI + LAr data~\cite{DeRomeri:2022twg} (red).}
\label{fig:res_NSI}
\end{figure}

We proceed with the study of new neutrino interactions, beginning our discussion with the case of conventional NSIs, for which we present the CONUS+ constraints assuming both flavor-preserving and flavor-changing couplings. Our results are illustrated in Fig.~\ref{fig:res_NSI}, where the contours are shown at 90\% C.L. In the upper-left panel, we depict the constraints in the flavor-preserving plane $(\epsilon_{ee}^{dV}, \epsilon_{ee}^{uV})$,  while in the upper-right panel, we instead show the flavor-changing $(\epsilon_{e\mu/\tau}^{dV}, \epsilon_{e\mu/\tau}^{uV})$ plane. Note that for the reactor antineutrino CONUS+ experiment, the constraints on the parameters $\epsilon_{e \mu}^{qV}$ and $\epsilon_{e \tau}^{qV}$ are identical.
We should note that the CONUS+ constraint in the $(\epsilon_{ee}^{dV}, \epsilon_{ee}^{uV})$ plane has also been obtained in~\cite{Alpizar-Venegas:2025wor}. We find our result to be in agreement, although we perform a spectral fit instead of a single-bin analysis. This is partly understood because NSIs do not induce any spectral features in the predicted rates.  The rest of the results in Fig.~\ref{fig:res_NSI} are reported in this work for the first time.

\begin{table}[t]
\centering
\begin{tabular}{|c|c|c|}
\hline
\textbf{~~~NSI~~~}               & \textbf{CONUS+ (This Work)}                    & \textbf{COHERENT (CsI+LAr)}~\cite{DeRomeri:2022twg})                          \\ \hline
$\varepsilon_{ee}^{uV}$    & $~~~[-0.037, 0.026] \cup [0.348, 0.411]~~~$ & $~~~[-0.024, 0.045] \cup [0.34, 0.43]~~~$         \\ \hline
$\varepsilon_{ee}^{dV}$    & $[-0.034, 0.024] \cup [0.322, 0.380]$ & $   [-0.027, 0.048] \cup [0.30, 0.39] $ \\ \hline
$\varepsilon_{e\mu}^{uV}$  & $[-0.123, 0.123]$                       & $[-0.081, 0.081]$                           \\ \hline
$\varepsilon_{e\mu}^{dV}$  & $[-0.114, 0.114]$                       & $[-0.071, 0.071]$                           \\ \hline
$\varepsilon_{e\tau}^{uV}$ & $[-0.123, 0.123]$                       & $[-0.13, 0.13]$                             \\ \hline
$\varepsilon_{e\tau}^{dV}$ & $[-0.114, 0.114]$                       & $[-0.12, 0.12]$                             \\ \hline
\end{tabular}
\caption{NSI bounds at 1$\sigma$ C.L. obtained from CONUS$+$ data, and their comparison with COHERENT CsI+LAr bounds from~\cite{DeRomeri:2022twg}.}
\label{tab:nsi}
\end{table}

For comparison, we also depict in the $(\epsilon_{ee}^{dV}, \epsilon_{ee}^{uV})$ plane the constraints obtained in Ref.~\cite{DeRomeri:2022twg} from the analysis of COHERENT CsI+LAr data.  The two experiments provide similar constraints in shape, although CONUS+ turns out to be slightly better in some cases (see also Table~\ref{tab:nsi}). As can be seen, the allowed contours in the $(\epsilon_{ee}^{dV}, \epsilon_{ee}^{uV})$ parameter space appear as two distinct bands, reflecting the cancellations that can occur between the SM and NSI couplings, as can be inferred from Eq.~\eqref{eq:weak:charge:NSI}. Since the target materials have different proton to neutron ratios, the allowed bands for CONUS+ (Ge) and COHERENT (CsI and LAr) have different slopes.
On the other hand, the constraints on the flavor-changing parameter space, as expected, appear as a single band since no interference effects are possible in this case. In the lower-left and lower-right panels, we show the corresponding constraints by combining one flavor-preserving and one flavor-changing coupling. Specifically, the respective results are given assuming NSI $d$ quarks only, in the $(\epsilon_{ee}^{dV}, \epsilon_{e\mu}^{dV})$ and $(\epsilon_{ee}^{dV}, \epsilon_{e\tau}^{dV})$ planes. We compare with the corresponding constraints extracted from the analysis of COHERENT data~\cite{DeRomeri:2022twg}. In this case, we find that the two experiments provide complementary results, since they constrain slightly different regions of the available parameter space. The CONUS+ bounds are stronger on $\epsilon_{ee}^{dV}$, while a bit less competitive for $\epsilon_{e\mu}^{dV}$, the reason being related to the large $\nu_\mu$ component of the COHERENT neutrino flux. While not explicitly shown here, we have verified that the corresponding constraints on NSI with $u$ quarks are slightly weaker (see also Table~\ref{tab:nsi}). This is expected since the number of neutrons is always larger than the one of protons,  translating into stronger exclusions for the the NSI coupling with $d$ quarks. For clarity, let us note again that, as far as CONUS+ is concerned, the constraints on $\epsilon_{e\mu}^{qV}$ and $\epsilon_{e\tau}^{qV}$ are identical, however we distinguish the axis labeling to be consistent with the constraints from COHERENT.

To summarize the NSI results, we provide in Table~\ref{tab:nsi} the one-dimensional constraints together with others existing in the literature. At this point, we should warn the reader that special care should be taken when comparing CONUS+ and COHERENT CsI + LAr~\cite{DeRomeri:2022twg} bounds. Indeed, the effect of the uncertainty on the nuclear radius on COHERENT NSI limits was not considered in Ref.~\cite{DeRomeri:2022twg} (for a relevant discussion focusing on the COHERENT-LAr data only, see Ref.~\cite{Miranda:2019wdy}).
On the other hand,  given the lower neutrino energies, the CONUS+ constraints are not affected by nuclear size uncertainties. We have explicitly verified that CONUS+ data are not capable of constraining the nuclear root mean square radii of germanium, since for the reactor experiments \cevns~occurs in a regime of (almost) full coherency.

\subsection{Neutrino generalized interactions}
\label{subsec:res-NGI}

\begin{figure}[t]
\centering
\includegraphics[width=0.48\textwidth]{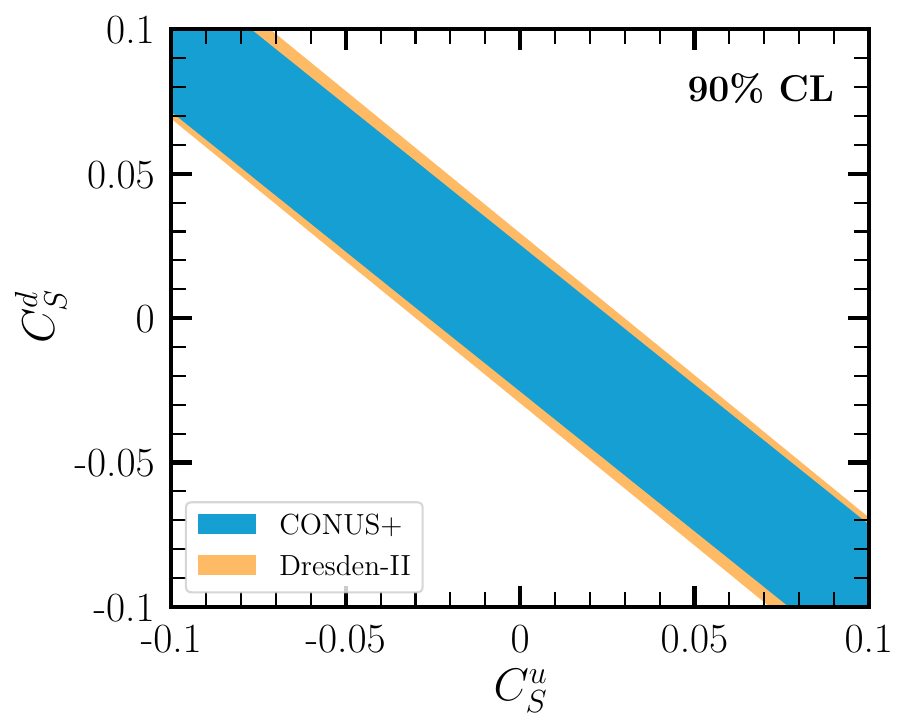}
\includegraphics[width=0.48\textwidth]{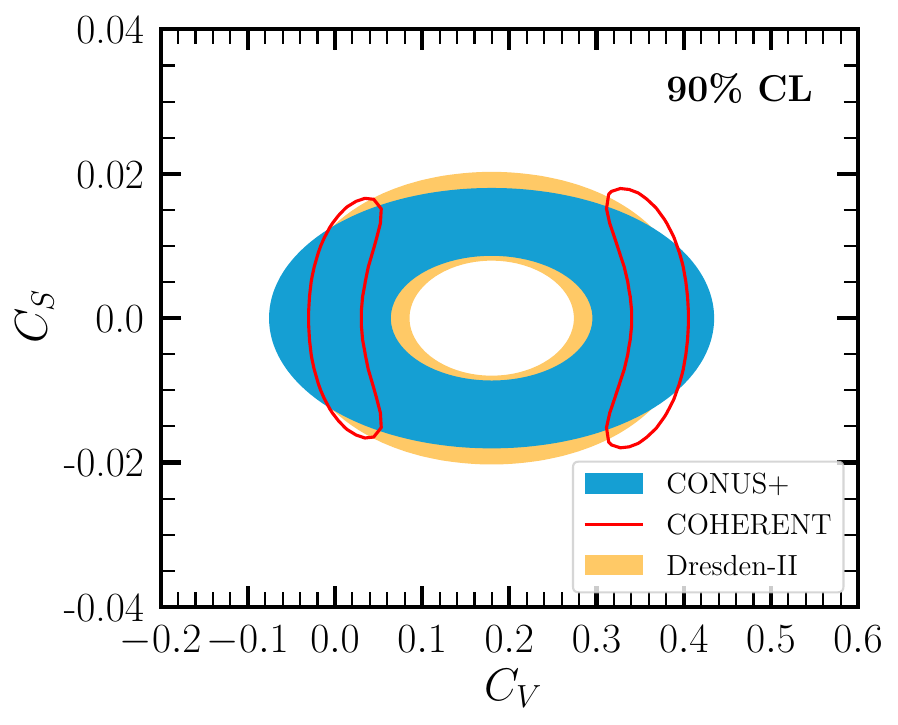}
\includegraphics[width=0.48\textwidth]{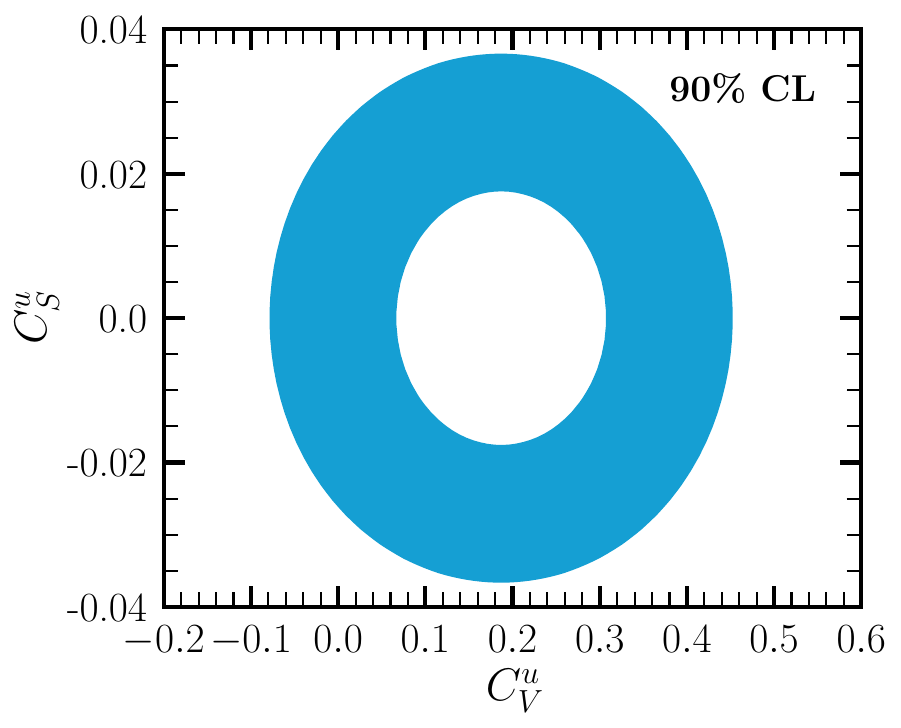}
\includegraphics[width=0.48\textwidth]{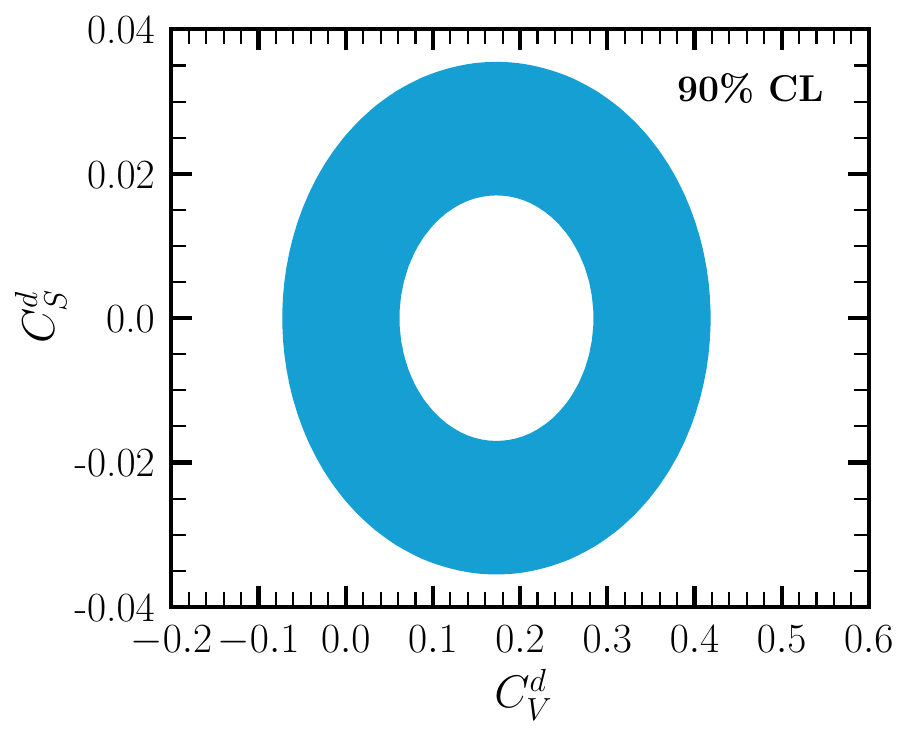}
\caption{$90 \%$ C.L. allowed regions on scalar and vector NGIs, from the CONUS+ analysis (blue), see the text for more details. The analysis  includes only \cevns~interactions. For comparison, we also show existing limits from COHERENT (CsI + LAr) data~\cite{DeRomeri:2022twg} (red) and from Dresden-II data~\cite{Majumdar:2022nby} (orange).}
\label{fig:res_NGI}
\end{figure}

Going beyond conventional vector-type NSIs, we now explore the impact of recent CONUS+ data on the NGI framework, which accommodates additional Lorentz-invariant interactions~\cite{AristizabalSierra:2018eqm}. As explained in Sec.~\ref{sec:theory}, the axial-vector, tensor, and pseudoscalar interactions can be safely neglected. Here, we focus on scalar and vector interactions separately, as well as on the combined effect of a simultaneous presence of both scalar and vector NGIs. Considering scalar interactions only, we present in the upper-left panel of Fig.~\ref{fig:res_NGI} the 90\% C.L. allowed regions, in the $(C_S^u, C_S^d)$ plane.\footnote{As explained in Sec.~\ref{sec:theory}, for simplicity we absorb the neutrino coupling $C_{\nu a}$ in the quark-dependent $C_a^q$ coupling, as implied by the definition $C_a^q =\sqrt{C_{\nu a} \cdot C_{q a}}$.} As expected, the contour is a single band due to the lack of interference with the SM \cevns~cross section. Then, we allow both scalar and vector interactions to be present at the same time and show the  $(C_V, C_S)$ contour region at 90\% C.L. in the upper right panel of Fig.~\ref{fig:res_NGI}. For simplicity, in this case we have assumed universal quark couplings for the scalar and vector cases, i.e., $C_S^u = C_S^d \equiv C_S$ and $C_V^u = C_V^d \equiv C_V$. The constrained region has a ring shape because of the interference effect taking place when the vector interaction is on. For completeness, in the lower-left and lower-right panels of  Fig.~\ref{fig:res_NGI}, we show results for scalar and vector interactions with $u$ and $d$ quarks only, respectively. As stressed previously in the NSI discussion, the NGIs  involving $d$ quarks are more constrained compared to those with $u$ quarks, due to the neutron dominance in the \cevns~cross section. 

It is interesting to compare our results obtained from the analysis of the CONUS+ data with existing results from the analysis of Dresden-II data from Ref.~\cite{Majumdar:2022nby}, and from the COHERENT CsI+LAr data in Ref.~\cite{DeRomeri:2022twg}. In the $(C_S^u, C_S^d)$ plane, we find that CONUS+ slightly improves the bounds obtained from the Dresden-II analysis~\cite{Majumdar:2022nby}. Focusing on the $(C_V, C_S)$ parameter space instead, the complementarity among the three experiments is evident. The allowed regions from  CONUS+ and Dresden-II have similar shapes, with the latter  being slightly more restrictive for the vector coupling. On the other hand,  
 for the case of scalar interactions, CONUS+ is slightly more constraining than both COHERENT and Dresden-II. Finally, the lower uncertainty of the COHERENT CsI dataset leads to the strongest constraints on this parameter space, appearing as two distinct regions,\footnote{We have rescaled the COHERENT constraints on the $C_V$ coupling, to account for a missing factor of 2 affecting the results shown in Ref.~\cite{DeRomeri:2022twg}.} as expected due to the possible destructive interference between the vector NGI and the SM contribution.

\subsection{Light mediators}
\label{subsec:res-lightmed}

\begin{figure}[!htb]
\centering
\includegraphics[width=0.48\textwidth]{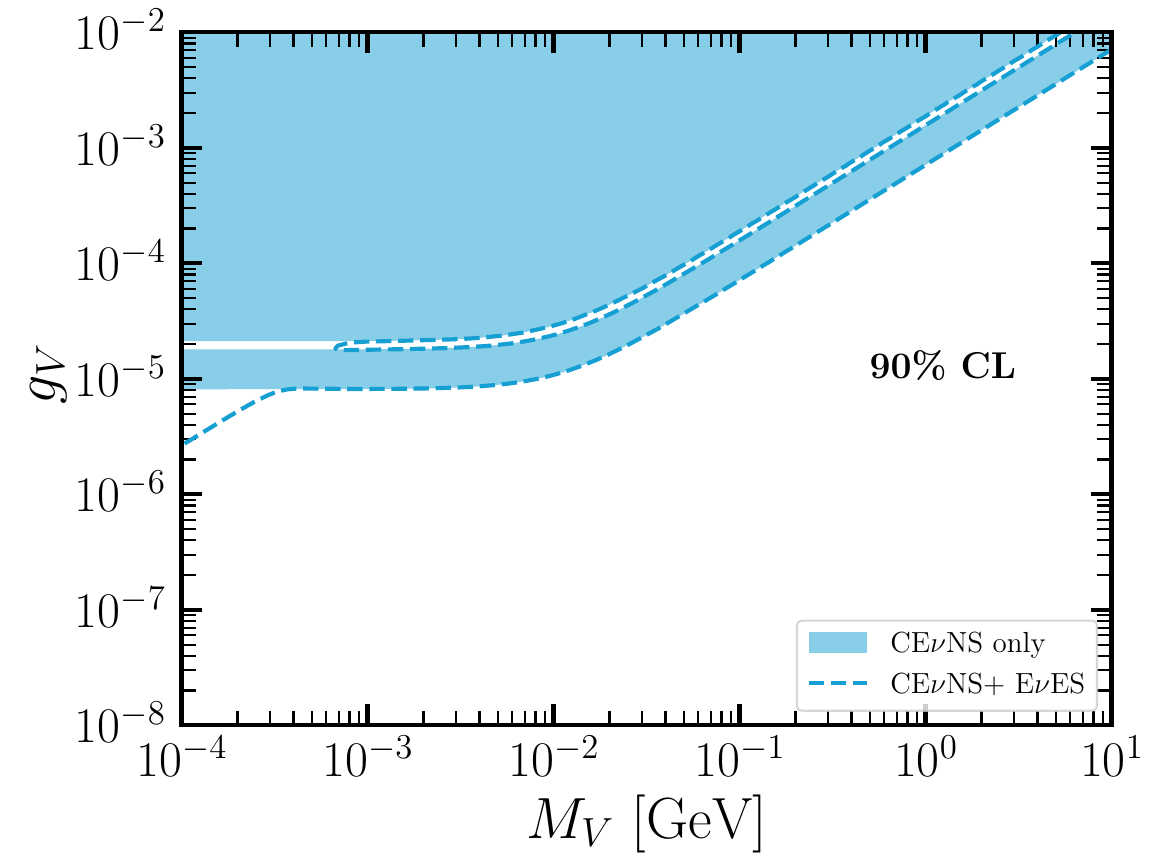}
\includegraphics[width=0.48\textwidth]{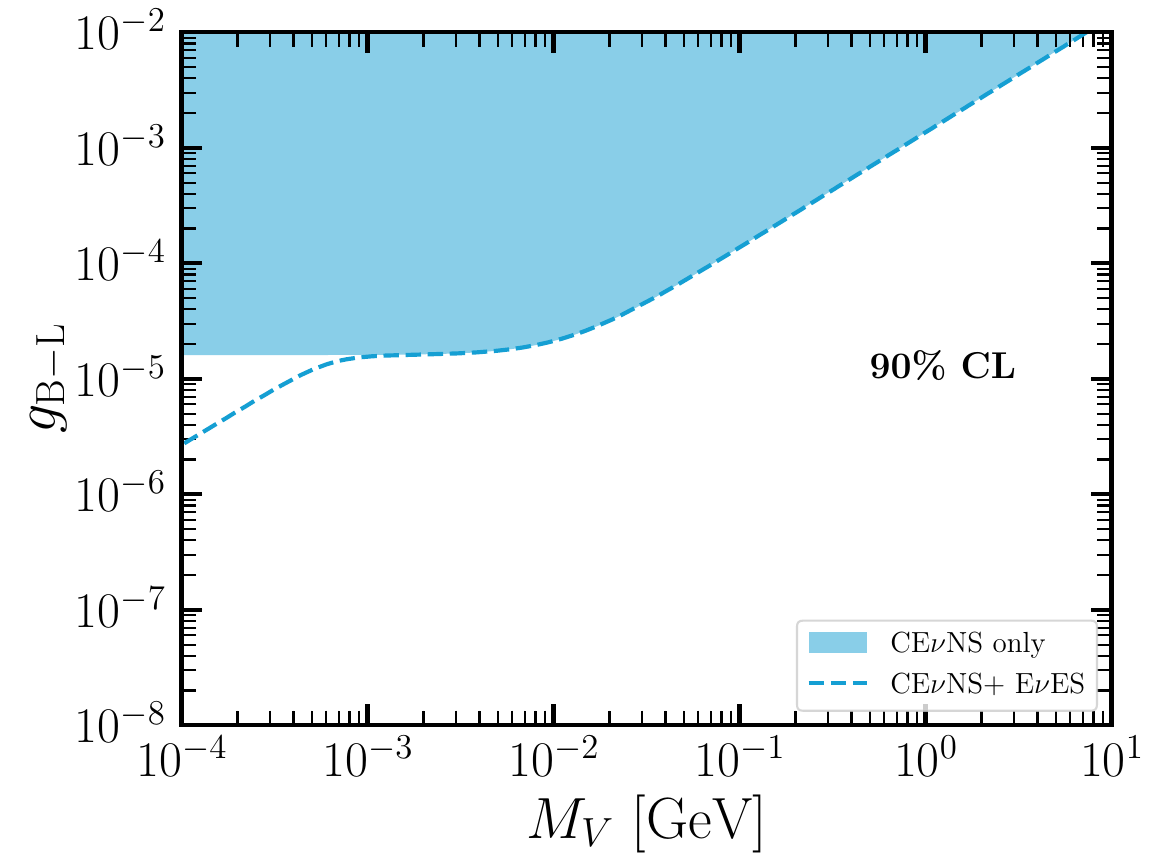}
\caption{90\% C.L. exclusion limits for new neutrino
interactions with a light vector mediator obtained from the analysis of CONUS+ data, for the universal scenario (left) and for the B-L model (right).}
\label{fig:res_LV}
\end{figure}

We now proceed with the discussion of new interactions with light scalar or vector mediators. We show in Fig.~\ref{fig:res_LV} the $90\%$ C.L.  from our CONUS+ statistical analysis, of only \cevns~data (light blue shaded area) and of \cevns+\eves~data (light blue dashed contour). The left panel corresponds to the universal scenario (not anomaly-free), which, we recall, assumes the new light vector mediator to couple with equal strengths to neutrinos, electrons, and first-generation quarks. The figure on the right, instead, depicts the exclusion limits for the B-L model. As evident from the figures, incorporating the \eves~events improves the bounds by over one order of magnitude for $M_V \lesssim 1$ MeV. In the universal model, a small allowed region appears due to the allowed destructive interference between the new vector interaction and the SM. This white band is reduced when taking into account the \eves~signal. 
All in all, the \cevns-only limit improves by a factor of about 2 upon the \cevns-only result from COHERENT CsI+LAr~\cite{DeRomeri:2022twg}  (see also \cite{AtzoriCorona:2022moj}), while at $M_V = 0.01$ MeV the CONUS+ bound is one order of magnitude stronger. Compared to the Dresden-II limit~\cite{AristizabalSierra:2022axl} obtained assuming the iron-filter quenching factor (see also \cite{Coloma:2022avw}), the \cevns-only CONUS+ bound is a factor of few less stringent. Nevertheless, let us highlight that the CONUS+ result~\cite{Ackermann:2025obx} is found to be in agreement with the Lindhard theory, and hence in conflict with~\cite{Colaresi:2022obx}. In all cases, the bounds obtained accounting for both \cevns~and \eves~events improve previous results from COHERENT and Dresden-II at low mediator masses. The reason is that, in that regime, the \eves~cross section is enhanced proportionally to $\propto E_{\rm er}^{-2}$. Notice also that the CONUS+ bounds completely rules out the Large Mixing Angle-Dark solution found in oscillations, in the electron flavor, as already pointed out in earlier papers~\cite{Denton:2018xmq,Denton:2022nol}. 

\begin{figure}[t]
\centering
\includegraphics[width=0.49\textwidth]{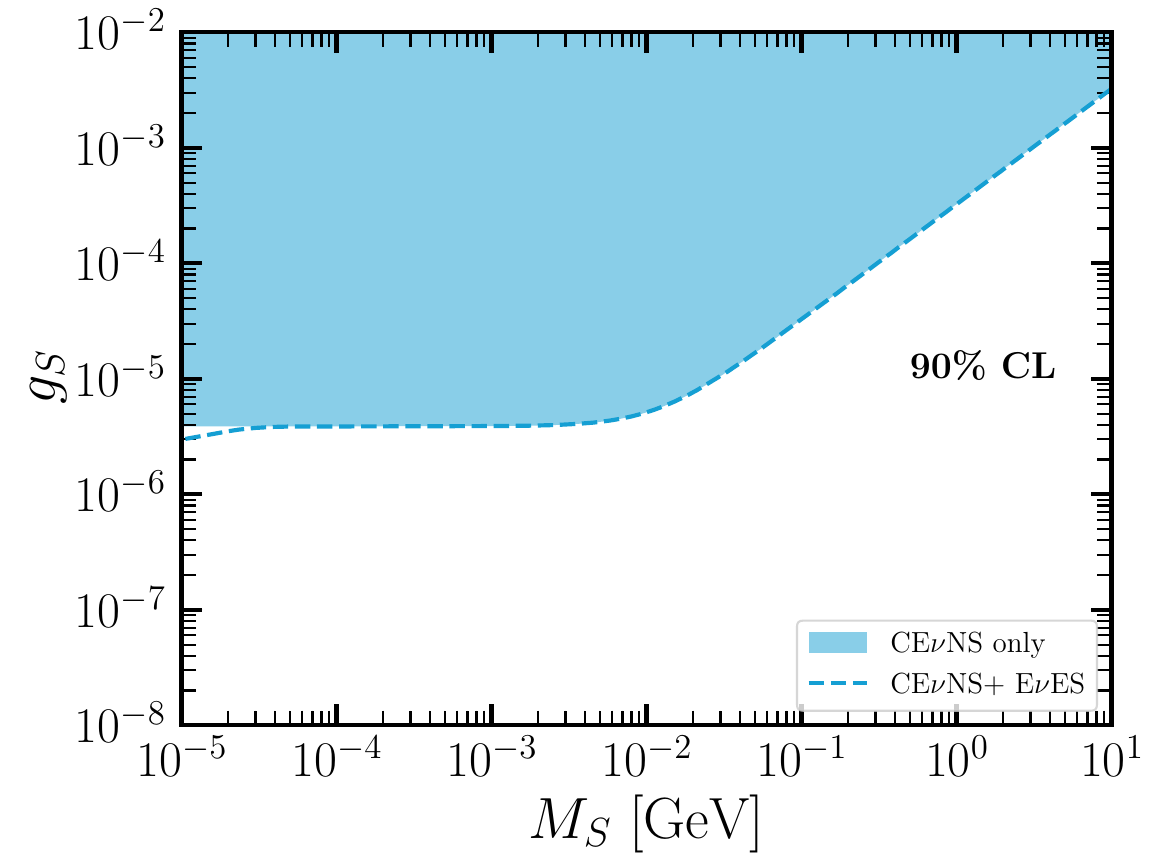}
\includegraphics[width=0.49\textwidth]{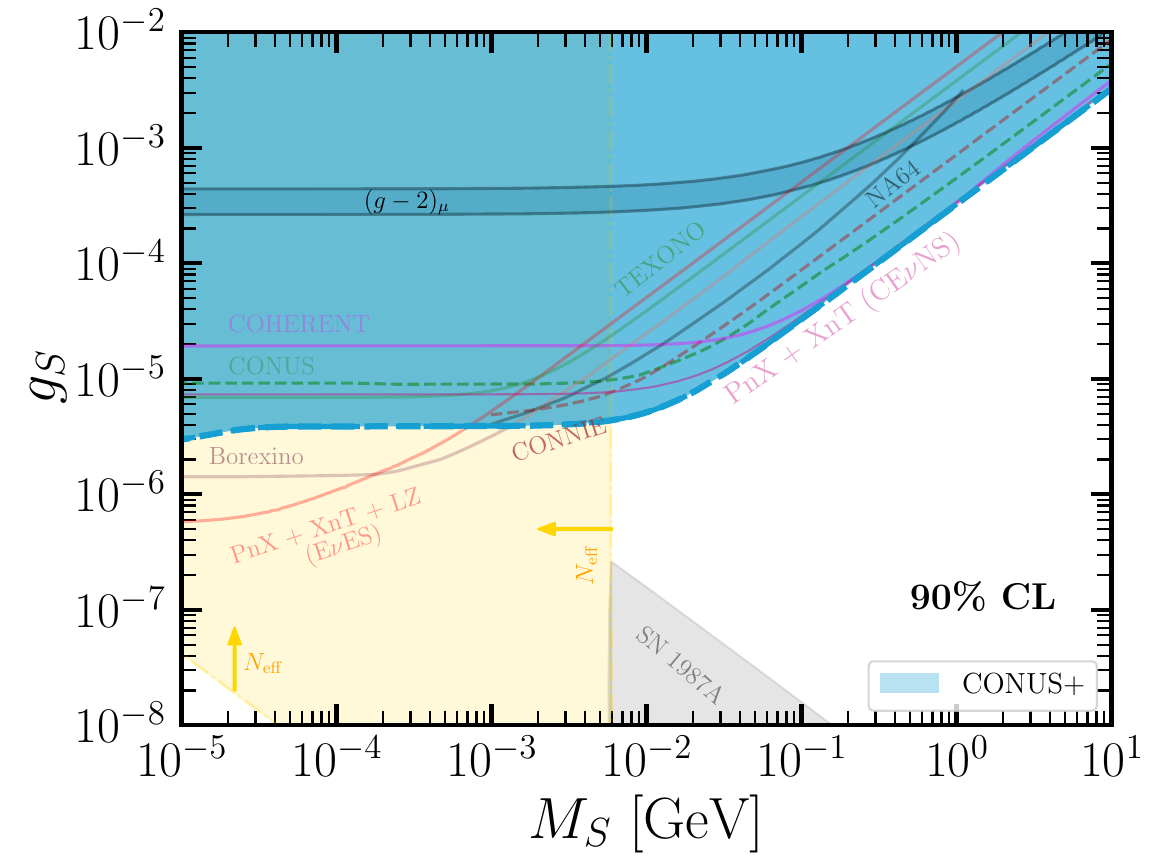}
\caption{Left panel: 90\% C.L. exclusion limits for new neutrino
interactions with a light scalar mediator obtained from the analysis of CONUS+ data. Right panel: 90\% C.L. exclusion limit, including both \cevns~and \eves~signals. Existing bounds from other searches are also shown for comparison.}
\label{fig:res_LS}
\end{figure}
Moving now to a new interaction with a light scalar mediator, we present our $90\%$ C.L.  from the CONUS+ analysis in the left panel of Fig.~\ref{fig:res_LS}. As before, we distinguish the results with and without \eves, although in this case its inclusion does not significantly affect the bounds. In fact, in this case, the \eves~cross section scales as $\propto E_{\rm er}^{-1}$. The impact of \eves~events starts to be visible at $M_V \lesssim 0.03$ MeV. In this case, the CONUS+ bound on the light scalar mediator improves the COHERENT CsI+LAr~\cite{DeRomeri:2022twg} bound (see also \cite{AtzoriCorona:2022moj}) by a factor of 4, while they are again less stringent than those from Dresden-II~\cite{AristizabalSierra:2022axl}, although with the different quenching factor assumption, as previously explained. Finally, let us note that our results are found to be in agreement with those presented in~\cite{Chattaraj:2025fvx}, focusing on the same (scalar and B-L) scenarios.

\begin{figure}[!htb]
\centering
\includegraphics[width=0.49\textwidth]{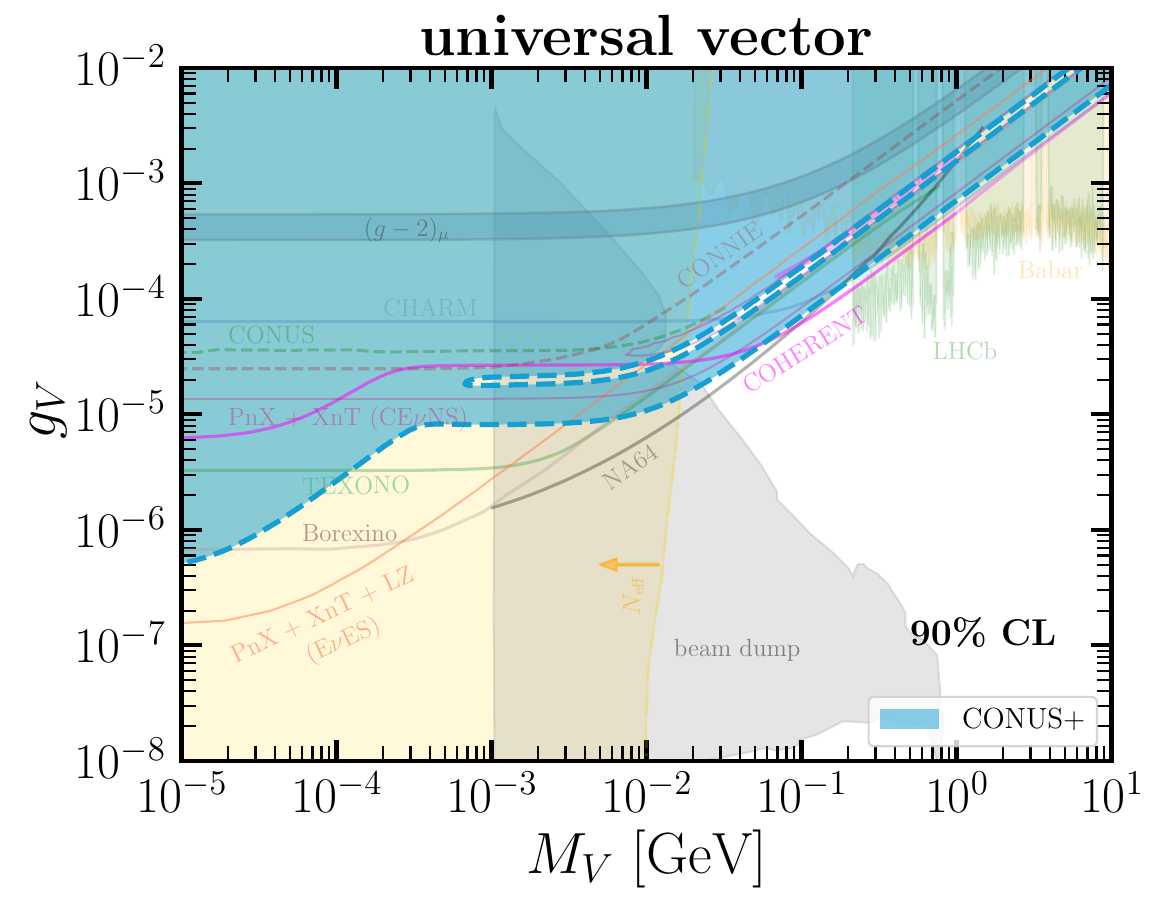}
\includegraphics[width=0.49\textwidth]{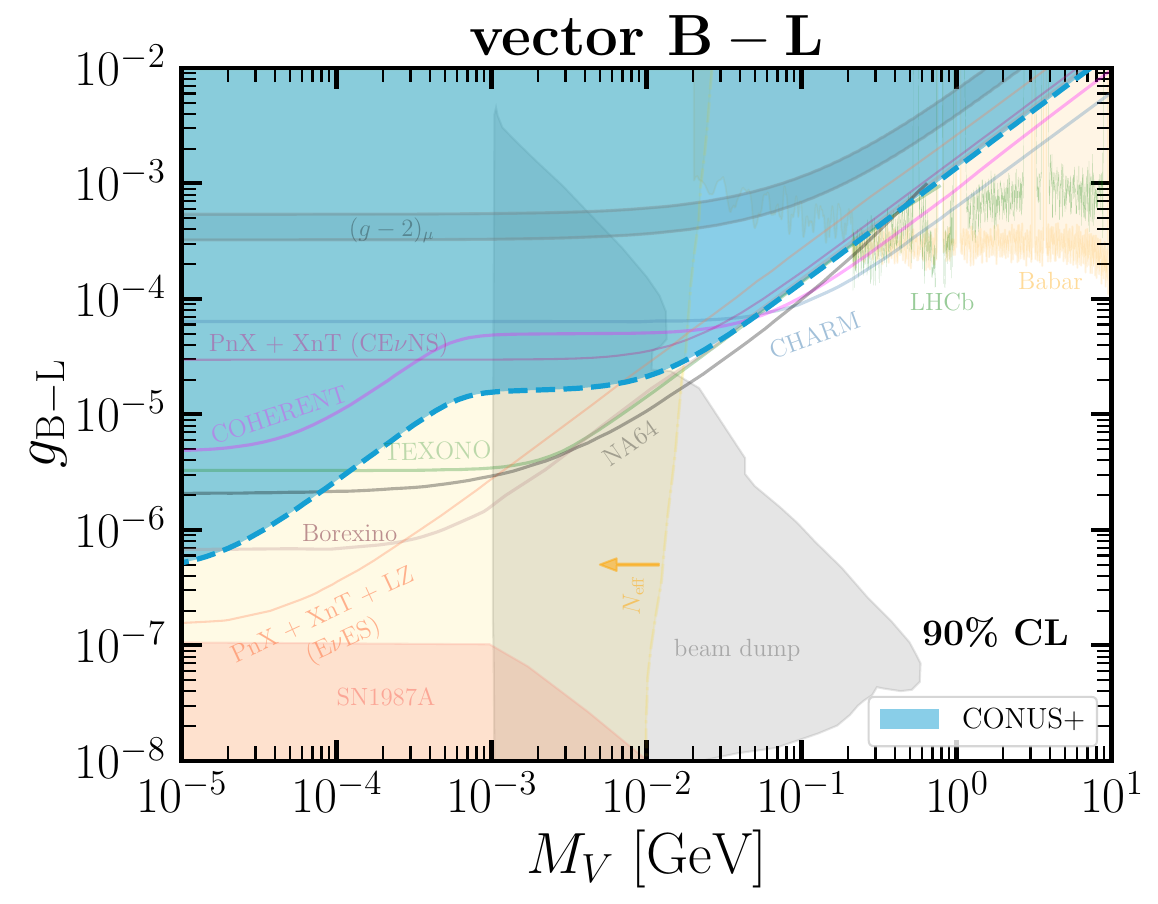}
\caption{90\% C.L. exclusion limits for new neutrino
interactions with a light vector mediator, for the universal (left) and B-L (right) models, obtained from the analysis of CONUS+ data. Both \cevns~and \eves~signals are included. Existing bounds from other searches are also shown for comparison.}
\label{fig:res_lightmed_all}
\end{figure}

For the sake of completeness, we present in Fig.~\ref{fig:res_LS} (right panel) and in Fig.~\ref{fig:res_lightmed_all} our results for the scalar mediator and for the vector mediators, compared to other existing bounds on the same interactions. We include limits from other \cevns~measurements,
COHERENT~\cite{DeRomeri:2022twg,AtzoriCorona:2022moj}, CONUS~\cite{CONUS:2021dwh,Lindner:2024eng}, and CONNIE~\cite{CONNIE:2019xid,CONNIE:2024pwt}, and from a combined analysis of PandaX-4T and XENONnT~\cite{DeRomeri:2024iaw}. We also include limits from elastic neutrino-electron scattering data at BOREXINO~\cite{Coloma:2022avw}, CHARM-II~\cite{Bauer:2018onh}, and TEXONO~\cite{TEXONO:2009knm,Bauer:2018onh}, and from a combined analysis of PandaX-4T, XENONnT, and LZ electron recoil data~\cite{A:2022acy,DeRomeri:2024dbv}.
We further show exclusions from NA64~\cite{NA64:2021xzo,NA64:2022yly,NA64:2023wbi} and other fixed-target and beam-dump experiments, which include  E137~\cite{Bjorken:1988as}, E141~\cite{Riordan:1987aw},
KEK~\cite{Konaka:1986cb},
E774~\cite{Bross:1989mp},
Orsay~\cite{Davier:1989wz,Bjorken:2009mm,Andreas:2012mt}, and
$\nu$-CAL~I~\cite{Blumlein:1990ay,Blumlein:1991xh,Blumlein:2011mv,Blumlein:2013cua}, from
CHARM~\cite{CHARM:1985anb,Gninenko:2012eq},
NOMAD~\cite{NOMAD:2001eyx},
PS191~\cite{Bernardi:1985ny,Gninenko:2011uv}, A1~\cite{Merkel:2014avp}, and APEX~\cite{APEX:2011dww}, and from
colliders
(BaBar~\cite{BaBar:2014zli,BaBar:2017tiz} and LHCb~\cite{LHCb:2017trq}).

The CONUS+ bound on the scalar interaction turns out to be the dominant one in the mass range $7~{\rm MeV} \lesssim M_S \lesssim 100$ MeV, surpassing the recent PandaX-4T and XENONnT~\cite{DeRomeri:2024iaw} result. For the vector interactions, the CONUS+ bound is not competitive versus the NA64 one (in the universal case) nor the Borexino, TEXONO, and CHARM limits (in the B-L model), but it does improve upon other \cevns~measurements, especially in the mass range $M_V \lesssim 100$ MeV. Very light mediators with $M_a \lesssim 10$ MeV are also, in general, in conflict with astrophysical and cosmological observations. Regarding the universal vector model, it seems clear that combining results from different \cevns~data, obtained with different target materials, can probe the degeneracy region.

\subsection{Light sterile neutrino oscillations}
\label{subsec:res-vsosc}

\begin{figure}[t]
\centering
\includegraphics[width=0.49\textwidth]{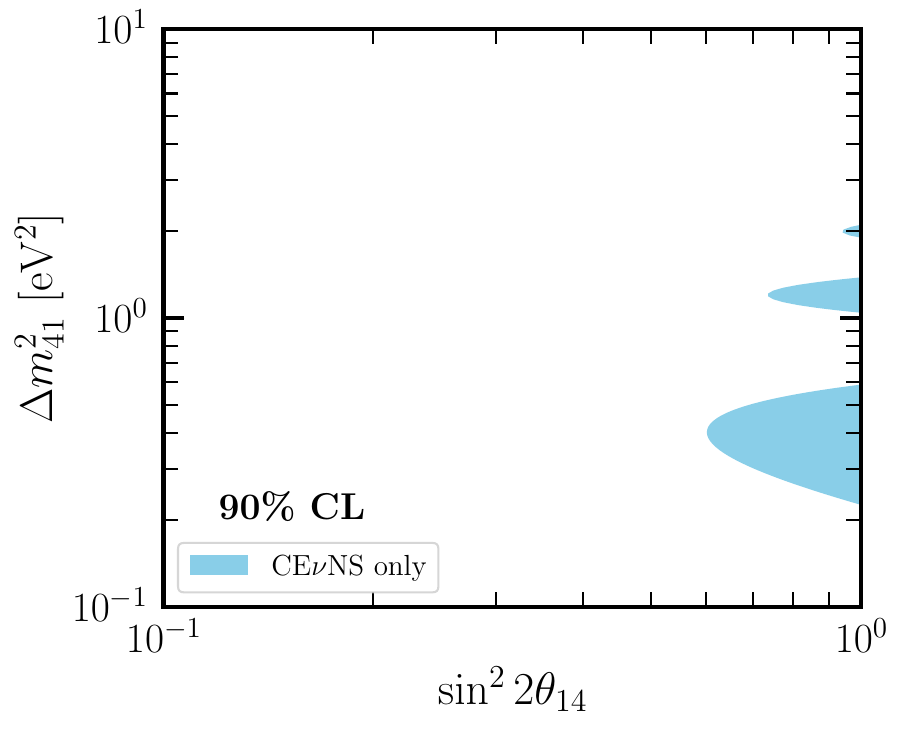}
\caption{90\% C.L. exclusion limits in the sterile oscillation parameter space $(\sin^2 2 \theta_{14}, \Delta m_{41}^2)$, obtained from the analysis of CONUS+ data.}
\label{fig:res_sterile_osc}
\end{figure}

Turning now to the sterile neutrino phenomenology, we begin our analysis by exploring the constraining power of the recent CONUS+ data regarding active-sterile neutrino oscillations. In Fig.~\ref{fig:res_sterile_osc}, we present the exclusion region at 90\% C.L. in the parameter space $(\sin^2 2 \theta_{14}, \Delta m_{41}^2)$. Evidently, the constraint is still very poor, far from the sensitivity reached by other experiments~\cite{Giunti:2022btk}. Notice that similar conclusions were reached in~\cite{DeRomeri:2022twg} by analyzing COHERENT CsI+LAr data.

\subsection{Sterile dipole portal}
\label{subsec:res-dipole}

\begin{figure}[t]
\centering
\includegraphics[width=0.49\textwidth]{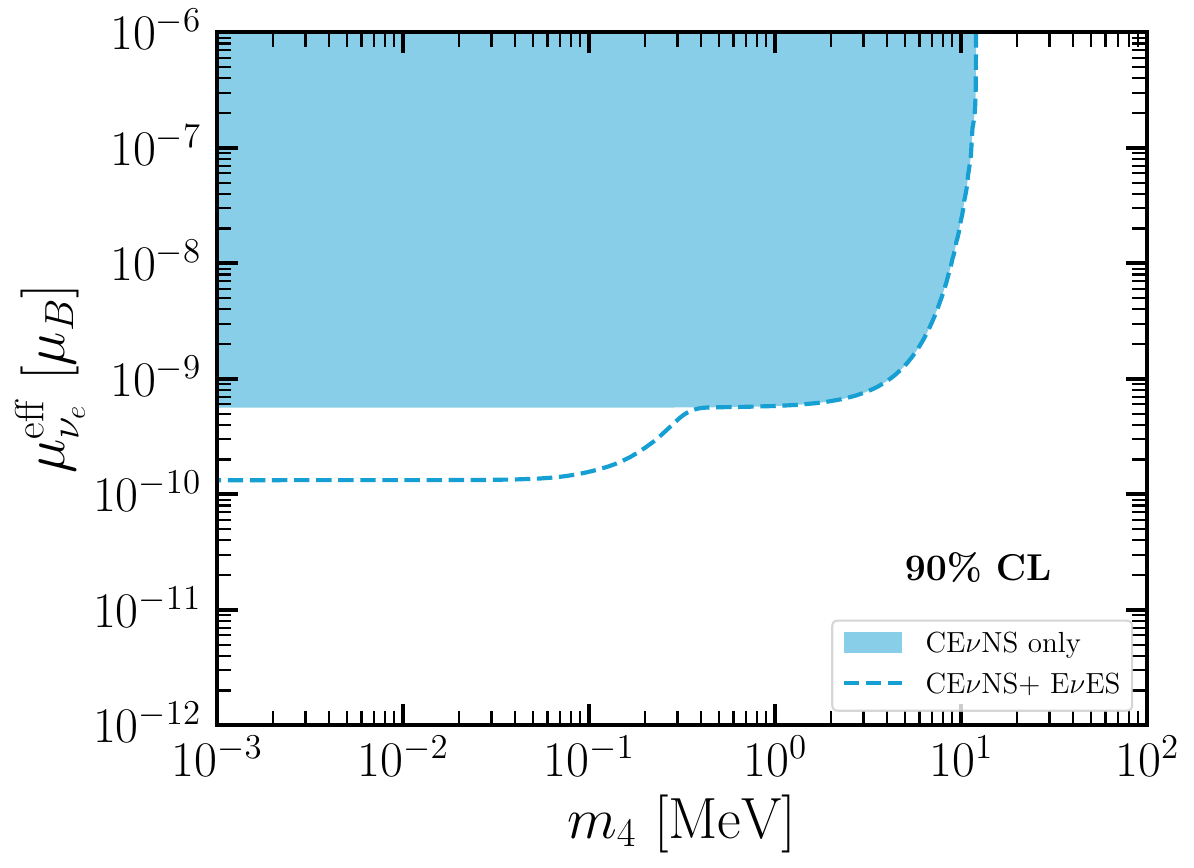}
\includegraphics[width=0.49\textwidth]{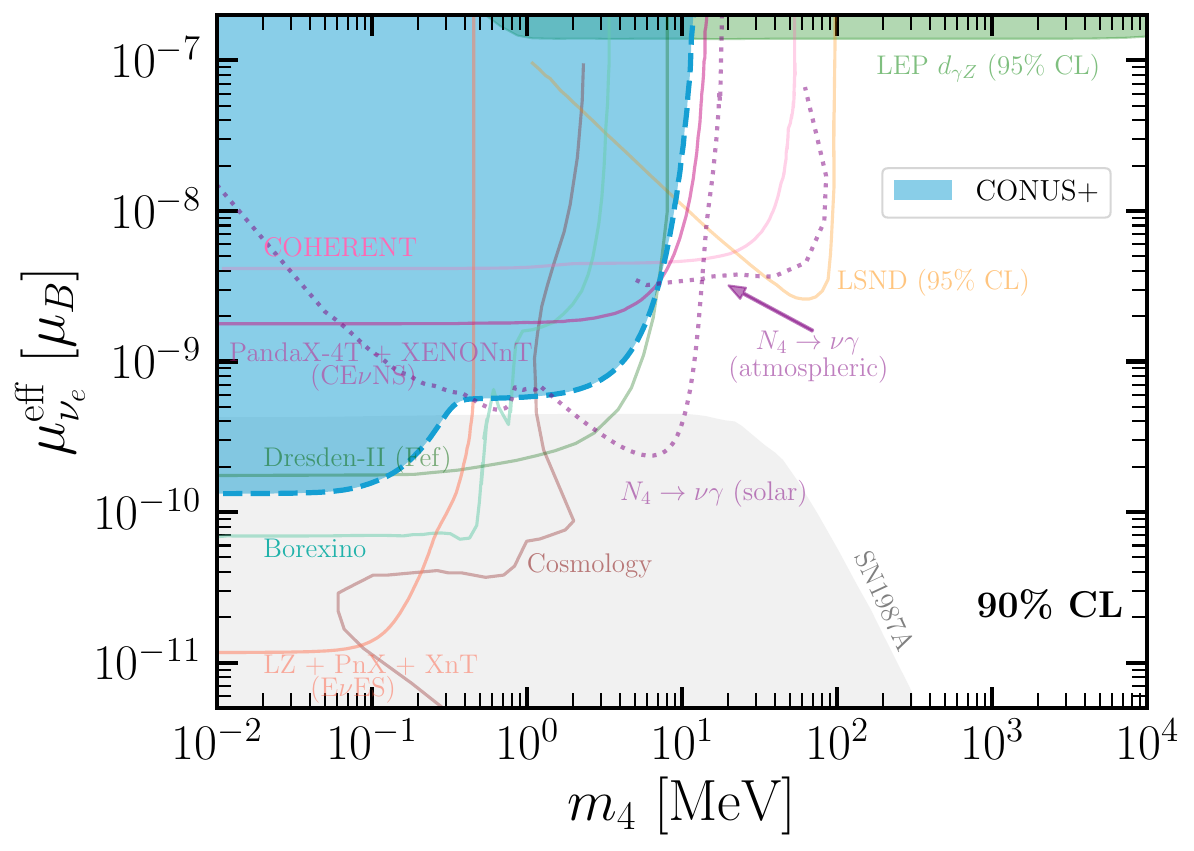}
\caption{Left: $90 \%$ CL exclusion regions for an effective neutrino magnetic moment inducing active-sterile transitions for electron neutrinos from the analysis of CONUS+ data. Right: our CONUS+ bound, including both \cevns~and \eves~signals, is compared to existing limits from other experiments. }
\label{fig:res_sterile_dipole}
\end{figure}

We now discuss the implications of the CONUS+ measurement on the sterile dipole portal scenario.
We present in Fig.~\ref{fig:res_sterile_dipole} (left) the 90\% C.L. exclusion contours, in terms of the effective magnetic moment inducing the transition $\nu_\ell + \mathcal{N} \to \nu_4 + \mathcal{N}$, and the mass of the sterile state, $m_4$. Since the \eves~cross section in this scenario is enhanced, we include the \eves~events in our analysis, depicted as light blue dashed lines. The light blue shaded area corresponds instead to the exclusion contour obtained when only \cevns~events are taken into account. Let us recall here that the typical neutrino energy of a nuclear reactor source leads to a kinematical constraint on the mass of the sterile state $ m_4^2 \lesssim 2m_\mathcal{N} T_\mathcal{N}\left(\sqrt{\frac{2}{m_\mathcal{N}  T_\mathcal{N}}}E_\nu -1\right) \lesssim 10$ MeV. Moreover, in the limit $m_4 \to 0$, the bound approaches the value obtained in the neutrino effective magnetic moment scenario (see Sec.~\ref{subsec:res-magmom}). 

In the right panel of Fig.~\ref{fig:res_sterile_dipole}, our 90\% C.L. exclusion bound from the CONUS+ analysis is superimposed on several existing limits from other searches. We show constraints from COHERENT CsI+LAr~\cite{Miranda:2021kre,DeRomeri:2022twg}, Dresden-II~\cite{AristizabalSierra:2022axl} [considering the iron-filter (Fef) quenching factor model], LSND~\cite{Magill:2018jla},
Borexino~\cite{Brdar:2020quo,Plestid:2020vqf}, 
LEP~\cite{Magill:2018jla}, $N_4 \to \nu \gamma$ decay from the analysis of solar (Borexino and Super-Kamiokande)~\cite{Plestid:2020vqf,Plestidlumsolnu} and atmospheric (Super-Kamiokande)~\cite{Gustafson:2022rsz} data,  a combined analysis~\cite{A:2022acy,Giunti:2023yha,DeRomeri:2024hvc} of electron recoil data from XENONnT, LUX-ZEPLIN (LZ), and PandaX-4T, and a combined analysis of \cevns~data from XENONnT and PandaX-4T~\cite{DeRomeri:2024hvc}.
Limits from SN1987A~\cite{Magill:2018jla,Chauhan:2024nfa}, and cosmological data, including BBN~\cite{Magill:2018jla,Brdar:2020quo} and CMB constraints on $\Delta N_\mathrm{eff}$~\cite{Brdar:2020quo}, are also shown for comparison. 
We find that the CONUS+ bounds, especially when including the \eves~signal, improve all other \cevns-inferred bounds, for $m_4 \lesssim 0.1$ MeV (despite falling in a region in tension with cosmology). In the mass range $0.2~{\rm MeV} \lesssim M_S \lesssim 8$~MeV, they are surpassed only by the Dresden-II results (though with a different quenching factor), by Borexino and Super Kamiokande limits from solar neutrino up-scattering inside the Earth ($N_4 \to \nu \gamma$).

\subsection{Up-scattering production of a sterile fermion}
\label{subsec:res-upscat}

The last BSM scenario considered is the up-scattering production of a sterile fermion,  $\nu_\ell \mathcal{N} \rightarrow \chi \mathcal{N}$ (and $\nu_\ell e \rightarrow \chi e$ for neutrino scattering on atomic electrons), mediated by a light scalar or vector particle. Also in this case, the mass of the produced fermion is constrained kinematically through $m_\chi \leq \sqrt{(m_\mathcal{N} (m_\mathcal{N}  + 2 E_{\nu}^{\mathrm{max}})} - m_\mathcal{N}$.

\begin{figure}[t]
\centering
\includegraphics[width=0.48\textwidth]{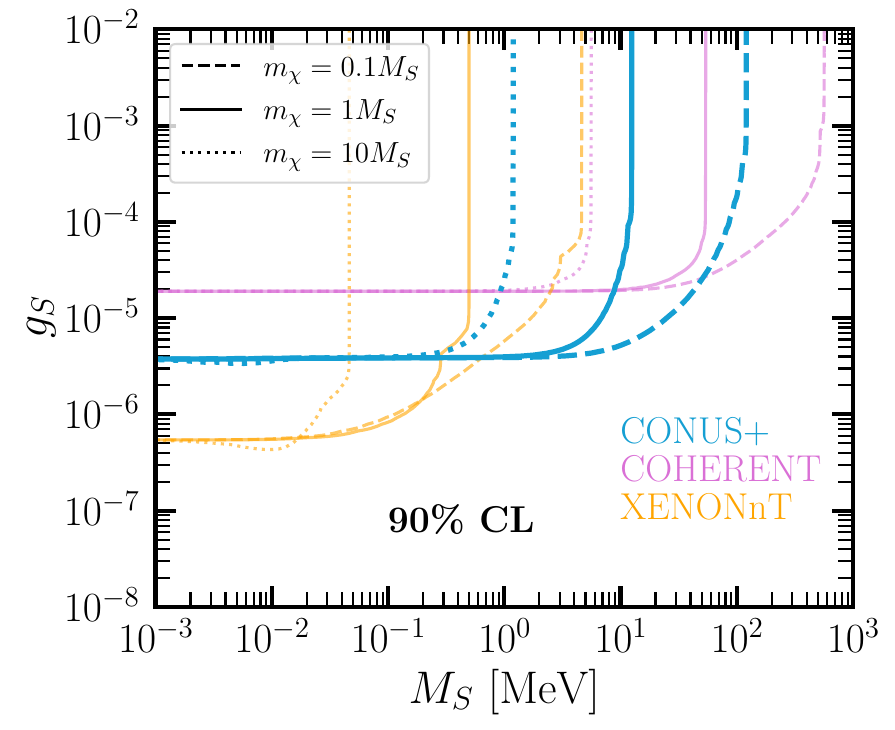}
\includegraphics[width=0.48\textwidth]{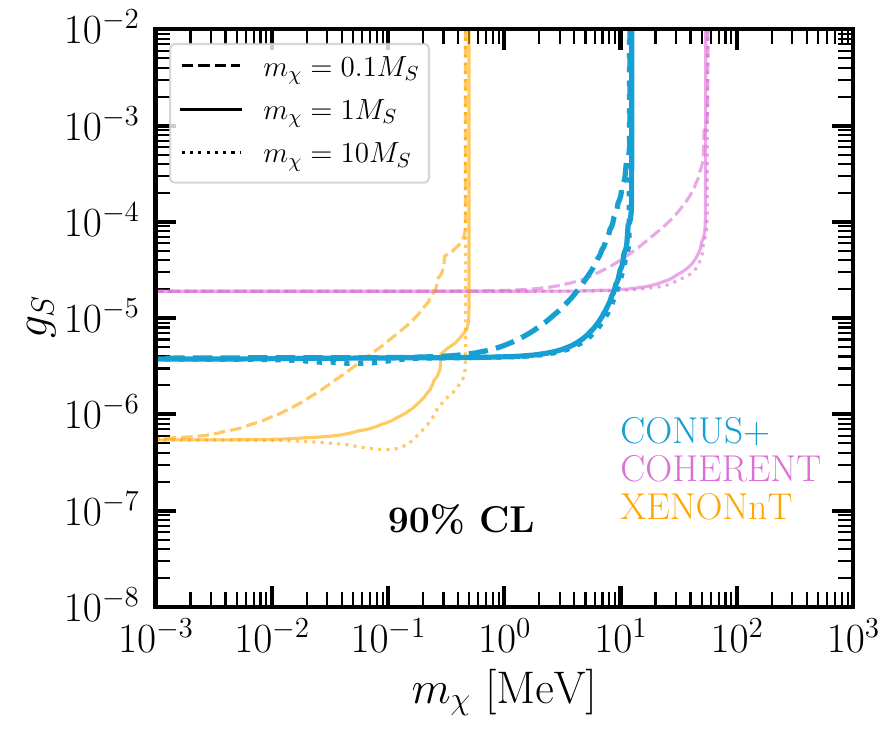}
\caption{90\% C.L. exclusion regions for the case of a scalar-mediated up-scattering production of a sterile fermion. The panel on the left shows the exclusion regions projecting on the mediator mass, while in the right panel the projection is done over the sterile fermion mass. We depict the limits for  three benchmark scenarios, $m_\chi= \{0.1, \,1, \,10\} \times \, M_S$.}
\label{fig:res_SNL_S}
\end{figure}

We show in Fig.~\ref{fig:res_SNL_S} the $90 \%$ C.L. exclusion regions from our CONUS+ analysis, in terms of the mediator mass (left panel) or the sterile fermion mass (right panel), for different benchmark ratios $m_\chi/M_S$. For the sake of comparison, we also show previously obtained limits from COHERENT CsI + LAr (magenta) data and XENONnT electron recoil data (orange)~\cite{Candela:2024ljb}. From this result, it is clear that CONUS+ data allow us to probe a region of parameter space currently unexplored, for $1~{\rm MeV} \lesssim M_S \lesssim 40$ MeV, for ratios  $m_\chi/M_S \lesssim 5$.

\begin{figure}[h]
\centering
\includegraphics[width=0.48\textwidth]{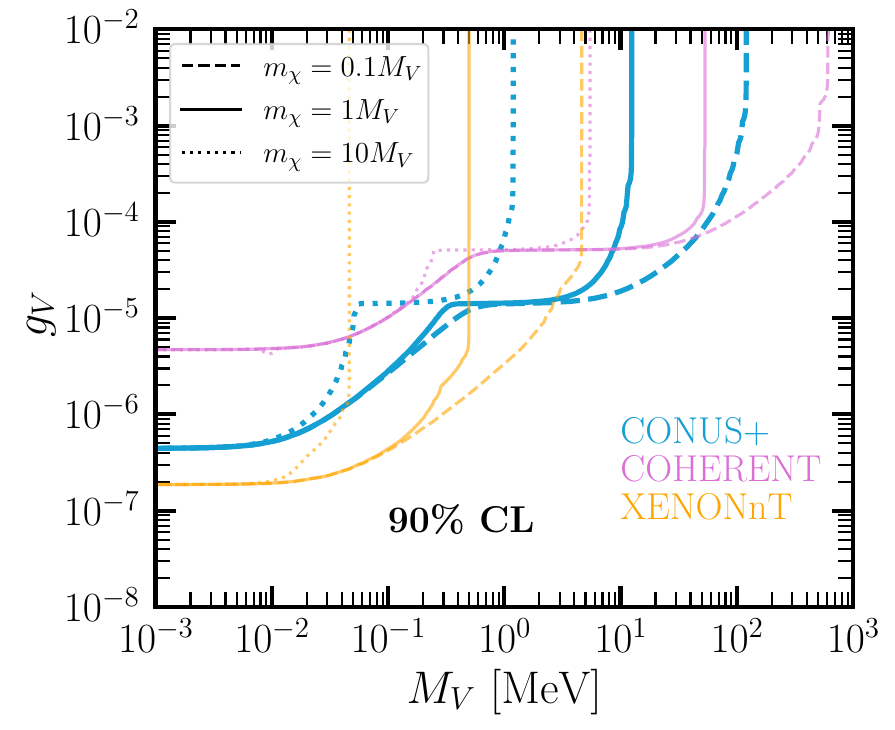}
\includegraphics[width=0.48\textwidth]{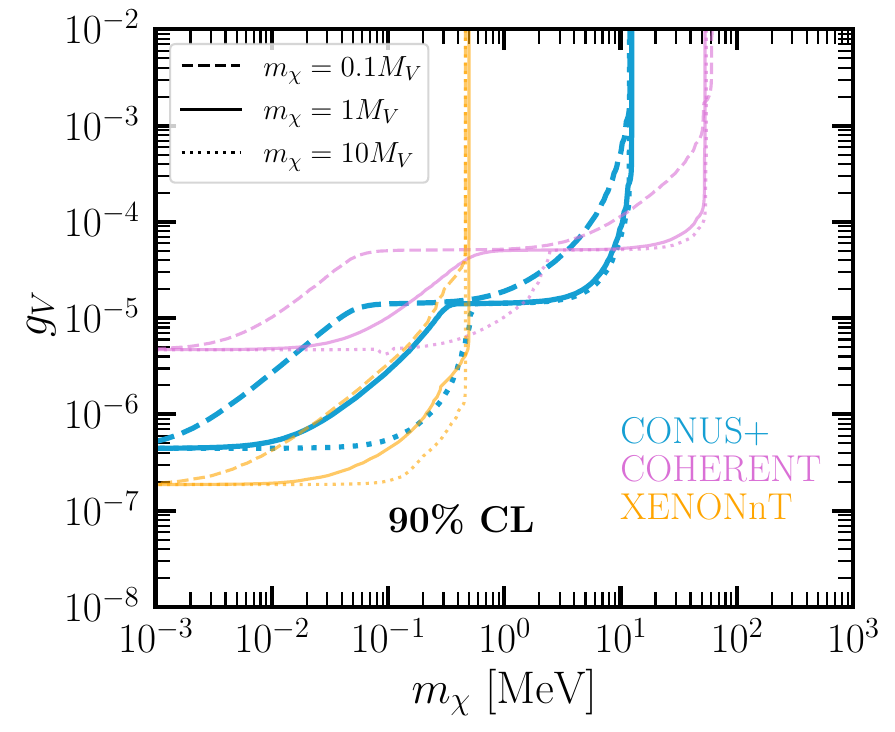}
\caption{90\% C.L. exclusion regions for the case of a vector-mediated up-scattering production of a sterile fermion. The panel on the left shows the exclusion regions projecting on the mediator mass, while in the right panel the projection is done over the sterile fermion mass. We depict the limits for  three benchmark scenarios, $m_\chi= \{0.1, \,1, \,10\} \times \, M_V$.}
\label{fig:res_SNL_V}
\end{figure}

Similarly, we analyze the up-scattering through a vector mediator. We show in Fig.~\ref{fig:res_SNL_V} the $90 \%$ C.L. exclusion regions from our CONUS+ analysis, in terms of the mediator mass (left panel) or the sterile fermion mass (right panel), for different benchmark ratios $m_\chi/M_V$. In this case, the CONUS+ improvement is a bit less pronounced, but still present in the intermediate mass region. 
These results show the complementarity between experiments using different neutrino sources (solar, reactor, and spallation source) in testing the up-scattering production via neutrino scattering on nuclei and atomic electrons.

\section{Conclusions}
\label{sec:conclusions}
The CONUS+ experiment has recently measured \cevns~induced from reactor neutrinos, using  germanium detectors with $\mathcal{O}(100)$ eV thresholds. The exposure used to obtain this  resulted in a statistical significance of $3.7 \sigma$, and the measurement was in agreement with the SM prediction. We have analyzed these data and investigated several implications in the context of both SM and BSM physics. First, we  obtained a determination of the weak mixing angle  at an energy scale of $\sim 10$ MeV. Then, we  explored implications of this observation concerning several BSM scenarios. We  considered new neutrino interactions in the form of NSIs, NGIs, and new scalar and vector light mediators. Next, we  studied active-sterile neutrino oscillations and the up-scattering production of sterile neutrinos through the sterile dipole portal, and through a light scalar or vector mediator. We  also explored neutrino magnetic moments, as an example of neutrino nontrivial electromagnetic properties that may arise in BSM theories. 

A reactor experiment like CONUS+ is specially suited to test spectral distortions at low-recoil energy. Consequently, the bounds obtained on neutrino magnetic moments and light mediators were rather strong and improved upon those inferred from spallation neutron source data, as expected. In this regard, they were somewhat comparable in reach to bounds previously obtained from the Dresden-II reactor experiment, though with some differences depending on the specific scenario and, most importantly, under different quenching factor modeling, a source of a tension between the two datasets. Some of the constraints were further improved through the inclusion of \eves~events in the statistical analysis, which were not distinguishable from nuclear recoils in the detectors. Forthcoming data with increased statistics will allow us to further improve these limits and probe currently unexplored regions of parameter space, in several cases. Additionally, the combination of reactor data with spallation neutron source and DM direct detection measurements, exploiting different sources and target materials, will be of pivotal importance to break the degeneracies that appear in some BSM scenarios (e.g., NSIs), and to extract more precise information on nuclear and SM physics of relevance for \cevns.

\section*{Acknowledgments}
 We are grateful to Anirban Majumdar for kindly sharing with us information regarding the resolution function of CONUS+. We are also deeply grateful to Christian Buck from  CONUS+ collaboration for sharing  useful information with us, and for helping us interpret their experimental results.
We are further indebted to Matteo Cadeddu, Francesca Dordei, Mattia Atzori-Corona, Nicola Cargioli, and Carlo Giunti for sharing useful information on the CONUS+ analysis and for helpful discussions, and to Mariam Tórtola for insightful discussions.
This work is supported by the Spanish grants CNS2023-144124 (MCIN/AEI/10.13039/501100011033 and “Next Generation EU”/PRTR), PID2023-147306NB-I00 and CEX2023-001292-S (MCIU/AEI/ 10.13039/501100011033), as well as by the grants CIDEXG/2022/20, CIAPOS/2022/254, and CIPROM/2021/054 from Generalitat Valenciana.

\bibliographystyle{utphys}
\bibliography{main}  

\providecommand{\href}[2]{#2}\begingroup\raggedright\begin{thebibliography}{100}

\bibitem{Freedman:1973yd}
D.~Z. Freedman, ``{Coherent Neutrino Nucleus Scattering as a Probe of the Weak
  Neutral Current},'' \href{http://dx.doi.org/10.1103/PhysRevD.9.1389}{{\em
  Phys. Rev. D} {\bfseries 9} (1974) 1389--1392}.

\bibitem{Kopeliovich:1974mv}
V.~B. Kopeliovich and L.~L. Frankfurt, ``{Isotopic and chiral structure of
  neutral current},'' {\em JETP Lett.} {\bfseries 19} (1974) 145--147.

\bibitem{Drukier:1984vhf}
A.~Drukier and L.~Stodolsky, ``{Principles and Applications of a Neutral
  Current Detector for Neutrino Physics and Astronomy},''
  \href{http://dx.doi.org/10.1103/PhysRevD.30.2295}{{\em Phys. Rev. D}
  {\bfseries 30} (1984) 2295}.

\bibitem{Abdullah:2022zue}
M.~Abdullah {\em et~al.}, ``{Coherent elastic neutrino-nucleus scattering:
  Terrestrial and astrophysical applications},''
  \href{http://arxiv.org/abs/2203.07361}{{\ttfamily arXiv:2203.07361
  [hep-ph]}}.

\bibitem{Cadeddu:2023tkp}
M.~Cadeddu, F.~Dordei, and C.~Giunti, ``{A view of coherent elastic
  neutrino-nucleus scattering},''
  \href{http://dx.doi.org/10.1209/0295-5075/ace7f0}{{\em EPL} {\bfseries 143}
  no.~3, (2023) 34001}, \href{http://arxiv.org/abs/2307.08842}{{\ttfamily
  arXiv:2307.08842 [hep-ph]}}.

\bibitem{COHERENT:2017ipa}
{\bfseries COHERENT} Collaboration, D.~Akimov {\em et~al.}, ``{Observation of
  Coherent Elastic Neutrino-Nucleus Scattering},''
  \href{http://dx.doi.org/10.1126/science.aao0990}{{\em Science} {\bfseries
  357} (2017) 1123--1126}, \href{http://arxiv.org/abs/1708.01294}{{\ttfamily
  arXiv:1708.01294 [nucl-ex]}}.

\bibitem{COHERENT:2021xmm}
{\bfseries COHERENT} Collaboration, D.~Akimov {\em et~al.}, ``{Measurement of
  the Coherent Elastic Neutrino-Nucleus Scattering Cross Section on CsI by
  COHERENT},'' \href{http://dx.doi.org/10.1103/PhysRevLett.129.081801}{{\em
  Phys. Rev. Lett.} {\bfseries 129} (2022) 081801},
  \href{http://arxiv.org/abs/2110.07730}{{\ttfamily arXiv:2110.07730
  [hep-ex]}}.

\bibitem{COHERENT:2020ybo}
{\bfseries COHERENT} Collaboration, D.~Akimov {\em et~al.}, ``{COHERENT
  Collaboration data release from the first detection of coherent elastic
  neutrino-nucleus scattering on argon},''
  \href{http://arxiv.org/abs/2006.12659}{{\ttfamily arXiv:2006.12659
  [nucl-ex]}}.

\bibitem{Adamski:2024yqt}
S.~Adamski {\em et~al.}, ``{First detection of coherent elastic
  neutrino-nucleus scattering on germanium},''
  \href{http://arxiv.org/abs/2406.13806}{{\ttfamily arXiv:2406.13806
  [hep-ex]}}.

\bibitem{Goodman:1984dc}
M.~W. Goodman and E.~Witten, ``{Detectability of Certain Dark Matter
  Candidates},'' \href{http://dx.doi.org/10.1103/PhysRevD.31.3059}{{\em Phys.
  Rev. D} {\bfseries 31} (1985) 3059}.

\bibitem{Schumann:2019eaa}
M.~Schumann, ``{Direct Detection of WIMP Dark Matter: Concepts and Status},''
  \href{http://dx.doi.org/10.1088/1361-6471/ab2ea5}{{\em J. Phys. G} {\bfseries
  46} no.~10, (2019) 103003}, \href{http://arxiv.org/abs/1903.03026}{{\ttfamily
  arXiv:1903.03026 [astro-ph.CO]}}.

\bibitem{Billard:2021uyg}
J.~Billard {\em et~al.}, ``{Direct detection of dark matter\textemdash{}APPEC
  committee report*},'' \href{http://dx.doi.org/10.1088/1361-6633/ac5754}{{\em
  Rept. Prog. Phys.} {\bfseries 85} no.~5, (2022) 056201},
  \href{http://arxiv.org/abs/2104.07634}{{\ttfamily arXiv:2104.07634
  [hep-ex]}}.

\bibitem{XENON:2024ijk}
{\bfseries XENON} Collaboration, E.~Aprile {\em et~al.}, ``{First Indication of
  Solar $^8B$~Neutrinos via Coherent Elastic Neutrino-Nucleus Scattering with
  XENONnT},'' \href{http://dx.doi.org/10.1103/PhysRevLett.133.191002}{{\em
  Phys. Rev. Lett.} {\bfseries 133} (2024) 191002},
  \href{http://arxiv.org/abs/2408.02877}{{\ttfamily arXiv:2408.02877
  [nucl-ex]}}.

\bibitem{PandaX:2024muv}
{\bfseries PandaX} Collaboration, Z.~Bo {\em et~al.}, ``{First Indication of
  Solar B8 Neutrinos through Coherent Elastic Neutrino-Nucleus Scattering in
  PandaX-4T},'' \href{http://dx.doi.org/10.1103/PhysRevLett.133.191001}{{\em
  Phys. Rev. Lett.} {\bfseries 133} no.~19, (2024) 191001},
  \href{http://arxiv.org/abs/2407.10892}{{\ttfamily arXiv:2407.10892
  [hep-ex]}}.

\bibitem{AristizabalSierra:2024nwf}
D.~Aristizabal~Sierra, N.~Mishra, and L.~Strigari, ``{Implications of first
  neutrino-induced nuclear recoil measurements in direct detection
  experiments},'' \href{http://arxiv.org/abs/2409.02003}{{\ttfamily
  arXiv:2409.02003 [hep-ph]}}.

\bibitem{Li:2024iij}
G.~Li, C.-Q. Song, F.-J. Tang, and J.-H. Yu, ``{Constraints on neutrino
  non-standard interactions from COHERENT and PandaX-4T},''
  \href{http://arxiv.org/abs/2409.04703}{{\ttfamily arXiv:2409.04703
  [hep-ph]}}.

\bibitem{Xia:2024ytb}
S.-y. Xia, ``{Measuring Solar neutrino Fluxes in Direct Detection Experiments
  in the Presence of Light Mediators},''
  \href{http://arxiv.org/abs/2410.01167}{{\ttfamily arXiv:2410.01167
  [hep-ph]}}.

\bibitem{Maity:2024aji}
T.~N. Maity and C.~Boehm, ``{First measurement of the weak mixing angle in
  direct detection experiments},''
  \href{http://arxiv.org/abs/2409.04385}{{\ttfamily arXiv:2409.04385
  [hep-ph]}}.

\bibitem{DeRomeri:2024iaw}
V.~De~Romeri, D.~K. Papoulias, and C.~A. Ternes, ``{Bounds on new neutrino
  interactions from the first CE$\nu$NS data at direct detection
  experiments},'' \href{http://arxiv.org/abs/2411.11749}{{\ttfamily
  arXiv:2411.11749 [hep-ph]}}.

\bibitem{DeRomeri:2024hvc}
V.~De~Romeri, D.~K. Papoulias, G.~Sanchez~Garcia, C.~A. Ternes, and
  M.~T\'ortola, ``{Neutrino electromagnetic properties and sterile dipole
  portal in light of the first solar CE$\nu$NS data},''
  \href{http://arxiv.org/abs/2412.14991}{{\ttfamily arXiv:2412.14991
  [hep-ph]}}.

\bibitem{Blanco-Mas:2024ale}
P.~Blanco-Mas, P.~Coloma, G.~Herrera, P.~Huber, J.~Kopp, I.~M. Shoemaker, and
  Z.~Tabrizi, ``{Clarity through the Neutrino Fog: Constraining New Forces in
  Dark Matter Detectors},'' \href{http://arxiv.org/abs/2411.14206}{{\ttfamily
  arXiv:2411.14206 [hep-ph]}}.

\bibitem{Kerman:2016jqp}
{\bfseries TEXONO} Collaboration, S.~Kerman, V.~Sharma, M.~Deniz, H.~T. Wong,
  J.~W. Chen, H.~B. Li, S.~T. Lin, C.~P. Liu, and Q.~Yue, ``{Coherency in
  Neutrino-Nucleus Elastic Scattering},''
  \href{http://dx.doi.org/10.1103/PhysRevD.93.113006}{{\em Phys. Rev. D}
  {\bfseries 93} no.~11, (2016) 113006},
  \href{http://arxiv.org/abs/1603.08786}{{\ttfamily arXiv:1603.08786
  [hep-ph]}}.

\bibitem{CONNIE:2021ggh}
{\bfseries CONNIE} Collaboration, A.~Aguilar-Arevalo {\em et~al.}, ``{Search
  for coherent elastic neutrino-nucleus scattering at a nuclear reactor with
  CONNIE 2019 data},'' \href{http://dx.doi.org/10.1007/JHEP05(2022)017}{{\em
  JHEP} {\bfseries 05} (2022) 017},
  \href{http://arxiv.org/abs/2110.13033}{{\ttfamily arXiv:2110.13033
  [hep-ex]}}.

\bibitem{CONUS:2020skt}
{\bfseries CONUS} Collaboration, H.~Bonet {\em et~al.}, ``{Constraints on
  elastic neutrino nucleus scattering in the fully coherent regime from the
  CONUS experiment},''
  \href{http://dx.doi.org/10.1103/PhysRevLett.126.041804}{{\em Phys. Rev.
  Lett.} {\bfseries 126} no.~4, (2021) 041804},
  \href{http://arxiv.org/abs/2011.00210}{{\ttfamily arXiv:2011.00210
  [hep-ex]}}.

\bibitem{nGeN:2022uje}
{\bfseries \ensuremath{\nu}GeN} Collaboration, I.~Alekseev {\em et~al.},
  ``{First results of the \ensuremath{\nu}GeN experiment on coherent elastic
  neutrino-nucleus scattering},''
  \href{http://dx.doi.org/10.1103/PhysRevD.106.L051101}{{\em Phys. Rev. D}
  {\bfseries 106} no.~5, (2022) L051101},
  \href{http://arxiv.org/abs/2205.04305}{{\ttfamily arXiv:2205.04305
  [nucl-ex]}}.

\bibitem{NUCLEUS:2019igx}
{\bfseries NUCLEUS} Collaboration, G.~Angloher {\em et~al.}, ``{Exploring
  $\hbox {CE}\nu \hbox {NS}$ with NUCLEUS at the Chooz nuclear power plant},''
  \href{http://dx.doi.org/10.1140/epjc/s10052-019-7454-4}{{\em Eur. Phys. J. C}
  {\bfseries 79} no.~12, (2019) 1018},
  \href{http://arxiv.org/abs/1905.10258}{{\ttfamily arXiv:1905.10258
  [physics.ins-det]}}.

\bibitem{MINER:2016igy}
{\bfseries MINER} Collaboration, G.~Agnolet {\em et~al.}, ``{Background Studies
  for the MINER Coherent Neutrino Scattering Reactor Experiment},''
  \href{http://dx.doi.org/10.1016/j.nima.2017.02.024}{{\em Nucl. Instrum. Meth.
  A} {\bfseries 853} (2017) 53--60},
  \href{http://arxiv.org/abs/1609.02066}{{\ttfamily arXiv:1609.02066
  [physics.ins-det]}}.

\bibitem{Ricochet:2021rjo}
{\bfseries Ricochet} Collaboration, C.~Augier {\em et~al.}, ``{Ricochet
  Progress and Status},''
  \href{http://dx.doi.org/10.1007/s10909-023-02971-5}{{\em J. Low Temp. Phys.}
  {\bfseries 212} (2023) 127--137},
  \href{http://arxiv.org/abs/2111.06745}{{\ttfamily arXiv:2111.06745
  [physics.ins-det]}}.

\bibitem{Akimov:2022xvr}
D.~Y. Akimov {\em et~al.}, ``{The RED-100 experiment},''
  \href{http://dx.doi.org/10.1088/1748-0221/17/11/T11011}{{\em JINST}
  {\bfseries 17} no.~11, (2022) T11011},
  \href{http://arxiv.org/abs/2209.15516}{{\ttfamily arXiv:2209.15516
  [physics.ins-det]}}.

\bibitem{NEON:2022hbk}
{\bfseries NEON} Collaboration, J.~J. Choi {\em et~al.}, ``{Exploring coherent
  elastic neutrino-nucleus scattering using reactor electron antineutrinos in
  the NEON experiment},''
  \href{http://dx.doi.org/10.1140/epjc/s10052-023-11352-x}{{\em Eur. Phys. J.
  C} {\bfseries 83} no.~3, (2023) 226},
  \href{http://arxiv.org/abs/2204.06318}{{\ttfamily arXiv:2204.06318
  [hep-ex]}}.

\bibitem{NEWS-G:2021mhf}
{\bfseries NEWS-G} Collaboration, L.~Balogh {\em et~al.}, ``{Quenching factor
  measurements of neon nuclei in neon gas},''
  \href{http://dx.doi.org/10.1103/PhysRevD.105.052004}{{\em Phys. Rev. D}
  {\bfseries 105} no.~5, (2022) 052004},
  \href{http://arxiv.org/abs/2109.01055}{{\ttfamily arXiv:2109.01055
  [physics.ins-det]}}.

\bibitem{SBC:2021yal}
{\bfseries SBC, CE\ensuremath{\nu}NS Theory Group at IF-UNAM} Collaboration,
  L.~J. Flores {\em et~al.}, ``{Physics reach of a low threshold scintillating
  argon bubble chamber in coherent elastic neutrino-nucleus scattering reactor
  experiments},'' \href{http://dx.doi.org/10.1103/PhysRevD.103.L091301}{{\em
  Phys. Rev. D} {\bfseries 103} no.~9, (2021) L091301},
  \href{http://arxiv.org/abs/2101.08785}{{\ttfamily arXiv:2101.08785
  [hep-ex]}}.

\bibitem{Colaresi:2022obx}
J.~Colaresi, J.~I. Collar, T.~W. Hossbach, C.~M. Lewis, and K.~M. Yocum,
  ``{Measurement of Coherent Elastic Neutrino-Nucleus Scattering from Reactor
  Antineutrinos},''
  \href{http://dx.doi.org/10.1103/PhysRevLett.129.211802}{{\em Phys. Rev.
  Lett.} {\bfseries 129} no.~21, (2022) 211802},
  \href{http://arxiv.org/abs/2202.09672}{{\ttfamily arXiv:2202.09672
  [hep-ex]}}.

\bibitem{AristizabalSierra:2022axl}
D.~Aristizabal~Sierra, V.~De~Romeri, and D.~K. Papoulias, ``{Consequences of
  the Dresden-II reactor data for the weak mixing angle and new physics},''
  \href{http://dx.doi.org/10.1007/JHEP09(2022)076}{{\em JHEP} {\bfseries 09}
  (2022) 076}, \href{http://arxiv.org/abs/2203.02414}{{\ttfamily
  arXiv:2203.02414 [hep-ph]}}.

\bibitem{Coloma:2022avw}
P.~Coloma, I.~Esteban, M.~C. Gonzalez-Garcia, L.~Larizgoitia, F.~Monrabal, and
  S.~Palomares-Ruiz, ``{Bounds on new physics with data of the Dresden-II
  reactor experiment and COHERENT},''
  \href{http://dx.doi.org/10.1007/JHEP05(2022)037}{{\em JHEP} {\bfseries 05}
  (2022) 037}, \href{http://arxiv.org/abs/2202.10829}{{\ttfamily
  arXiv:2202.10829 [hep-ph]}}.

\bibitem{Liao:2022hno}
J.~Liao, H.~Liu, and D.~Marfatia, ``{Implications of the first evidence for
  coherent elastic scattering of reactor neutrinos},''
  \href{http://dx.doi.org/10.1103/PhysRevD.106.L031702}{{\em Phys. Rev. D}
  {\bfseries 106} no.~3, (2022) L031702},
  \href{http://arxiv.org/abs/2202.10622}{{\ttfamily arXiv:2202.10622
  [hep-ph]}}.

\bibitem{Majumdar:2022nby}
A.~Majumdar, D.~K. Papoulias, R.~Srivastava, and J.~W.~F. Valle, ``{Physics
  implications of recent Dresden-II reactor data},''
  \href{http://dx.doi.org/10.1103/PhysRevD.106.093010}{{\em Phys. Rev. D}
  {\bfseries 106} no.~9, (2022) 093010},
  \href{http://arxiv.org/abs/2208.13262}{{\ttfamily arXiv:2208.13262
  [hep-ph]}}.

\bibitem{AtzoriCorona:2022qrf}
M.~Atzori~Corona, M.~Cadeddu, N.~Cargioli, F.~Dordei, C.~Giunti, Y.~F. Li,
  C.~A. Ternes, and Y.~Y. Zhang, ``{Impact of the Dresden-II and COHERENT
  neutrino scattering data on neutrino electromagnetic properties and
  electroweak physics},'' \href{http://dx.doi.org/10.1007/JHEP09(2022)164}{{\em
  JHEP} {\bfseries 09} (2022) 164},
  \href{http://arxiv.org/abs/2205.09484}{{\ttfamily arXiv:2205.09484
  [hep-ph]}}.

\bibitem{Denton:2022nol}
P.~B. Denton and J.~Gehrlein, ``{New constraints on the dark side of
  non-standard interactions from reactor neutrino scattering data},''
  \href{http://dx.doi.org/10.1103/PhysRevD.106.015022}{{\em Phys. Rev. D}
  {\bfseries 106} no.~1, (2022) 015022},
  \href{http://arxiv.org/abs/2204.09060}{{\ttfamily arXiv:2204.09060
  [hep-ph]}}.

\bibitem{Ackermann:2025obx}
N.~Ackermann {\em et~al.}, ``{First observation of reactor antineutrinos by
  coherent scattering},'' \href{http://arxiv.org/abs/2501.05206}{{\ttfamily
  arXiv:2501.05206 [hep-ex]}}.

\bibitem{CONUS:2024lnu}
{\bfseries CONUS+} Collaboration, N.~Ackermann {\em et~al.},
  ``{CONUS+~Experiment},''
  \href{http://dx.doi.org/10.1140/epjc/s10052-024-13551-6}{{\em Eur. Phys. J.
  C} {\bfseries 84} no.~12, (2024) 1265},
  \href{http://arxiv.org/abs/2407.11912}{{\ttfamily arXiv:2407.11912
  [hep-ex]}}. [Erratum: Eur.Phys.J.C 85, 19 (2025)].

\bibitem{CONUS:2021dwh}
{\bfseries CONUS} Collaboration, H.~Bonet {\em et~al.}, ``{Novel constraints on
  neutrino physics beyond the standard model from the CONUS experiment},''
  \href{http://dx.doi.org/10.1007/JHEP05(2022)085}{{\em JHEP} {\bfseries 05}
  (2022) 085}, \href{http://arxiv.org/abs/2110.02174}{{\ttfamily
  arXiv:2110.02174 [hep-ph]}}.

\bibitem{CONUS:2022qbb}
{\bfseries CONUS} Collaboration, H.~Bonet {\em et~al.}, ``{First upper limits
  on neutrino electromagnetic properties from the CONUS experiment},''
  \href{http://dx.doi.org/10.1140/epjc/s10052-022-10722-1}{{\em Eur. Phys. J.
  C} {\bfseries 82} no.~9, (2022) 813},
  \href{http://arxiv.org/abs/2201.12257}{{\ttfamily arXiv:2201.12257
  [hep-ex]}}.

\bibitem{Lindner:2024eng}
M.~Lindner, T.~Rink, and M.~Sen, ``{Light vector bosons and the weak mixing
  angle in the light of future germanium-based reactor CE\ensuremath{\nu}NS
  experiments},'' \href{http://dx.doi.org/10.1007/JHEP08(2024)171}{{\em JHEP}
  {\bfseries 08} (2024) 171}, \href{http://arxiv.org/abs/2401.13025}{{\ttfamily
  arXiv:2401.13025 [hep-ph]}}.

\bibitem{Alpizar-Venegas:2025wor}
M.~Alp\'\i{}zar-Venegas, L.~J. Flores, E.~Peinado, and E.~V\'azquez-J\'auregui,
  ``{Exploring the Standard Model and Beyond from the Evidence of CE$\nu$NS
  with Reactor Antineutrinos in CONUS+},''
  \href{http://arxiv.org/abs/2501.10355}{{\ttfamily arXiv:2501.10355
  [hep-ph]}}.

\bibitem{Chattaraj:2025fvx}
A.~Chattaraj, A.~Majumdar, and R.~Srivastava, ``{Probing Standard Model and
  Beyond with Reactor CE$\nu$NS Data of CONUS+ experiment},''
  \href{http://arxiv.org/abs/2501.12441}{{\ttfamily arXiv:2501.12441
  [hep-ph]}}.

\bibitem{AtzoriCorona:2025ygn}
M.~Atzori~Corona, M.~Cadeddu, N.~Cargioli, F.~Dordei, and C.~Giunti, ``{Reactor
  antineutrinos CE$\nu$NS on germanium: CONUS+ and TEXONO as a new gateway to
  SM and BSM physics},'' \href{http://arxiv.org/abs/2501.18550}{{\ttfamily
  arXiv:2501.18550 [hep-ph]}}.

\bibitem{Barranco:2005yy}
J.~Barranco, O.~G. Miranda, and T.~I. Rashba, ``{Probing new physics with
  coherent neutrino scattering off nuclei},''
  \href{http://dx.doi.org/10.1088/1126-6708/2005/12/021}{{\em JHEP} {\bfseries
  12} (2005) 021}, \href{http://arxiv.org/abs/hep-ph/0508299}{{\ttfamily
  arXiv:hep-ph/0508299}}.

\bibitem{Cadeddu:2020lky}
M.~Cadeddu, F.~Dordei, C.~Giunti, Y.~F. Li, E.~Picciau, and Y.~Y. Zhang,
  ``{Physics results from the first COHERENT observation of coherent elastic
  neutrino-nucleus scattering in argon and their combination with cesium-iodide
  data},'' \href{http://dx.doi.org/10.1103/PhysRevD.102.015030}{{\em Phys. Rev.
  D} {\bfseries 102} no.~1, (2020) 015030},
  \href{http://arxiv.org/abs/2005.01645}{{\ttfamily arXiv:2005.01645
  [hep-ph]}}.

\bibitem{ParticleDataGroup:2024cfk}
{\bfseries Particle Data Group} Collaboration, S.~Navas {\em et~al.}, ``{Review
  of particle physics},''
  \href{http://dx.doi.org/10.1103/PhysRevD.110.030001}{{\em Phys. Rev. D}
  {\bfseries 110} no.~3, (2024) 030001}.

\bibitem{Klein:1999qj}
S.~Klein and J.~Nystrand, ``{Exclusive vector meson production in relativistic
  heavy ion collisions},''
  \href{http://dx.doi.org/10.1103/PhysRevC.60.014903}{{\em Phys. Rev. C}
  {\bfseries 60} (1999) 014903},
  \href{http://arxiv.org/abs/hep-ph/9902259}{{\ttfamily arXiv:hep-ph/9902259}}.

\bibitem{Giunti:2014ixa}
C.~Giunti and A.~Studenikin, ``{Neutrino electromagnetic interactions: a window
  to new physics},'' \href{http://dx.doi.org/10.1103/RevModPhys.87.531}{{\em
  Rev. Mod. Phys.} {\bfseries 87} (2015) 531},
  \href{http://arxiv.org/abs/1403.6344}{{\ttfamily arXiv:1403.6344 [hep-ph]}}.

\bibitem{AristizabalSierra:2021fuc}
D.~Aristizabal~Sierra, O.~G. Miranda, D.~K. Papoulias, and G.~S. Garcia,
  ``{Neutrino magnetic and electric dipole moments: From measurements to
  parameter space},'' \href{http://dx.doi.org/10.1103/PhysRevD.105.035027}{{\em
  Phys. Rev. D} {\bfseries 105} no.~3, (2022) 035027},
  \href{http://arxiv.org/abs/2112.12817}{{\ttfamily arXiv:2112.12817
  [hep-ph]}}.

\bibitem{Vogel:1989iv}
P.~Vogel and J.~Engel, ``{Neutrino Electromagnetic Form-Factors},''
  \href{http://dx.doi.org/10.1103/PhysRevD.39.3378}{{\em Phys. Rev. D}
  {\bfseries 39} (1989) 3378}.

\bibitem{Ohlsson:2012kf}
T.~Ohlsson, ``{Status of non-standard neutrino interactions},''
  \href{http://dx.doi.org/10.1088/0034-4885/76/4/044201}{{\em Rept. Prog.
  Phys.} {\bfseries 76} (2013) 044201},
  \href{http://arxiv.org/abs/1209.2710}{{\ttfamily arXiv:1209.2710 [hep-ph]}}.

\bibitem{Miranda:2015dra}
O.~G. Miranda and H.~Nunokawa, ``{Non standard neutrino interactions: current
  status and future prospects},''
  \href{http://dx.doi.org/10.1088/1367-2630/17/9/095002}{{\em New J. Phys.}
  {\bfseries 17} no.~9, (2015) },
  \href{http://arxiv.org/abs/1505.06254}{{\ttfamily arXiv:1505.06254
  [hep-ph]}}.

\bibitem{Farzan:2017xzy}
Y.~Farzan and M.~Tortola, ``{Neutrino oscillations and Non-Standard
  Interactions},'' \href{http://dx.doi.org/10.3389/fphy.2018.00010}{{\em Front.
  in Phys.} {\bfseries 6} (2018) 10},
  \href{http://arxiv.org/abs/1710.09360}{{\ttfamily arXiv:1710.09360
  [hep-ph]}}.

\bibitem{Schechter:1980gr}
J.~Schechter and J.~W.~F. Valle, ``{Neutrino Masses in SU(2) x U(1)
  Theories},'' \href{http://dx.doi.org/10.1103/PhysRevD.22.2227}{{\em
  Phys.Rev.D} {\bfseries 22} (1980) 2227}.

\bibitem{Valle:1987gv}
J.~Valle, ``{Resonant Oscillations of Massless Neutrinos in Matter},''
  \href{http://dx.doi.org/10.1016/0370-2693(87)90947-6}{{\em Phys.Lett.B}
  {\bfseries 199} (1987) 432--436}.

\bibitem{Giunti:2019xpr}
C.~Giunti, ``{General COHERENT constraints on neutrino nonstandard
  interactions},'' \href{http://dx.doi.org/10.1103/PhysRevD.101.035039}{{\em
  Phys. Rev. D} {\bfseries 101} no.~3, (2020) 035039},
  \href{http://arxiv.org/abs/1909.00466}{{\ttfamily arXiv:1909.00466
  [hep-ph]}}.

\bibitem{Lee:1956qn}
T.~D. Lee and C.-N. Yang, ``{Question of Parity Conservation in Weak
  Interactions},'' \href{http://dx.doi.org/10.1103/PhysRev.104.254}{{\em Phys.
  Rev.} {\bfseries 104} (1956) 254--258}.

\bibitem{Lindner:2016wff}
M.~Lindner, W.~Rodejohann, and X.-J. Xu, ``{Coherent Neutrino-Nucleus
  Scattering and new Neutrino Interactions},''
  \href{http://dx.doi.org/10.1007/JHEP03(2017)097}{{\em JHEP} {\bfseries 03}
  (2017) 097}, \href{http://arxiv.org/abs/1612.04150}{{\ttfamily
  arXiv:1612.04150 [hep-ph]}}.

\bibitem{AristizabalSierra:2018eqm}
D.~Aristizabal~Sierra, V.~De~Romeri, and N.~Rojas, ``{COHERENT analysis of
  neutrino generalized interactions},''
  \href{http://dx.doi.org/10.1103/PhysRevD.98.075018}{{\em Phys. Rev. D}
  {\bfseries 98} (2018) 075018},
  \href{http://arxiv.org/abs/1806.07424}{{\ttfamily arXiv:1806.07424
  [hep-ph]}}.

\bibitem{Cirelli:2013ufw}
M.~Cirelli, E.~Del~Nobile, and P.~Panci, ``{Tools for model-independent bounds
  in direct dark matter searches},''
  \href{http://dx.doi.org/10.1088/1475-7516/2013/10/019}{{\em JCAP} {\bfseries
  10} (2013) 019}, \href{http://arxiv.org/abs/1307.5955}{{\ttfamily
  arXiv:1307.5955 [hep-ph]}}.

\bibitem{DelNobile:2021wmp}
E.~Del~Nobile, ``{The Theory of Direct Dark Matter Detection: A Guide to
  Computations},'' \href{http://arxiv.org/abs/2104.12785}{{\ttfamily
  arXiv:2104.12785 [hep-ph]}}.

\bibitem{Langacker:2008yv}
P.~Langacker, ``{The Physics of Heavy $Z^\prime$ Gauge Bosons},''
  \href{http://dx.doi.org/10.1103/RevModPhys.81.1199}{{\em Rev. Mod. Phys.}
  {\bfseries 81} (2009) 1199--1228},
  \href{http://arxiv.org/abs/0801.1345}{{\ttfamily arXiv:0801.1345 [hep-ph]}}.

\bibitem{Okada:2018ktp}
S.~Okada, ``{$Z'$ Portal Dark Matter in the Minimal $B-L$ Model},''
  \href{http://dx.doi.org/10.1155/2018/5340935}{{\em Adv. High Energy Phys.}
  {\bfseries 2018} (2018) 5340935},
  \href{http://arxiv.org/abs/1803.06793}{{\ttfamily arXiv:1803.06793
  [hep-ph]}}.

\bibitem{Link:2019pbm}
J.~M. Link and X.-J. Xu, ``{Searching for BSM neutrino interactions in dark
  matter detectors},'' \href{http://dx.doi.org/10.1007/JHEP08(2019)004}{{\em
  JHEP} {\bfseries 08} (2019) 004},
  \href{http://arxiv.org/abs/1903.09891}{{\ttfamily arXiv:1903.09891
  [hep-ph]}}.

\bibitem{Abdullahi:2022jlv}
A.~M. Abdullahi {\em et~al.}, ``{The present and future status of heavy neutral
  leptons},'' \href{http://dx.doi.org/10.1088/1361-6471/ac98f9}{{\em J. Phys.
  G} {\bfseries 50} no.~2, (2023) 020501},
  \href{http://arxiv.org/abs/2203.08039}{{\ttfamily arXiv:2203.08039
  [hep-ph]}}.

\bibitem{Minkowski:1977sc}
P.~Minkowski, ``{$\mu \to e\gamma$ at a Rate of One Out of $10^{9}$ Muon
  Decays?},''
\href{http://dx.doi.org/10.1016/0370-2693(77)90435-X}{{\em Phys. Lett.}
  {\bfseries 67B} (1977) 421--428}.

\bibitem{Yanagida:1979as}
T.~Yanagida, ``{Horizontal gauge symmetry and masses of neutrinos},'' {\em
  Conf. Proc. C} {\bfseries 7902131} (1979) 95--99.

\bibitem{GellMann:1980vs}
M.~Gell-Mann, P.~Ramond, and R.~Slansky, ``{Complex Spinors and Unified
  Theories},'' {\em Conf. Proc.} {\bfseries C790927} (1979) 315--321,
  \href{http://arxiv.org/abs/1306.4669}{{\ttfamily arXiv:1306.4669 [hep-th]}}.

\bibitem{Mohapatra:1979ia}
R.~N. Mohapatra and G.~Senjanovic, ``{Neutrino Mass and Spontaneous Parity
  Nonconservation},'' \href{http://dx.doi.org/10.1103/PhysRevLett.44.912}{{\em
  Phys. Rev. Lett.} {\bfseries 44} (1980) 912}.

\bibitem{McKeen:2010rx}
D.~McKeen and M.~Pospelov, ``{Muon Capture Constraints on Sterile Neutrino
  Properties},'' \href{http://dx.doi.org/10.1103/PhysRevD.82.113018}{{\em Phys.
  Rev. D} {\bfseries 82} (2010) 113018},
  \href{http://arxiv.org/abs/1011.3046}{{\ttfamily arXiv:1011.3046 [hep-ph]}}.

\bibitem{Magill:2018jla}
G.~Magill, R.~Plestid, M.~Pospelov, and Y.-D. Tsai, ``{Dipole Portal to Heavy
  Neutral Leptons},'' \href{http://dx.doi.org/10.1103/PhysRevD.98.115015}{{\em
  Phys. Rev. D} {\bfseries 98} no.~11, (2018) 115015},
  \href{http://arxiv.org/abs/1803.03262}{{\ttfamily arXiv:1803.03262
  [hep-ph]}}.

\bibitem{Gninenko:1998nn}
S.~N. Gninenko and N.~V. Krasnikov, ``{Limits on the magnetic moment of sterile
  neutrino and two photon neutrino decay},''
  \href{http://dx.doi.org/10.1016/S0370-2693(99)00130-6}{{\em Phys. Lett. B}
  {\bfseries 450} (1999) 165--172},
  \href{http://arxiv.org/abs/hep-ph/9808370}{{\ttfamily arXiv:hep-ph/9808370}}.

\bibitem{Grimus:2000tq}
W.~Grimus and T.~Schwetz, ``{Elastic neutrino electron scattering of solar
  neutrinos and potential effects of magnetic and electric dipole moments},''
  \href{http://dx.doi.org/10.1016/S0550-3213(00)00451-X}{{\em Nucl. Phys. B}
  {\bfseries 587} (2000) 45--66},
  \href{http://arxiv.org/abs/hep-ph/0006028}{{\ttfamily arXiv:hep-ph/0006028}}.

\bibitem{Grimus:1997aa}
W.~Grimus and P.~Stockinger, ``{Effects of neutrino oscillations and neutrino
  magnetic moments on elastic neutrino - electron scattering},''
  \href{http://dx.doi.org/10.1103/PhysRevD.57.1762}{{\em Phys. Rev. D}
  {\bfseries 57} (1998) 1762--1768},
  \href{http://arxiv.org/abs/hep-ph/9708279}{{\ttfamily arXiv:hep-ph/9708279}}.

\bibitem{Miranda:2021kre}
O.~G. Miranda, D.~K. Papoulias, O.~Sanders, M.~T\'ortola, and J.~W.~F. Valle,
  ``{Low-energy probes of sterile neutrino transition magnetic moments},''
  \href{http://dx.doi.org/10.1007/JHEP12(2021)191}{{\em JHEP} {\bfseries 12}
  (2021) 191}, \href{http://arxiv.org/abs/2109.09545}{{\ttfamily
  arXiv:2109.09545 [hep-ph]}}.

\bibitem{Brdar:2018qqj}
V.~Brdar, W.~Rodejohann, and X.-J. Xu, ``{Producing a new Fermion in Coherent
  Elastic Neutrino-Nucleus Scattering: from Neutrino Mass to Dark Matter},''
  \href{http://dx.doi.org/10.1007/JHEP12(2018)024}{{\em JHEP} {\bfseries 12}
  (2018) 024}, \href{http://arxiv.org/abs/1810.03626}{{\ttfamily
  arXiv:1810.03626 [hep-ph]}}.

\bibitem{Chang:2020jwl}
W.-F. Chang and J.~Liao, ``{Constraints on light singlet fermion interactions
  from coherent elastic neutrino-nucleus scattering},''
  \href{http://dx.doi.org/10.1103/PhysRevD.102.075004}{{\em Phys. Rev. D}
  {\bfseries 102} no.~7, (2020) 075004},
  \href{http://arxiv.org/abs/2002.10275}{{\ttfamily arXiv:2002.10275
  [hep-ph]}}.

\bibitem{Chao:2021bvq}
W.~Chao, T.~Li, J.~Liao, and M.~Su, ``{Loop effects with a vector mediator in
  coherent neutrino-nucleus scattering},''
  \href{http://dx.doi.org/10.1103/PhysRevD.104.095017}{{\em Phys. Rev. D}
  {\bfseries 104} no.~9, (2021) 095017},
  \href{http://arxiv.org/abs/2108.02341}{{\ttfamily arXiv:2108.02341
  [hep-ph]}}.

\bibitem{Candela:2023rvt}
P.~M. Candela, V.~De~Romeri, and D.~K. Papoulias, ``{COHERENT production of a
  dark fermion},'' \href{http://dx.doi.org/10.1103/PhysRevD.108.055001}{{\em
  Phys. Rev. D} {\bfseries 108} no.~5, (2023) 055001},
  \href{http://arxiv.org/abs/2305.03341}{{\ttfamily arXiv:2305.03341
  [hep-ph]}}.

\bibitem{Alonso-Gonzalez:2023tgm}
D.~Alonso-Gonz\'alez, D.~W.~P. Amaral, A.~Bariego-Quintana, D.~Cerde\~no, and
  M.~de~los Rios, ``{Measuring the sterile neutrino mass in spallation source
  and direct detection experiments},''
  \href{http://dx.doi.org/10.1007/JHEP12(2023)096}{{\em JHEP} {\bfseries 12}
  (2023) 096}, \href{http://arxiv.org/abs/2307.05176}{{\ttfamily
  arXiv:2307.05176 [hep-ph]}}.

\bibitem{Candela:2024ljb}
P.~M. Candela, V.~De~Romeri, P.~Melas, D.~K. Papoulias, and N.~Saoulidou,
  ``{Up-scattering production of a sterile fermion at DUNE: complementarity
  with spallation source and direct detection experiments},''
  \href{http://dx.doi.org/10.1007/JHEP10(2024)032}{{\em JHEP} {\bfseries 10}
  (2024) 032}, \href{http://arxiv.org/abs/2404.12476}{{\ttfamily
  arXiv:2404.12476 [hep-ph]}}.

\bibitem{Kopeikin:2012zz}
V.~I. Kopeikin, ``{Flux and spectrum of reactor antineutrinos},''
  \href{http://dx.doi.org/10.1134/S1063778812020123}{{\em Phys. Atom. Nucl.}
  {\bfseries 75} (2012) 143--152}.

\bibitem{Mueller:2011nm}
T.~A. Mueller {\em et~al.}, ``{Improved Predictions of Reactor Antineutrino
  Spectra},'' \href{http://dx.doi.org/10.1103/PhysRevC.83.054615}{{\em Phys.
  Rev. C} {\bfseries 83} (2011) 054615},
  \href{http://arxiv.org/abs/1101.2663}{{\ttfamily arXiv:1101.2663 [hep-ex]}}.

\bibitem{osti_4701226}
J.~Lindhard, V.~Nielsen, M.~Scharff, and P.~V. Thomsen, ``Integral equations
  governing radiation effects. (notes on atomic collisions, iii),'' {\em Kgl.
  Danske Videnskab., Selskab. Mat. Fys. Medd.} {\bfseries Vol: 33: No. 10} (01,
  1963) . \url{https://www.osti.gov/biblio/4701226}.

\bibitem{Bonhomme:2022lcz}
A.~Bonhomme {\em et~al.}, ``{Direct measurement of the ionization quenching
  factor of nuclear recoils in germanium in the keV energy range},''
  \href{http://dx.doi.org/10.1140/epjc/s10052-022-10768-1}{{\em Eur. Phys. J.
  C} {\bfseries 82} no.~9, (2022) 815},
  \href{http://arxiv.org/abs/2202.03754}{{\ttfamily arXiv:2202.03754
  [physics.ins-det]}}.

\bibitem{DeRomeri:2022twg}
V.~De~Romeri, O.~G. Miranda, D.~K. Papoulias, G.~Sanchez~Garcia, M.~T\'ortola,
  and J.~W.~F. Valle, ``{Physics implications of a combined analysis of
  COHERENT CsI and LAr data},''
  \href{http://dx.doi.org/10.1007/JHEP04(2023)035}{{\em JHEP} {\bfseries 04}
  (2023) 035}, \href{http://arxiv.org/abs/2211.11905}{{\ttfamily
  arXiv:2211.11905 [hep-ph]}}.

\bibitem{Qweak:2018tjf}
{\bfseries Qweak} Collaboration, D.~Androi\'c {\em et~al.}, ``{Precision
  measurement of the weak charge of the proton},''
  \href{http://dx.doi.org/10.1038/s41586-018-0096-0}{{\em Nature} {\bfseries
  557} no.~7704, (2018) 207--211},
  \href{http://arxiv.org/abs/1905.08283}{{\ttfamily arXiv:1905.08283
  [nucl-ex]}}.

\bibitem{SLACE158:2005uay}
{\bfseries SLAC E158} Collaboration, P.~L. Anthony {\em et~al.}, ``{Precision
  measurement of the weak mixing angle in Moller scattering},''
  \href{http://dx.doi.org/10.1103/PhysRevLett.95.081601}{{\em Phys. Rev. Lett.}
  {\bfseries 95} (2005) 081601},
  \href{http://arxiv.org/abs/hep-ex/0504049}{{\ttfamily arXiv:hep-ex/0504049}}.

\bibitem{PVDIS:2014cmd}
{\bfseries PVDIS} Collaboration, D.~Wang {\em et~al.}, ``{Measurement of parity
  violation in electron\textendash{}quark scattering},''
  \href{http://dx.doi.org/10.1038/nature12964}{{\em Nature} {\bfseries 506}
  no.~7486, (2014) 67--70}.

\bibitem{NuTeV:2001whx}
{\bfseries NuTeV} Collaboration, G.~P. Zeller {\em et~al.}, ``{A Precise
  Determination of Electroweak Parameters in Neutrino Nucleon Scattering},''
  \href{http://dx.doi.org/10.1103/PhysRevLett.88.091802}{{\em Phys. Rev. Lett.}
  {\bfseries 88} (2002) 091802},
  \href{http://arxiv.org/abs/hep-ex/0110059}{{\ttfamily arXiv:hep-ex/0110059}}.
  [Erratum: Phys.Rev.Lett. 90, 239902 (2003)].

\bibitem{AtzoriCorona:2024vhj}
M.~Atzori~Corona, M.~Cadeddu, N.~Cargioli, F.~Dordei, and C.~Giunti, ``{Refined
  determination of the weak mixing angle at low energy},''
  \href{http://dx.doi.org/10.1103/PhysRevD.110.033005}{{\em Phys. Rev. D}
  {\bfseries 110} no.~3, (2024) 033005},
  \href{http://arxiv.org/abs/2405.09416}{{\ttfamily arXiv:2405.09416
  [hep-ph]}}.

\bibitem{Cadeddu:2024baq}
M.~Cadeddu, N.~Cargioli, J.~Erler, M.~Gorchtein, J.~Piekarewicz, X.~Roca-Maza,
  and H.~Spiesberger, ``{Simultaneous extraction of the weak radius and the
  weak mixing angle from parity-violating electron scattering on C12},''
  \href{http://dx.doi.org/10.1103/PhysRevC.110.035501}{{\em Phys. Rev. C}
  {\bfseries 110} no.~3, (2024) 035501},
  \href{http://arxiv.org/abs/2407.09743}{{\ttfamily arXiv:2407.09743
  [hep-ph]}}.

\bibitem{Coloma:2022umy}
P.~Coloma, M.~C. Gonzalez-Garcia, M.~Maltoni, J.~a.~P. Pinheiro, and S.~Urrea,
  ``{Constraining new physics with Borexino Phase-II spectral data},''
  \href{http://dx.doi.org/10.1007/JHEP07(2022)138}{{\em JHEP} {\bfseries 07}
  (2022) 138}, \href{http://arxiv.org/abs/2204.03011}{{\ttfamily
  arXiv:2204.03011 [hep-ph]}}. [Erratum: JHEP 11, 138 (2022)].

\bibitem{TEXONO:2006xds}
{\bfseries TEXONO} Collaboration, H.~T. Wong {\em et~al.}, ``{A Search of
  Neutrino Magnetic Moments with a High-Purity Germanium Detector at the
  Kuo-Sheng Nuclear Power Station},''
  \href{http://dx.doi.org/10.1103/PhysRevD.75.012001}{{\em Phys. Rev. D}
  {\bfseries 75} (2007) 012001},
  \href{http://arxiv.org/abs/hep-ex/0605006}{{\ttfamily arXiv:hep-ex/0605006}}.

\bibitem{Beda:2012zz}
A.~G. Beda, V.~B. Brudanin, V.~G. Egorov, D.~V. Medvedev, V.~S. Pogosov, M.~V.
  Shirchenko, and A.~S. Starostin, ``{The results of search for the neutrino
  magnetic moment in GEMMA experiment},''
  \href{http://dx.doi.org/10.1155/2012/350150}{{\em Adv. High Energy Phys.}
  {\bfseries 2012} (2012) 350150}.

\bibitem{A:2022acy}
S.~K. A., A.~Majumdar, D.~K. Papoulias, H.~Prajapati, and R.~Srivastava,
  ``{Implications of first LZ and XENONnT results: A comparative study of
  neutrino properties and light mediators},''
  \href{http://dx.doi.org/10.1016/j.physletb.2023.137742}{{\em Phys. Lett. B}
  {\bfseries 839} (2023) 137742},
  \href{http://arxiv.org/abs/2208.06415}{{\ttfamily arXiv:2208.06415
  [hep-ph]}}.

\bibitem{Giunti:2023yha}
C.~Giunti and C.~A. Ternes, ``{Testing neutrino electromagnetic properties at
  current and future dark matter experiments},''
  \href{http://dx.doi.org/10.1103/PhysRevD.108.095044}{{\em Phys. Rev. D}
  {\bfseries 108} no.~9, (2023) 095044},
  \href{http://arxiv.org/abs/2309.17380}{{\ttfamily arXiv:2309.17380
  [hep-ph]}}.

\bibitem{Miranda:2019wdy}
O.~G. Miranda, D.~K. Papoulias, M.~T\'ortola, and J.~W.~F. Valle, ``{Probing
  neutrino transition magnetic moments with coherent elastic neutrino-nucleus
  scattering},'' \href{http://dx.doi.org/10.1007/JHEP07(2019)103}{{\em JHEP}
  {\bfseries 07} (2019) 103}, \href{http://arxiv.org/abs/1905.03750}{{\ttfamily
  arXiv:1905.03750 [hep-ph]}}.

\bibitem{AtzoriCorona:2022moj}
M.~Atzori~Corona, M.~Cadeddu, N.~Cargioli, F.~Dordei, C.~Giunti, Y.~F. Li,
  E.~Picciau, C.~A. Ternes, and Y.~Y. Zhang, ``{Probing light mediators and (g
  \ensuremath{-} 2)$_{μ}$ through detection of coherent elastic neutrino
  nucleus scattering at COHERENT},''
  \href{http://dx.doi.org/10.1007/JHEP05(2022)109}{{\em JHEP} {\bfseries 05}
  (2022) 109}, \href{http://arxiv.org/abs/2202.11002}{{\ttfamily
  arXiv:2202.11002 [hep-ph]}}.

\bibitem{Denton:2018xmq}
P.~B. Denton, Y.~Farzan, and I.~M. Shoemaker, ``{Testing large non-standard
  neutrino interactions with arbitrary mediator mass after COHERENT data},''
  \href{http://dx.doi.org/10.1007/JHEP07(2018)037}{{\em JHEP} {\bfseries 07}
  (2018) 037}, \href{http://arxiv.org/abs/1804.03660}{{\ttfamily
  arXiv:1804.03660 [hep-ph]}}.

\bibitem{CONNIE:2019xid}
{\bfseries CONNIE} Collaboration, A.~Aguilar-Arevalo {\em et~al.}, ``{Search
  for light mediators in the low-energy data of the CONNIE reactor neutrino
  experiment},'' \href{http://dx.doi.org/10.1007/JHEP04(2020)054}{{\em JHEP}
  {\bfseries 04} (2020) 054}, \href{http://arxiv.org/abs/1910.04951}{{\ttfamily
  arXiv:1910.04951 [hep-ex]}}.

\bibitem{CONNIE:2024pwt}
{\bfseries CONNIE} Collaboration, A.~A. Aguilar-Arevalo {\em et~al.},
  ``{Searches for CE\ensuremath{\nu}NS and Physics beyond the Standard Model
  using Skipper-CCDs at CONNIE},''
  \href{http://arxiv.org/abs/2403.15976}{{\ttfamily arXiv:2403.15976
  [hep-ex]}}.

\bibitem{Bauer:2018onh}
M.~Bauer, P.~Foldenauer, and J.~Jaeckel, ``{Hunting All the Hidden Photons},''
  \href{http://dx.doi.org/10.1007/JHEP07(2018)094}{{\em JHEP} {\bfseries 07}
  (2018) 094}, \href{http://arxiv.org/abs/1803.05466}{{\ttfamily
  arXiv:1803.05466 [hep-ph]}}.

\bibitem{TEXONO:2009knm}
{\bfseries TEXONO} Collaboration, M.~Deniz {\em et~al.}, ``{Measurement of
  Nu(e)-bar -Electron Scattering Cross-Section with a CsI(Tl) Scintillating
  Crystal Array at the Kuo-Sheng Nuclear Power Reactor},''
  \href{http://dx.doi.org/10.1103/PhysRevD.81.072001}{{\em Phys. Rev. D}
  {\bfseries 81} (2010) 072001},
  \href{http://arxiv.org/abs/0911.1597}{{\ttfamily arXiv:0911.1597 [hep-ex]}}.

\bibitem{DeRomeri:2024dbv}
V.~De~Romeri, D.~K. Papoulias, and C.~A. Ternes, ``{Light vector mediators at
  direct detection experiments},''
  \href{http://dx.doi.org/10.1007/JHEP05(2024)165}{{\em JHEP} {\bfseries 05}
  (2024) 165}, \href{http://arxiv.org/abs/2402.05506}{{\ttfamily
  arXiv:2402.05506 [hep-ph]}}.

\bibitem{NA64:2021xzo}
{\bfseries NA64} Collaboration, Y.~M. Andreev {\em et~al.}, ``{Constraints on
  New Physics in Electron $g-2$ from a Search for Invisible Decays of a Scalar,
  Pseudoscalar, Vector, and Axial Vector},''
  \href{http://dx.doi.org/10.1103/PhysRevLett.126.211802}{{\em Phys. Rev.
  Lett.} {\bfseries 126} no.~21, (2021) 211802},
  \href{http://arxiv.org/abs/2102.01885}{{\ttfamily arXiv:2102.01885
  [hep-ex]}}.

\bibitem{NA64:2022yly}
{\bfseries NA64} Collaboration, Y.~M. Andreev {\em et~al.}, ``{Search for a New
  B-L Z' Gauge Boson with the NA64 Experiment at CERN},''
  \href{http://dx.doi.org/10.1103/PhysRevLett.129.161801}{{\em Phys. Rev.
  Lett.} {\bfseries 129} no.~16, (2022) 161801},
  \href{http://arxiv.org/abs/2207.09979}{{\ttfamily arXiv:2207.09979
  [hep-ex]}}.

\bibitem{NA64:2023wbi}
{\bfseries NA64} Collaboration, Y.~M. Andreev {\em et~al.}, ``{Search for Light
  Dark Matter with NA64 at CERN},''
  \href{http://dx.doi.org/10.1103/PhysRevLett.131.161801}{{\em Phys. Rev.
  Lett.} {\bfseries 131} no.~16, (2023) 161801},
  \href{http://arxiv.org/abs/2307.02404}{{\ttfamily arXiv:2307.02404
  [hep-ex]}}.

\bibitem{Bjorken:1988as}
J.~D. Bjorken, S.~Ecklund, W.~R. Nelson, A.~Abashian, C.~Church, B.~Lu, L.~W.
  Mo, T.~A. Nunamaker, and P.~Rassmann, ``{Search for Neutral Metastable
  Penetrating Particles Produced in the SLAC Beam Dump},''
  \href{http://dx.doi.org/10.1103/PhysRevD.38.3375}{{\em Phys. Rev. D}
  {\bfseries 38} (1988) 3375}.

\bibitem{Riordan:1987aw}
E.~M. Riordan {\em et~al.}, ``{A Search for Short Lived Axions in an Electron
  Beam Dump Experiment},''
  \href{http://dx.doi.org/10.1103/PhysRevLett.59.755}{{\em Phys. Rev. Lett.}
  {\bfseries 59} (1987) 755}.

\bibitem{Konaka:1986cb}
A.~Konaka {\em et~al.}, ``{Search for Neutral Particles in Electron Beam Dump
  Experiment},'' \href{http://dx.doi.org/10.1103/PhysRevLett.57.659}{{\em Phys.
  Rev. Lett.} {\bfseries 57} (1986) 659}.

\bibitem{Bross:1989mp}
A.~Bross, M.~Crisler, S.~H. Pordes, J.~Volk, S.~Errede, and J.~Wrbanek, ``{A
  Search for Shortlived Particles Produced in an Electron Beam Dump},''
  \href{http://dx.doi.org/10.1103/PhysRevLett.67.2942}{{\em Phys. Rev. Lett.}
  {\bfseries 67} (1991) 2942--2945}.

\bibitem{Davier:1989wz}
M.~Davier and H.~Nguyen~Ngoc, ``{An Unambiguous Search for a Light Higgs
  Boson},'' \href{http://dx.doi.org/10.1016/0370-2693(89)90174-3}{{\em Phys.
  Lett. B} {\bfseries 229} (1989) 150--155}.

\bibitem{Bjorken:2009mm}
J.~D. Bjorken, R.~Essig, P.~Schuster, and N.~Toro, ``{New Fixed-Target
  Experiments to Search for Dark Gauge Forces},''
  \href{http://dx.doi.org/10.1103/PhysRevD.80.075018}{{\em Phys. Rev. D}
  {\bfseries 80} (2009) 075018},
  \href{http://arxiv.org/abs/0906.0580}{{\ttfamily arXiv:0906.0580 [hep-ph]}}.

\bibitem{Andreas:2012mt}
S.~Andreas, C.~Niebuhr, and A.~Ringwald, ``{New Limits on Hidden Photons from
  Past Electron Beam Dumps},''
  \href{http://dx.doi.org/10.1103/PhysRevD.86.095019}{{\em Phys. Rev. D}
  {\bfseries 86} (2012) 095019},
  \href{http://arxiv.org/abs/1209.6083}{{\ttfamily arXiv:1209.6083 [hep-ph]}}.

\bibitem{Blumlein:1990ay}
J.~Blumlein {\em et~al.}, ``{Limits on neutral light scalar and pseudoscalar
  particles in a proton beam dump experiment},''
  \href{http://dx.doi.org/10.1007/BF01548556}{{\em Z. Phys. C} {\bfseries 51}
  (1991) 341--350}.

\bibitem{Blumlein:1991xh}
J.~Blumlein {\em et~al.}, ``{Limits on the mass of light (pseudo)scalar
  particles from Bethe-Heitler e+ e- and mu+ mu- pair production in a proton -
  iron beam dump experiment},''
  \href{http://dx.doi.org/10.1142/S0217751X9200171X}{{\em Int. J. Mod. Phys. A}
  {\bfseries 7} (1992) 3835--3850}.

\bibitem{Blumlein:2011mv}
J.~Blumlein and J.~Brunner, ``{New Exclusion Limits for Dark Gauge Forces from
  Beam-Dump Data},''
  \href{http://dx.doi.org/10.1016/j.physletb.2011.05.046}{{\em Phys. Lett. B}
  {\bfseries 701} (2011) 155--159},
  \href{http://arxiv.org/abs/1104.2747}{{\ttfamily arXiv:1104.2747 [hep-ex]}}.

\bibitem{Blumlein:2013cua}
J.~Bl\"umlein and J.~Brunner, ``{New Exclusion Limits on Dark Gauge Forces from
  Proton Bremsstrahlung in Beam-Dump Data},''
  \href{http://dx.doi.org/10.1016/j.physletb.2014.02.029}{{\em Phys. Lett. B}
  {\bfseries 731} (2014) 320--326},
  \href{http://arxiv.org/abs/1311.3870}{{\ttfamily arXiv:1311.3870 [hep-ph]}}.

\bibitem{CHARM:1985anb}
{\bfseries CHARM} Collaboration, F.~Bergsma {\em et~al.}, ``{Search for Axion
  Like Particle Production in 400-{GeV} Proton - Copper Interactions},''
  \href{http://dx.doi.org/10.1016/0370-2693(85)90400-9}{{\em Phys. Lett. B}
  {\bfseries 157} (1985) 458--462}.

\bibitem{Gninenko:2012eq}
S.~N. Gninenko, ``{Constraints on sub-GeV hidden sector gauge bosons from a
  search for heavy neutrino decays},''
  \href{http://dx.doi.org/10.1016/j.physletb.2012.06.002}{{\em Phys. Lett. B}
  {\bfseries 713} (2012) 244--248},
  \href{http://arxiv.org/abs/1204.3583}{{\ttfamily arXiv:1204.3583 [hep-ph]}}.

\bibitem{NOMAD:2001eyx}
{\bfseries NOMAD} Collaboration, P.~Astier {\em et~al.}, ``{Search for heavy
  neutrinos mixing with tau neutrinos},''
  \href{http://dx.doi.org/10.1016/S0370-2693(01)00362-8}{{\em Phys. Lett. B}
  {\bfseries 506} (2001) 27--38},
  \href{http://arxiv.org/abs/hep-ex/0101041}{{\ttfamily arXiv:hep-ex/0101041}}.

\bibitem{Bernardi:1985ny}
G.~Bernardi {\em et~al.}, ``{Search for Neutrino Decay},''
  \href{http://dx.doi.org/10.1016/0370-2693(86)91602-3}{{\em Phys. Lett. B}
  {\bfseries 166} (1986) 479--483}.

\bibitem{Gninenko:2011uv}
S.~N. Gninenko, ``{Stringent limits on the $\pi^0 \to \gamma X, X \to e+e-$
  decay from neutrino experiments and constraints on new light gauge bosons},''
  \href{http://dx.doi.org/10.1103/PhysRevD.85.055027}{{\em Phys. Rev. D}
  {\bfseries 85} (2012) 055027},
  \href{http://arxiv.org/abs/1112.5438}{{\ttfamily arXiv:1112.5438 [hep-ph]}}.

\bibitem{Merkel:2014avp}
H.~Merkel {\em et~al.}, ``{Search at the Mainz Microtron for Light Massive
  Gauge Bosons Relevant for the Muon g-2 Anomaly},''
  \href{http://dx.doi.org/10.1103/PhysRevLett.112.221802}{{\em Phys. Rev.
  Lett.} {\bfseries 112} no.~22, (2014) 221802},
  \href{http://arxiv.org/abs/1404.5502}{{\ttfamily arXiv:1404.5502 [hep-ex]}}.

\bibitem{APEX:2011dww}
{\bfseries APEX} Collaboration, S.~Abrahamyan {\em et~al.}, ``{Search for a New
  Gauge Boson in Electron-Nucleus Fixed-Target Scattering by the APEX
  Experiment},'' \href{http://dx.doi.org/10.1103/PhysRevLett.107.191804}{{\em
  Phys. Rev. Lett.} {\bfseries 107} (2011) 191804},
  \href{http://arxiv.org/abs/1108.2750}{{\ttfamily arXiv:1108.2750 [hep-ex]}}.

\bibitem{BaBar:2014zli}
{\bfseries BaBar} Collaboration, J.~P. Lees {\em et~al.}, ``{Search for a Dark
  Photon in $e^+e^-$ Collisions at BaBar},''
  \href{http://dx.doi.org/10.1103/PhysRevLett.113.201801}{{\em Phys. Rev.
  Lett.} {\bfseries 113} no.~20, (2014) 201801},
  \href{http://arxiv.org/abs/1406.2980}{{\ttfamily arXiv:1406.2980 [hep-ex]}}.

\bibitem{BaBar:2017tiz}
{\bfseries BaBar} Collaboration, J.~P. Lees {\em et~al.}, ``{Search for
  Invisible Decays of a Dark Photon Produced in ${e}^{+}{e}^{-}$ Collisions at
  BaBar},'' \href{http://dx.doi.org/10.1103/PhysRevLett.119.131804}{{\em Phys.
  Rev. Lett.} {\bfseries 119} no.~13, (2017) 131804},
  \href{http://arxiv.org/abs/1702.03327}{{\ttfamily arXiv:1702.03327
  [hep-ex]}}.

\bibitem{LHCb:2017trq}
{\bfseries LHCb} Collaboration, R.~Aaij {\em et~al.}, ``{Search for Dark
  Photons Produced in 13 TeV $pp$ Collisions},''
  \href{http://dx.doi.org/10.1103/PhysRevLett.120.061801}{{\em Phys. Rev.
  Lett.} {\bfseries 120} no.~6, (2018) 061801},
  \href{http://arxiv.org/abs/1710.02867}{{\ttfamily arXiv:1710.02867
  [hep-ex]}}.

\bibitem{Giunti:2022btk}
C.~Giunti, Y.~F. Li, C.~A. Ternes, O.~Tyagi, and Z.~Xin, ``{Gallium Anomaly:
  critical view from the global picture of \ensuremath{\nu}$_{e}$ and $
  {\overline{\nu}}_e $ disappearance},''
  \href{http://dx.doi.org/10.1007/JHEP10(2022)164}{{\em JHEP} {\bfseries 10}
  (2022) 164}, \href{http://arxiv.org/abs/2209.00916}{{\ttfamily
  arXiv:2209.00916 [hep-ph]}}.

\bibitem{Brdar:2020quo}
V.~Brdar, A.~Greljo, J.~Kopp, and T.~Opferkuch, ``{The Neutrino Magnetic Moment
  Portal: Cosmology, Astrophysics, and Direct Detection},''
  \href{http://dx.doi.org/10.1088/1475-7516/2021/01/039}{{\em JCAP} {\bfseries
  01} (2021) 039}, \href{http://arxiv.org/abs/2007.15563}{{\ttfamily
  arXiv:2007.15563 [hep-ph]}}.

\bibitem{Plestid:2020vqf}
R.~Plestid, ``{Luminous solar neutrinos I: Dipole portals},''
  \href{http://dx.doi.org/10.1103/PhysRevD.104.075027}{{\em Phys. Rev. D}
  {\bfseries 104} (2021) 075027},
  \href{http://arxiv.org/abs/2010.04193}{{\ttfamily arXiv:2010.04193
  [hep-ph]}}.

\bibitem{Plestidlumsolnu}
R.~Plestid, ``Luminous solar neutrinos: The notebooks.''
  \url{https://github.com/ryanplestid/luminous-solar-nu}, 2020.

\bibitem{Gustafson:2022rsz}
R.~A. Gustafson, R.~Plestid, and I.~M. Shoemaker, ``{Neutrino portals,
  terrestrial upscattering, and atmospheric neutrinos},''
  \href{http://dx.doi.org/10.1103/PhysRevD.106.095037}{{\em Phys. Rev. D}
  {\bfseries 106} no.~9, (2022) 095037},
  \href{http://arxiv.org/abs/2205.02234}{{\ttfamily arXiv:2205.02234
  [hep-ph]}}.

\bibitem{Chauhan:2024nfa}
G.~Chauhan, S.~Horiuchi, P.~Huber, and I.~M. Shoemaker, ``{Probing the sterile
  neutrino dipole portal with SN1987A and low-energy supernovae},''
  \href{http://dx.doi.org/10.1103/PhysRevD.110.015007}{{\em Phys. Rev. D}
  {\bfseries 110} no.~1, (2024) 015007},
  \href{http://arxiv.org/abs/2402.01624}{{\ttfamily arXiv:2402.01624
  [hep-ph]}}.

\end{thebibliography}\endgroup

\end{document}